\documentclass[prd,superscriptaddress,preprint,aps,amsmath,amssymb,showpacs,nofootinbib]{revtex4-2}
\usepackage{graphicx,enumerate,xcolor,enumitem,xspace,braket,comment,multirow,booktabs}
\usepackage[normalem]{ulem}
\usepackage[colorlinks=true,citecolor=blue]{hyperref}

\newcommand{\tr}{\operatorname{Tr}}
\begin{document}

\title{Quantum Tomography of Fermion Pairs \\ in $e^+e^-$ Collisions: Longitudinal Beam Polarization Effects}

\author{Yu-Chen Guo}
\email{ycguo@lnnu.edu.cn}
\affiliation{ Center for Theoretical and Experimental High Energy Physics, Department of Physics,
Liaoning Normal University, Dalian 116029, China}
\affiliation{PITT PACC, Department of Physics and Astronomy,\\ University of Pittsburgh, 3941 O’Hara St., Pittsburgh, PA 15260, USA}

\author{Tao Han}
\email{than@pitt.edu}
\affiliation{PITT PACC, Department of Physics and Astronomy,\\ University of Pittsburgh, 3941 O’Hara St., Pittsburgh, PA 15260, USA}

\author{Matthew Low}
\email{matthew.w.low@gmail.com}
\affiliation{PITT PACC, Department of Physics and Astronomy,\\ University of Pittsburgh, 3941 O’Hara St., Pittsburgh, PA 15260, USA}

\author{Youle Su}
\email{yos76@pitt.edu}
\affiliation{PITT PACC, Department of Physics and Astronomy,\\ University of Pittsburgh, 3941 O’Hara St., Pittsburgh, PA 15260, USA}

\date{\today}
{\hfill PITT-PACC-2601}

\begin{abstract}
We present a quantum tomography study of fermion pair production at future $e^+e^-$ colliders, emphasizing how longitudinal beam polarization controls the two-qubit spin density matrix.
We study the processes $e^+ e^- \to t\bar{t},\ e^+e^-\to \mu^+\mu^-$ and Bhabha scattering $e^+e^-\to e^+e^-$, representing the mass threshold behavior, the $Z$ pole resonance and the $s/t$-channel interplay. We choose to focus on three key concepts: quantum entanglement via the concurrence $\mathcal{C}$, Bell nonlocality via the optimal Clauser Horne Shimony Holt (CHSH) parameter $\mathcal{B}$,
and non-stabilizerness (``magic'') via the second stabilizer R\'enyi entropy $\mathcal{M}_2$. For the $s$-channel-dominated channels, longitudinal polarization mainly reshapes single-spin polarizations while leaving the spin-correlation matrix largely unchanged, rendering $\mathcal{C}$ and $\mathcal{B}$ comparatively robust, but inducing a pronounced variation of $\mathcal{M}_2$.
In contrast, in Bhabha scattering, polarization modifies the relative contributions of the $s$-channel and $t$-channel and can strongly affect all three observables.
The observability of entanglement, Bell nonlocality, and magic exceeds the $5\sigma$ level when both statistical and systematic uncertainties are included, establishing the fermion pair systems as ideal laboratories for quantum-information studies in high energy leptonic collisions.
With optimized beam polarization, future $e^+e^-$ colliders will provide a unique opportunity to experimentally explore and influence quantum resources in particle interactions.
\end{abstract}
\maketitle
\newpage
{
\hypersetup{linkcolor=blue}
\tableofcontents
}
%%%%%%%%%%%%%%%%%%%%%%%%%%%%%%%%%%%%%%%%%%%%%%%%%%%%%%%%%%%%%%%%%%%%%%%%%%%%%%%%%%%%%%%%%%%%%%%%%%%%%%
\section{Introduction}
\label{sec:intro}

One of the most striking features of quantum mechanics is the existence of correlations between particles that defy the classical explanation. Phenomena such as quantum entanglement and Bell nonlocality have been extensively demonstrated in table-top experiments with photons, ions, and atoms, but their study in the high energy regime, where the fundamental constituents of matter are produced, is only beginning \cite{Afik:2025ejh,Barr:2024djo}. This regime presents unique challenges and opportunities, as entanglement is generated via Standard Model (SM) interactions, and is governed by relativistic kinematics and chiral couplings. Particle colliders such as the Large Hadron Collider (LHC) provide a unique opportunity to probe quantum mechanical principles at the shortest distance scales, the highest energies, and in environments far removed from the low-energy laboratory settings where quantum foundations are typically explored. In turn, there is vast quantum information (QI) to uncover.

The first explicit collider proposal to measure spin entanglement focused on $t\bar t$ production, exploiting leptonic top quark decays as spin analyzers~\cite{Afik:2020onf}. Soon after, entanglement measured by ATLAS~\cite{ATLAS:2023fsd} and CMS~\cite{CMS:2024pts}, with further improvements expected in the semileptonic channel~\cite{Dong:2023xiw,Han:2023fci,CMS:2024zkc}.
In general, the $t\bar{t}$ system is an interesting system for investigating unique aspects of entanglement in high energy collisions~\cite{Aguilar-Saavedra:2023hss,Cheng:2023qmz,Aguilar-Saavedra:2024hwd,Cheng:2024btk}.  Other final states in which entanglement can be studied at the LHC include the $\tau^+ \tau^-$~\cite{Fabbrichesi:2022ovb,Zhang:2025mmm}, diboson~\cite{Aguilar-Saavedra:2022wam,Aguilar-Saavedra:2022mpg,Ashby-Pickering:2022umy,Fabbrichesi:2023cev,Morales:2023gow,Aoude:2023hxv,Bernal:2023ruk,Bi:2023uop,Aguilar-Saavedra:2024whi,Bernal:2024xhm,Subba:2024mnl,DelGratta:2025qyp,Goncalves:2025qem,Ruzi:2025jql,Goncalves:2025xer}, $b\bar{b}$~\cite{Afik:2025grr} states.
Lepton colliders are excellent environments for probe of entanglement in final states such as the $\tau^+\tau^-$~\cite{Altakach:2022ywa, Ma:2023yvd, Ehataht:2023zzt, LoChiatto:2024dmx, Han:2025ewp}, $t\bar{t}$~\cite{Maltoni:2024csn}, $\mu^+ \mu^-$~\cite{Ruzi:2024iqu}, diboson~\cite{Ruzi:2024cbt, Wu:2024ovc, Ding:2025mzj}, $Y\bar{Y}$~\cite{Wu:2024asu}, $\phi\phi$~\cite{Gabrielli:2024kbz}, $\Lambda\bar{\Lambda}$~\cite{Pei:2025yvr, Pei:2025ito}, and $q\bar{q}$~\cite{Cao:2025qua} systems.
While a wide range of systems at hadron and lepton colliders have been studied based on spin correlations~\cite{Fabbrichesi:2022ovb, Zhang:2025mmm, Aguilar-Saavedra:2022wam,Aguilar-Saavedra:2022mpg, Ashby-Pickering:2022umy, Fabbrichesi:2023cev, Morales:2023gow,Aoude:2023hxv, Bernal:2023ruk, Bi:2023uop, Aguilar-Saavedra:2024whi, Bernal:2024xhm,Subba:2024mnl, DelGratta:2025qyp, Goncalves:2025qem, Ruzi:2025jql, Goncalves:2025xer,Afik:2025grr, Altakach:2022ywa, Ma:2023yvd, Ehataht:2023zzt, LoChiatto:2024dmx, Han:2025ewp, Maltoni:2024csn, Wu:2024asu, Gabrielli:2024kbz, Ruzi:2024cbt, Wu:2024ovc, Ding:2025mzj, Ruzi:2024iqu, Cheng:2025zcf, Cheng:2025cuv, Shi:2016bvo, Shi:2019mlf, Shi:2019kjf},
entanglement between flavor quantum numbers can also be studied in neutral meson systems via quantum tomography~\cite{Cheng:2025zcf}.

Bell nonlocality can be quantified by Bell-inequality violation \cite{Bell:1964kc}, for example, through the optimal Clauser-Horne-Shimony-Holt (CHSH) parameter constructed from the spin-correlation matrix \cite{Clauser:1969ny}.  In the $t\bar{t}$ system, it is necessary to look for tops with high transverse momentum, which makes the observation at the LHC challenging~\cite{Fabbrichesi:2021npl,Severi:2021cnj,Afik:2022kwm,Han:2023fci,Cheng:2023qmz,Cheng:2024btk}.
Perhaps a more promising final state at the LHC to look for Bell nonlocality is $\tau^+ \tau^-$~\cite{Zhang:2025mmm} where a $5\sigma$ observation is predicted to be achievable in the current dataset.\footnote{We note that these studies do not test local realism~\cite{Abel:1992kz,Low:2025aqq}, but rather study quantum mechanical predictions in light of quantum information.}

In quantum computing, the concept of magic quantifies the computational advantage that a quantum computer would have over a classical computer \cite{Gottesman:1998hu}. The $t\bar{t}$ system at LHC~\cite{White:2024nuc} was shown to exhibit a range of magic values, as well as the $\tau^+ \tau^-$ system at a lepton collider~\cite{Fabbrichesi:2025ywl}.  It can also be used to set limits on the coefficients of effective field theory \cite{Aoude:2025jzc}.

A number of other quantum measures and concepts have been explored, including the nature of fictitious states~\cite{Afik:2022kwm,Cheng:2024btk}, the kinematic approach to quantum tomography~\cite{Cheng:2024rxi}, the use of distance measures to constrain new physics~\cite{Fabbrichesi:2025ywl}, decoherence effects~\cite{Burgess:2024heo, Aoude:2025ovu, Gu:2025ijz}, contextuality~\cite{Fabbrichesi:2025rsg}, quantum discord~\cite{Afik:2022dgh,Han:2024ugl}, the nonlocal advantage of quantum coherence~\cite{Rai:2025qke} and weak measurements~\cite{Barr:2025avs}.  For reviews, see Refs.~\cite{Barr:2024djo,Afik:2025ejh,Fang:2024ple}.

Due to the clean experimental environments, future $e^+e^-$ colliders provide an ideal system in which quantum correlations can be explored. Previous studies have extensively studied quantum entanglement and Bell nonlocality using the production of $\tau^+\tau^-$~\cite{Altakach:2022ywa, Ma:2023yvd, Ehataht:2023zzt, Barr:2024djo, Fabbrichesi:2024wcd, LoChiatto:2024dmx, Han:2025ewp, Zhang:2025mmm}, $t\bar{t}$~\cite{Maltoni:2024csn, Cheng:2024btk}, and $q\bar{q}$~\cite{Cheng:2025cuv, Cao:2025qua} at proposed $e^+e^-$ colliders.
Future $e^+e^-$ colliders are designed to be precision machines with the capability to employ polarized beams~\cite{CEPC, CEPCStudyGroup:2025kmw, Ai:2025cpj, ILC, Behnke:2013xla, Roloff:2018dqu, CLIC}. Beam polarizations may play a crucial role in precise measurements of a wide range of observables, including those based on quantum-information concepts, and can improve the statistical sensitivity of the measurement of quantum observables, while also allowing the disentangling of contributions from different helicity amplitudes.

% our work
In this work, we develop a uniform quantum tomography description of fermion-pair production with longitudinally polarized $e^+e^-$ beams~\cite{Moortgat-Pick:2005jsx, Vauth:2016pgg, Karl:2017xra, Blondel:2019jmp, List:2020wns, Nikitin:2020ood, Duan:2023cgu, Duan:2023lyp, CEPCStudyGroup:2023quu}. We employ Fano-Bloch decomposition to perform full two-qubit quantum tomography and quantify three key observables: concurrence ($\mathcal{C}$), Bell nonlocality ($\mathcal{B}$), and the quantum computational resource of magic ($\mathcal{M}_2$), for the $t\bar{t}$, $\mu^+\mu^-$ and $e^+e^-$ final states. We demonstrate the fundamental distinction in polarization dependence: for $s$-channel dominated processes ($t\bar{t}$ and $\mu^+\mu^-$), beam polarization primarily manipulates the single-particle polarization ($B^\pm_i$), thus strongly affecting $\mathcal{M}_2$, while $\mathcal{C}$ and $\mathcal{B}$ remain robust, as they depend mainly on the spin correlation matrix $C_{ij}$. We perform the first quantum analysis of high energy Bhabha scattering $e^+e^-\to e^+e^-$, where polarization acts as a unique channel modulator of $s/t$-channel interference, leading to unique behavior across all QI observables.
We also present realistic projections for the observability and precision of these quantum observables at future $e^+e^-$ colliders, establishing optimal polarization and kinematic regions for measurement.

The remainder of the paper is organized as follows. In Section \ref{sec:observables}, we present the framework of quantum tomography and introduce the three QI observables. Section \ref{sec:eecollider} analyzes the quantum behavior for the three final states, focusing on the physical features and the beam polarization effects. Section \ref{sec:Significance} discusses the observability of those variables from the experimental aspects.  Finally, Section \ref{sec:summary} provides a summary and conclusions.

%%%%%%%%%%%%%%%%%%%%%%%%%%%%%%%%%%%%%%%%%%%%%%%%%%%%%%%%%%%
%%%%%%%%%%%%%%%%%%%%%%%%%%%%%%%%%%%%%%%%%%%%%%%%%%%%%%%%%%%
\section{Quantum Tomography and Quantum Information Observables}
\label{sec:observables}
%%%%%%%%%%%%%%%%%%%%%%%%%%%%%%%%%%%%%%%%%%%%%%%%%%%%%%%%%%%
\subsection{Quantum Tomography}
\label{subsec:Tomography}

For a pure quantum state denoted by a state vector $|\psi\rangle$, an equivalent description is the associated density matrix in the form of a projection operator $\rho = |\psi\rangle \langle \psi|$.

A general quantum state can be written as
\begin{equation}
\rho=\sum_{\alpha}^n w_\alpha\,\ket{\alpha}\bra{\alpha},\qquad \sum_{\alpha}^n w_\alpha =1,
\label{eq:rho}
\end{equation}
where $w_a$ is the fraction of the ensemble for the sub-state $|\alpha\rangle$. If the state contains only one term $n=1$, it is a pure state $w_1=1$. Otherwise, the state is called a mixed state.
Whether the quantum state is mixed or pure can be quantified by the purity given by
\begin{equation}
    P = \mathrm{Tr}(\rho^2).
    \label{eq:P}
\end{equation}
For a pure state $P=1$ and for a mixed state  $P<1$.
An expectation value for an observable operator $O$ can be evaluated by $\langle O\rangle = {\rm Tr}(\rho\,O)$.

A two-qubit quantum state is described by a density matrix $\rho$ which can be decomposed by the Fano-Bloch  decomposition  \cite{Fano:1983zz} as
\begin{equation}
\label{eq:fano}
\rho = \frac{1}{4}\left(
\mathbb{I}_2 \otimes \mathbb{I}_2
+ \sum_i B^-_i \sigma_i \otimes \mathbb{I}_2
+ \sum_i B^+_i \mathbb{I}_2 \otimes \sigma_i
+ \sum_{ij} C_{ij} \sigma_i \otimes \sigma_j
\right).
\end{equation}
Here $\mathbb{I}_2$ denotes the identity $2\times2$ matrix, and $\sigma_i$ ($i=1,2,3$) are Pauli operators defined with respect to a chosen orthonormal spin basis $\{\hat e_i\}$, i.e., \ $\sigma_i\equiv\vec\sigma\cdot\hat e_i$.
The three components of $B_i^-$ ($B_i^+$) encode the polarization of the fermion $f$ (antifermion $\bar f$), given by the expectation values $B_i^-\equiv\langle\sigma_i\otimes\mathbb{I}_2\rangle$ and $B_i^+\equiv\langle\mathbb{I}_2\otimes\sigma_i\rangle$, while the $3\times 3$ coefficients $C_{ij}\equiv\langle\sigma_i\otimes\sigma_j\rangle$ encode the spin-spin correlations.
Together, $B_i^\pm$ and $C_{ij}$ constitute 15 real Fano coefficients that fully specify the two-spin density matrix.
Determining these coefficients from the data is the task of two-qubit quantum tomography.

\begin{figure}[tb]
    \centering
    \includegraphics[width=0.4\linewidth]{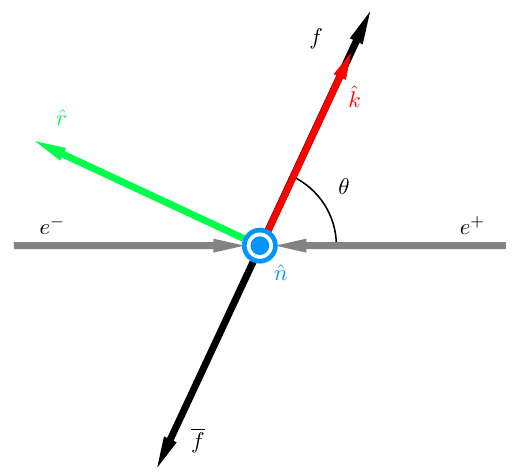}
    \caption{Definition of the helicity basis $\{\hat{r},\,\hat{n},\,\hat{k}\}$ in the $f\bar f$ center-of-mass frame. The unit vector $\hat k$ points along the fermion momentum, $\hat n$ is normal to the scattering plane, and $\hat r=\hat n\times\hat k$ completes a right-handed triad (equivalently $\hat n=\hat k\times\hat r$). }
    \label{fig:helicity-basis}
\end{figure}

When performing spin-correlation or entanglement measurements, a basis for spin quantization must be specified.
A simple choice is the fixed beam basis $(\hat{x},\,\hat{y},\,\hat{z})$ in the $e^-e^+$ lab frame, with the $\hat{z}$-axis aligned with the incoming $e^-$ beam. At high energies, a more convenient choice is the event-by-event helicity basis $\{\hat{r},\,\hat{n},\,\hat{k}\}$, defined in the $f\bar f$ center-of-momentum (c.~m.) frame as illustrated in Fig.~\ref{fig:helicity-basis}, where $\hat k$ is the fermion flight direction, $\hat n$ is normal to the scattering plane, and $\hat r=\hat n\times\hat k$ completes a right-handed triad.

The density matrix $\rho$ of the system $f\bar f$ can be reconstructed in two complementary ways. The decay approach infers $B_i^\pm$ and $C_{ij}$ from the angular distributions of spin analyzers in the decays of $f$ and $\bar f$~\cite{Afik:2020onf}. This is the method that ATLAS and CMS have used in their entanglement measurements~\cite{ATLAS:2023fsd,CMS:2024pts,CMS:2024zkc}. The kinematic approach reconstructs the spin density matrix directly from the measured production  kinematics \cite{Cheng:2024rxi}.
In this method, only the velocity $\beta$ and the scattering angle $\theta$ of the final state particles need to be determined, with or without their decay.
This method is particularly suitable for analyzing quantum correlation properties in systems involving stable particles.

%%%%%%%%%%%%%%%%%%%%%%%%%%%%%%
\subsection{Entanglement}
\label{subsec:concurrence}

Quantum entanglement is the phenomenon of two subsystems that cannot be fully described independently of each other~\cite{Einstein:1935rr,Horodecki:2009zz}. For a system of two qubits, the entanglement can be quantified by the concurrence~\cite{Wootters:1997id}
\begin{equation}
\label{eq:concurrence}
\mathcal{C} = \max\!\left\{0,\lambda_1-\lambda_2-\lambda_3-\lambda_4\right\},
\end{equation}
where $\lambda_i$ are the eigenvalues, ordered from largest to smallest, of the auxiliary matrix $R$
\begin{equation}
R \equiv \sqrt{\sqrt{\rho} \tilde{\rho} \sqrt{\rho}}\ ,
\qquad
\tilde{\rho} \equiv (\sigma_2 \otimes \sigma_2) \rho^* (\sigma_2 \otimes \sigma_2).
\label{eq:R}
\end{equation}
Here $\rho^*$ denotes complex conjugation in the computational basis.
The concurrence satisfies $0 \le \mathcal{C} \le 1$, with $\mathcal{C}=0$ for separable states and $\mathcal{C} = 1$ for maximally entangled Bell states. As seen in Eq.~(\ref{eq:R}), it is basis-choice independent under spin rotations.

%%%%%%%%%%%%%%%%%%%%%%%%%%%%%%%%%%%%%%%%%%%%
\subsection{Bell Nonlocality}
\label{subsec:BI}

Another quantum correlation is Bell nonlocality, which is evidenced by states that violate Bell's inequality~\cite{Bell:1964kc}. In two-qubit systems, this inequality is given by the CHSH inequality which is~\cite{Clauser:1969ny}
\begin{equation} \label{eq:CHSH}
| \langle \vec{a}_1 \cdot \vec{\sigma} \otimes \vec{b}_1 \cdot \vec{\sigma} \rangle
- \langle \vec{a}_1 \cdot \vec{\sigma} \otimes \vec{b}_2 \cdot \vec{\sigma} \rangle
+ \langle \vec{a}_2 \cdot \vec{\sigma} \otimes \vec{b}_1 \cdot \vec{\sigma} \rangle
+\langle \vec{a}_2 \cdot \vec{\sigma} \otimes \vec{b}_2 \cdot \vec{\sigma} \rangle |
\leq 2,
\end{equation}
where $\vec{a}_{1}$ and $\vec{a}_{2}$ specify the ``detector settings'' or spin axes of one qubit and $\vec{b}_{1}$ and $\vec{b}_{2}$ specify the ``detector settings'' or spin axes of the other qubit.

The Bell variable $\mathcal{B}(\vec{a}_{1},\vec{a}_{2},\vec{b}_{1},\vec{b}_{2})$ measures the left-hand side of the CHSH inequality:
\begin{equation}
\mathcal{B}(\vec{a}_1, \vec{a}_2, \vec{b}_1,\vec{b}_2) = |\langle \vec{a}_1 \cdot \vec{\sigma} \otimes \vec{b}_1 \cdot \vec{\sigma} \rangle
- \langle \vec{a}_1 \cdot \vec{\sigma} \otimes \vec{b}_2 \cdot \vec{\sigma} \rangle
+ \langle \vec{a}_2 \cdot \vec{\sigma} \otimes \vec{b}_1 \cdot \vec{\sigma} \rangle
+\langle \vec{a}_2 \cdot \vec{\sigma} \otimes \vec{b}_2 \cdot \vec{\sigma} \rangle|.
\end{equation}
The value of $\mathcal{B}(\vec{a}_1, \vec{a}_2, \vec{b}_1,\vec{b}_2)$ depends on the four choices of spin axes.  The optimal set of choices that maximizes $\mathcal{B}(\vec{a}_1, \vec{a}_2, \vec{b}_1,\vec{b}_2)$ yields a value~\cite{Horodecki:1995nsk,Fabbrichesi:2021npl}
\begin{equation}
\label{eq:bellnonlocality}
\mathcal{B} = 2 \sqrt{m_1 + m_2},
\end{equation}
where $m_1$ and $m_2$ are the two largest eigenvalues of the matrix $C^T C$, where $C$ is the spin correlation matrix from Eq.~\eqref{eq:fano}.
For Bell local states, one has $\mathcal{B}\le 2$, while $\mathcal{B}>2$ indicates that a state is Bell nonlocal.  The maximum value of the Bell variable reachable by a quantum mechanical system is $2\sqrt{2}$~\cite{Cirelson:1980ry,Low:2025aqq}.

%%%%%%%%%%%%%%%%%%%%%%%%%%%%%%%%%%%%%%%%%%%%
\subsection{Magic}
\label{subsec:magic}

In contrast to classical computers, which operate on bits, quantum computers act on qubits.  Quantum computers, in some cases, can solve problems substantially faster than classical computers.  This speed-up results from quantum correlations. However, more Bell-like quantum correlations do not necessarily result in an advantage to quantum computers.  For example, both separable states and Bell states are efficient in classical simulation~\cite{Gottesman:1998hu}.

A more direct indication of the advantage that is achieved by a quantum computer is given by ``magic.''  A state that has large magic cannot be efficiently classically simulated.  States, on the other hand, that can be efficiently classically simulated, are called stabilizer states.  These are defined as states that are reached by applying Pauli strings operators to the state $\ket{0} \otimes \ket{0} \otimes \ldots \otimes \ket{0}$.  Pauli strings for $n$ qubits are defined as
\begin{equation}
\label{eq:Pndef}
{\cal P}_n=P_1\otimes P_2\otimes\ldots\otimes P_n\ ,
\qquad
  P_a\in\{\sigma_1^{(a)},\sigma_2^{(a)},\sigma_3^{(a)}, \mathbb{I}_2^{(a)}\}\ .
\end{equation}
According to the Gottesman-Knill theorem~\cite{Gottesman:1998hu}, for quantum systems containing only stabilizer states, there is a classical computer that is just as efficient as a quantum computer.

The magic of a state specifies how far the state is from a stabilizer state.  Stabilizer R\'{e}nyi entropies (SREs) quantify magic for pure states~\cite{Leone:2021rzd},
\begin{equation}
\label{eq:Mqdef}
\mathcal{M}_q=\frac{1}{1-q}\log_2(\zeta_q)\ ,
\qquad
\zeta_q\equiv
\sum_{P\in{\cal P}_n} \frac{\langle\psi|P|\psi\rangle^{2q}}{2^n},
\end{equation}
with integer $q\ge2$. For our two-qubit systems ($n=2$), we focus on the case $q=2$, denoted the second SRE (SSRE) $\mathcal{M}_2$. For a mixed state $\rho$, we use the mixed-state extension of Ref.~\cite{Leone:2021rzd},
\begin{equation}
\label{eq:m2}
\mathcal{M}_2=-\log_2{\frac{\sum_{P\in{\cal P}_n} \tr(\rho P)^4}{\sum_{P\in{\cal P}_n} \tr(\rho P)^2}}.
\end{equation}
In terms of Fano coefficients ($B_i^\pm, C_{ij}$), $\mathcal{M}_2$ is a nonlinear functional of the $15$ quantum tomographic parameters:
\begin{equation}
\label{eq:m2fano}
\mathcal{M}_2=-\log_2{\frac{1+\sum_i{(B_i^+)^4}+\sum_j{(B_j^-)^4}+\sum_{ij}{(C_{ij})^4}}{1+\sum_i{(B_i^+)^2}+\sum_j{(B_j^-)^2}+\sum_{ij}{(C_{ij})^2}}}.
\end{equation}
 SSRE $\mathcal{M}_2$ has been shown to be useful in the $t\bar{t}$ system at LHC~\cite{White:2024nuc} and has been measured by the CMS collaboration~\cite{CMS:2025cim}.

The minimal SSRE value is $\mathcal{M}_2=0$, which corresponds to a stabilizer state.
In the Fano-Bloch representation, this zero-magic condition ($\mathcal{M}_2=0$) is achieved if and only if all Fano coefficients ($B^\pm_i, C_{ij}$) are restricted to the three discrete values $\{-1, 0, +1\}$. This constraint defines the stabilizer sector. In contrast, magic is maximized when the coefficients deviate maximally from these extremal values. The maximal SSRE value for a pure state of a two-qubit system is given by~\cite{Liu:2025frx}
\begin{equation}\label{eq:maxm2}
\max \left[ \mathcal{M}_{2} (\ket{\psi}) \right] = \log_2 \frac{16}{7} \approx 1.192.
\end{equation}
It should be noted that the maximal value quoted in the literature~\cite{Liu:2025frx} is $\log(16/7) \approx 0.827$, where the natural logarithm is used. In this work, we instead employ the base-2 logarithm $\log_{2}$, which is the standard normalization convention in quantum information theory.

Entanglement and magic are distinct quantum resources. For example, maximally entangled Bell states exhibit strong nonclassical correlations, yet they are stabilizer states with vanishing magic ($\mathcal{M}_2=0$) and admit efficient classical simulation within the stabilizer formalism. Conversely, nonzero magic can arise even when entanglement is small or vanishes, reflecting nonstabilizer structure beyond two-particle nonlocal correlations.
We will reiterate those points in Sec~\ref{subsec:SecIII-summary}.

%%%%%%%%%%%%%%%%%%%%%%%%%%%%%%%%%%%%%%%%%%%%%%%%%%%%%%%%%%%
%%%%%%%%%%%%%%%%%%%%%%%%%%%%%%%%%%%%%%%%%%%%%%%%%%%%%%%%%%%
\section{ Quantum Behavior of \texorpdfstring{$f\bar f$}{ff} final states with Longitudinal Beam Polarization}
\label{sec:eecollider}
%%%%%%%%%%%%%%%%%%%%%%%%%%%%%%%%%%%%%%%%%%%%%%%%%%%%%%%%%%%

Under the condition of full beam polarization, the initial state of the scattering process, $|\psi_{\mathrm{in}}\rangle$, is a well-defined pure state with a fixed spin configuration. The corresponding initial-state density matrix is simply the projection operator:
\begin{equation}
\rho_{\rm in}=\ket{\psi_{\rm in}}\bra{\psi_{\rm in}}.
\end{equation}
The transition operator $T$, originating from the $S$-matrix decomposition $S = \mathrm{1} + iT$,  maps the initial state to an outgoing state
\begin{equation} |\psi_{\mathrm{out}}\rangle = \frac{T\ket{\psi_{\mathrm{in}}}}{\sqrt{\bra{\psi_{\mathrm{in}}}T^\dagger T\ket{\psi_{\mathrm{in}}}}},
\end{equation}
where the denominator ensures the normalization of the state. Consequently, the final-state density matrix is given by
\begin{equation}
\rho_{\mathrm{out}} = \frac{T \ket{\psi_{\rm in}}\bra{\psi_{\rm in}} T^\dagger}{\bra{\psi_{\mathrm{in}}}T^\dagger T\ket{\psi_{\mathrm{in}}}}.
\end{equation}
The complete polarization of the colliding beams ensures that the transition involves a coherent superposition of amplitudes rather than a classical mixing. This preservation of quantum coherence guarantees that the resulting final state density matrix remains pure.

For unpolarized or partially polarized beams, the initial state is an incoherent statistical mixture of different helicity configurations, leading to a density matrix for the mixed state.
The mixed state can be expressed as a mixing of the four helicity configurations, $LR$, $RL$, $LL$, and $RR$,
\begin{equation}
\rho_{\mathrm{unpolarized}} = \rho_{LR} + \rho_{RL} + \rho_{LL} + \rho_{RR},
\label{eq:rho_unpol}
\end{equation}
where $L$ is for left-handed and $R$ is for right-handed longitudinal beam polarization.
For longitudinally-polarized beams, each beam can be described by a single-particle spin density matrix,
\begin{equation}
\rho_{e^-} = \frac{1}{2}(1+P_{e^-}\sigma_z),
\qquad\qquad\qquad
\rho_{e^+} = \frac{1}{2}(1+P_{e^+}\sigma_z),
\end{equation}
where $P_{e^\pm}\in[-1,1]$ denote the
%$[{\rm left,right}]$
percentage longitudinal polarizations and $\sigma_z$ is the Pauli matrix along the beam axis, with $P_{e^\pm}=+1~(-1)$ corresponding to a purely right-handed (left-handed) beam.

The combined initial spin density matrix is given by their tensor product,
\begin{equation}
\rho_{\mathrm{in}}
   = \rho_{e^-}\otimes\rho_{e^+}
   = \frac{1}{4}\big(1+P_{e^-}\sigma_z\big)
     \otimes\big(1+P_{e^+}\sigma_z\big).
\label{eq:rho_in}
\end{equation}
The three most common limiting cases are:
\begin{align}
&\text{unpolarized:}&& P_{e^-}=P_{e^+}=0\;\Rightarrow\; \rho_{\rm in}=\tfrac{1}{4}\,\mathbb{I}_2\otimes\mathbb{I}_2,
\label{eq:unpol_limit}\\
&\text{fully polarized ($e^-_Le^+_R$):}&& (P_{e^-},P_{e^+})=(-1,+1)\;\Rightarrow\; \rho_{\rm in}=\ket{LR}\bra{LR},\\
&\text{fully polarized ($e^-_Re^+_L$):}&& (P_{e^-},P_{e^+})=(+1,-1)\;\Rightarrow\; \rho_{\rm in}=\ket{RL}\bra{RL}.
\end{align}
Explicitly, $\rho_{\rm in}$ can be written as an incoherent mixture of the four helicity configurations
$\alpha\in\{LR,RL,LL,RR\}$. In terms of the form for a mixed state in Eq.~(\ref{eq:rho_in}), we have
\begin{align}
w_{LR}&=\frac{(1-P_{e^-})(1+P_{e^+})}{4},\;\qquad w_{RL}=\frac{(1+P_{e^-})(1-P_{e^+})}{4}, \nonumber\\
w_{LL}&=\frac{(1-P_{e^-})(1-P_{e^+})}{4},\;\qquad w_{RR}=\frac{(1+P_{e^-})(1+P_{e^+})}{4}. \nonumber
\end{align}
For fully polarized beams, $\rho_{\rm in}$ projects onto a single initial helicity, yielding a pure state production  before any phase-space averaging.

Let $\mathcal{M}_{\lambda_i\to\lambda_f}$ denote the helicity amplitudes of
$e^-e^+\to f\bar f$ in Eq.~\eqref{eq:rho_unpol}, with $\lambda_i=(\lambda_{e^-},\lambda_{e^+})$
and $\lambda_f=(\lambda_f,\lambda_{\bar f})$.
The spin density matrix of the production of the system $f\bar f$ is obtained by evolving the initial ensemble $\rho_{\rm in}$ according to the matrix element $\mathcal{M}$:
\begin{equation}
\rho_{f\bar f}\;=\;\frac{1}{\mathcal{N}}
\sum_{\lambda_i,\lambda_i'}
\mathcal{M}_{\lambda_i\to\lambda_f}\,
\rho_{\rm in}(\lambda_i,\lambda_i')\,
\mathcal{M}^{\dagger}_{\lambda_i'\to\lambda_f'}\,,
\qquad
\mathcal{N}=\mathrm{Tr}\!\left[\sum_{\lambda_i,\lambda_i'}
\mathcal{M}\,\rho_{\rm in}\,\mathcal{M}^\dagger\right].
\label{eq:rho_prod}
\end{equation}
In the unpolarized limit in Eq.~\eqref{eq:unpol_limit}, Eq.~\eqref{eq:rho_prod}
reduces to the familiar sum over initial helicities with equal weights. Details of the tree-level calculation for $e^-e^+\to f\bar f$ are provided in Appendix~\ref{sec:AppendixA}.

We adopt a common spin quantization axis $\hat k\equiv \vec p_f/|\vec p_f|$, applied to both the fermion $f$ and the antifermion $\bar{f}$.
In this convention, we denote
\begin{equation}
|{\uparrow}\rangle_f\equiv |f_R\rangle,\qquad |{\downarrow}\rangle_f\equiv |f_L\rangle,
\qquad
|{\uparrow}\rangle_{\bar f}\equiv |\bar f_L\rangle,\qquad |{\downarrow}\rangle_{\bar f}\equiv |\bar f_R\rangle,
\end{equation}
where the assignment for $\bar f$ follows because $\bar f$ moves along $-\hat k$.

For the fully-polarized initial configurations $LR$ ($e^-_Le^+_R$) or $RL$ ($e^-_Re^+_L$), the total angular momentum is aligned with the $z$-axis, and the final state $\ket{f\bar{f}}$ is a pure two-qubit state. A general pure state $\ket{\psi}$ of two qubits can be written as
\begin{equation}
\label{eq:psi}
\ket{\psi} = a\ket{\uparrow\uparrow} + b\ket{\downarrow\downarrow} +c\ket{\uparrow\downarrow} + d\ket{\downarrow\uparrow},
\end{equation}
where $a$, $b$, $c$, and $d$ are complex coefficients satisfying $|a|^2+|b|^2+|c|^2+|d|^2=1$.
For a general two-qubit pure state, the necessary and sufficient condition for separability is $ab = cd$, which ensures that $\ket{\psi}$ can be written as a product state $\ket{\phi_1}\otimes\ket{\phi_2}$. Whenever $ab\neq cd$, the state is entangled. Two special cases are particularly relevant in the collider setting~\cite{Nielsen_Chuang_2010}:\\
\noindent
(1)
$c=d=0$.\\
The coefficients $a$ and $b$ characterize the components $\ket{\uparrow\uparrow}$ and $\ket{\downarrow\downarrow}$, corresponding to the final state helicity configurations $f_R \bar{f}_L$ and $f_L \bar{f}_R$, respectively. For vector/axial-vector gage interactions, helicity is conserved in the ultrarelativistic limit $m_f^2\ll s$, so the produced state lies dominantly in the subspace $\{|{\uparrow}{\uparrow}\rangle,|{\downarrow}{\downarrow}\rangle\}$. The Bell limit $|a|=|b|$ is reached at specific scattering angles, resulting in $|\Phi^\pm\rangle=(|{\uparrow}{\uparrow}\rangle\pm|{\downarrow}{\downarrow}\rangle)/\sqrt2$.
Consequently, the dominance of the $a$ and $b$ coefficients indicates $s$-channel exchange of spin-1 gauge bosons. Specifically, QED typically generates the $\ket{\Phi^{\pm}}$ states through coherent superpositions of these helicity-conserving amplitudes.\\
\noindent
(2) $a=b=0$.\\
The coefficients $c$ and $d$ correspond to the states $\ket{\uparrow\downarrow}$ and $\ket{\downarrow\uparrow}$,
with the helicity configurations $f_R \bar{f}_R$ and $f_L \bar{f}_L$. They can also form the Bell states that are maximally entangled with $|c|=|d|$. However, such same-helicity final states require a chirality flip, which is forbidden for massless fermion gauge interactions, but is the signature of scalar or pseudoscalar interactions. The antisymmetric combination $\ket{\Psi^-} = (\ket{\uparrow\downarrow} - \ket{\downarrow\uparrow})/\sqrt{2}$ corresponds to a total spin $S_z=0$ singlet state. This state is naturally generated by a Yukawa interaction, such as the SM Higgs boson, where the couplings to left- and right-handed fermions are symmetric. The symmetric combination $\ket{\Psi^+} = (\ket{\uparrow\downarrow} + \ket{\downarrow\uparrow})/\sqrt{2}$ corresponds to the $S_z=0$ component of the spin triplet. This state arises from pseudoscalar currents ({\it e.g.}, a CP-odd Higgs boson $A^0$), where the relative sign between chiral couplings introduces a phase shift in the spin basis. The observation of non-vanishing $c$ or $d$ terms at high energies would imply  for new physics in the scalar sector with significant mass-insertion effects, distinct from the chirality-conserving gauge interactions characterized by $\ket{\Phi^{\pm}}$.

The expressions of $B^\pm_{i}$ and $C_{ij}$ for a pure state in terms of $a, b, c, d$ in Eq.~\eqref{eq:psi} are given by
\begin{equation}
B^+_i = \begin{pmatrix}
d a^{*} + c b^{*} + b c^{*} + a d^{*} \\
- i\, d a^{*} + i\, c b^{*} - i\, b c^{*} + i\, a d^{*} \\
a a^{*} - b b^{*} + c c^{*} - d d^{*}
\end{pmatrix}, \quad
B^-_j = \begin{pmatrix}
c a^{*} + d b^{*} + a c^{*} + b d^{*} \\
- i\, c a^{*} + i\, d b^{*} + i\, a c^{*} - i\, b d^{*} \\
a a^{*} - b b^{*} - c c^{*} + d d^{*}
\end{pmatrix} ,
\label{eq:Bs}
\end{equation}
\begin{equation}
C_{ij} = \begin{pmatrix}
b a^{*} + a b^{*} + d c^{*} + c d^{*} &
- i\, b a^{*} + i\, a b^{*} + i\, d c^{*} - i\, c d^{*} &
d a^{*} - c b^{*} - b c^{*} + a d^{*} \\[6pt]

- i\, b a^{*} + i\, a b^{*} - i\, d c^{*} + i\, c d^{*} &
- b a^{*} - a b^{*} + d c^{*} + c d^{*} &
- i\, d a^{*} - i\, c b^{*} + i\, b c^{*} + i\, a d^{*} \\[6pt]

c a^{*} - d b^{*} + a c^{*} - b d^{*} &
- i\, c a^{*} - i\, d b^{*} + i\, a c^{*} + i\, b d^{*} &
a a^{*} + b b^{*} - c c^{*} - d d^{*}
\end{pmatrix}.
\label{eq:Cs}
\end{equation}

For the process $e^-_{R/L}e_{L/R}^+\to \gamma^*,Z^* \to f\bar{f}$, the coefficients $a$, $b$, $c$, $d$ in Eq.~\eqref{eq:psi} take the form
\begin{subequations}
\label{eq:ttab}
\begin{align}
a &= e^2 \,\big(f_V^{R/L} + f_A^{R/L} \beta\big)\, (1\pm \cos\theta ), \\
b &= e^2 \,\big(f_V^{R/L} - f_A^{R/L} \beta\big)\, ( 1\mp\cos\theta ), \\
c &= d = \mp\, e^2 f_V^{R/L} \sqrt{1-\beta^2}\,\sin\theta ,
\end{align}
\end{subequations}
where $\beta=\sqrt{1-4m_f^2/s}$, the upper (lower) signs correspond to the beam polarization configuration $RL$ ($LR$). The coupling strengths are given by
\begin{subequations}
\label{eq:fafv}
\begin{align}
    f_{V}^L&= Q_e Q_f +  \frac{g_L^e (g_L^f + g_R^f)}{2c_W^2 s_W^2} \frac{s}{s-m_Z^2+is\Gamma_Z/m_Z},\\
    f_{A}^L&= -  \frac{g_L^e (g_L^f - g_R^f)}{2c_W^2 s_W^2} \frac{s}{s-m_Z^2+is\Gamma_Z/m_Z},\\
    f_{V}^R&= Q_e Q_f  + \frac{g_R^e (g_L^f + g_R^f)}{2c_W^2 s_W^2}\frac{s}{s-m_Z^2+is\Gamma_Z/m_Z},\\
    f_{A}^R&= -\frac{g_R^e (g_L^f - g_R^f)}{2c_W^2 s_W^2} \frac{s}{s-m_Z^2+is\Gamma_Z/m_Z},
\end{align}
\end{subequations}
where  $\Gamma_Z$ is the $Z$ boson width at $Z$-pole and the fermion chiral couplings are
\begin{equation}
    g_L^f=I^3_f-Q_f s_W^2\quad  {\rm and}\quad g_R^f= -Q_f s_W^2.
\end{equation}
$Q_f$ is the electric charge of the fermion $f$ ($Q_{e}=-1$), and $I_{e,f}^3$ is the third component of the weak isospin.

The full expressions of the spin correlation matrix $C_{ij}$ and the net polarization of the qubit $B^\pm_{i}$ for the $s$-channel of the unpolarized process $e^+e^- \to f\bar{f}$ are given in Appendix~\ref{sec:AppendixA}.
The spin correlation matrix $C_{ij}$ for the $s$-channel $e^+e^- \to f\bar{f}$ is parametrized by the effective vector and axial couplings ($f_V$, $f_A$) and exhibits a clear dependence on the collision energy $\beta = \sqrt{1 - 4m_f^2/s}$ and the scattering angle $\theta$. The full structure of $C_{ij}$ for a polarized pure state is given by
\begin{equation}
\label{eq:cij4tt}
C_{ij}=\frac{1}{\mathcal{D}} \begin{pmatrix}
s^2_\theta \left(f_V^2(2-\beta^2)- f_A^2 \beta^2 \right) & 0 & -2 (f_V^2c_\theta \pm f_Vf_A\beta) s_\theta \sqrt{1-\beta^2}\\
0&(f_A^2-f_V^2)\beta^2 s^2_\theta&0\\
-2 (f_V^2c_\theta \pm f_Vf_A\beta) s_\theta \sqrt{1-\beta^2} &0 & f_V^2(2c^2_\theta+\beta^2 s^2_\theta) + f_A^2\beta^2(1+c^2_\theta) \pm 4 f_Vf_A\beta
\end{pmatrix},
\end{equation}
where $s_\theta = \sin\theta$, $c_\theta = \cos\theta$, and the denominator is $\mathcal{D} = f_V^2(2-\beta^2s_{\theta}^2)\pm 4\beta f_V f_A c_{\theta} + \beta^2 f_A^2 (1+c_{\theta}^2)$. The upper (lower) signs correspond to the $RL$ ($LR$) beam polarization configurations, with the couplings $f_{V,A}$ taking the values of $f_{V,A}^R$ ($f_{V,A}^L$). For the fully-polarized process, concurrence $\mathcal{C}$ of a pure state can be simplified and parametrized analytically
\begin{equation}
\mathcal{C}=\frac{\beta^2s_{\theta}^2(f_V^2-f_A^2)}{\mathcal{D}},
\label{eq:C}
\end{equation}
and the Bell variable is given by
\begin{equation}
\mathcal B =2\sqrt{1+\mathcal C^2}.
\label{eq:B_vs_C}
\end{equation}
This simple relation between $\mathcal B$ and $\mathcal C$ is valid not only for pure states, but also for certain mixed states that are formed in an invariant rotational sub-set, such as in
($\ket{\uparrow\uparrow},\bra{\downarrow\downarrow}$), or in ($\ket{\uparrow\downarrow},\bra{\downarrow\uparrow}$).

%%%%%%%%%%%%%%%%%%%%%%%%%

The entanglement vanishes near the kinematic threshold $\beta \to 0$, and in the forward/backward regions $\theta\to 0$ and $\pi$.
The maximal entanglement for the $RL/LR$ polarizations reaches
\begin{equation}
\mathcal{C}_{\rm Max} =\frac{\beta^2 (f_V^2-f_A^2)}{f_V^2(2-\beta^2)-\beta^2f_A^2}
\label{eq:CMax}
\end{equation}
at
\begin{equation}
  \cos\theta = \mp \beta\ \frac{f_A}{f_V}.
  \label{eq:ct}
\end{equation}

In the massless limit $\beta \to 1$, the transverse polarization components vanish $B_{r,n}^\pm=0$, while the longitudinal polarization $B_{k}^\pm$ takes the form of
\begin{equation}
B_{k}^\pm= \frac{2(f_V\pm f_Ac_{\theta})(f_A\pm f_Vc_{\theta})}{(f_V^2+f_A^2)(1+c_{\theta}^2)\pm 4f_V f_A c_{\theta}},
\label{eq:Bk}
\end{equation}
The spin correlation matrix simplifies to a diagonal form
\begin{align}
  \label{eq:cij4mm}
    &C_{ij}=\frac{1}{(f_V^2+f_A^2)(1+c_{\theta}^2)\pm 4f_V f_A c_{\theta}} \times \\
    &\begin{pmatrix}
    \left(f_V^2- f_A^2 \right) s^2_\theta  & 0 & 0\\
    0&(f_A^2-f_V^2) s^2_\theta&0\\
    0 &0 & (f_V^2+f_A^2) (1+c_\theta^2)\pm 4f_Af_V
    \end{pmatrix}. \nonumber
\end{align}

The concurrence $\mathcal{C}$ in Eq.~(\ref{eq:C}) then reduces to
\begin{equation}
\mathcal{C}=\frac{(f_V^2-f_A^2)s_{\theta}^2}{(f_V^2+f_A^2)(1+c_{\theta}^2)\pm 4f_V f_A c_{\theta}}\ ,
\end{equation}
reaching the maximum $\mathcal{C}\to 1$, $\mathcal{B}\to 2\sqrt{2}$ at $\cos\theta = \mp f_A/ f_V$.
These expressions apply to general polarized processes in $s$-channel production $e^+e^- \to f\bar{f}$; for $t\bar{t}$ one substitutes $Q_t=2/3$ and $I^3_t=1/2$.

%%%%%%%%%%%%%%%%%%%%%%%%%%%%%%%%%%%%%%%%%%%%%%%%%%%%%%%%%%%
\subsection{Top-Quark Pair Production}
\label{subsec:tt}
%%%%%%%%%%%%%%%%%%%%%%%%%%%%%%%%%%%%%%%%%%%%%%%%%%%%%%%%%%%
\subsubsection{The Quantum State for \texorpdfstring{$t\bar{t}$}{tt}}
%%%%%%%%%%%%%%%%%%%%%%%%%%%%%%%%%%%%%%%%%%%%%%%%%%%%%%%%%%%

\begin{figure}[tb]
  \centering
  \includegraphics[width=0.45\linewidth]{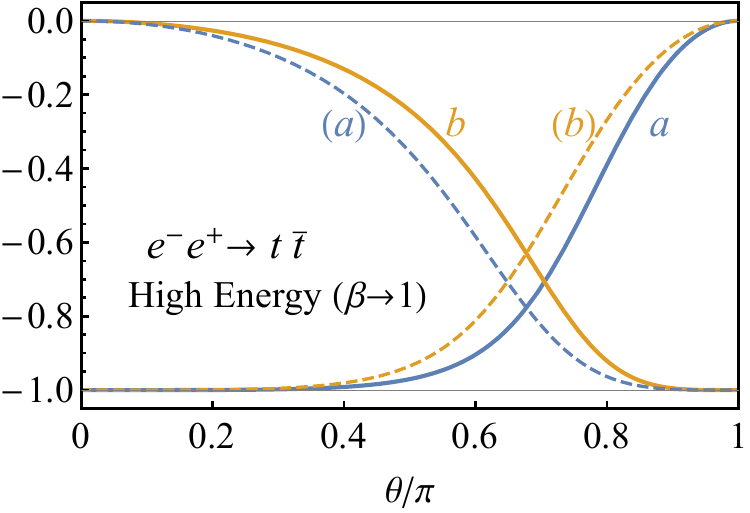}
  \caption{ Coefficients $a$ and $b$ in Eq.~(\ref{eq:psi}) for the quantum state of the $t\bar t$ system for  $RL$ ($LR$) beam polarizations as a  function of the scattering angle $\theta$ at a velocity $\beta \to 1$.}
  \label{fig:state_coefficient_tt}
\end{figure}

A $t\bar t$ state is produced via $s$-channel $Z/\gamma$ exchange.
In the high energy limit $\beta \to 1$, chirality aligns with helicity. Considering an initial state with net helicity
($\pm 1$) along the beam axis (corresponding to $e^-_R e^+_L$ and $e^-_L e^+_R$, respectively), angular momentum conservation ensures the final $t\bar{t}$ spins are aligned as $\ket{\uparrow\uparrow}$ and $\ket{\downarrow\downarrow}$, corresponding to $c=d=0$ in Eq.~(\ref{eq:psi}). The resulting $t\bar t$ state from the initial $e^+e^-$ state for a fixed helicity can be expressed as
\begin{subequations}
\begin{align}
e^-_R e^+_L:& \quad \ket{\psi} \propto (f^R_V+f^R_A) d^{1}_{1,1}(\theta) \ket{\uparrow\uparrow} + (f^R_V-f^R_A) d^{1}_{-1,1}(\theta)  \ket{\downarrow\downarrow} , \\
e^-_L e^+_R:& \quad \ket{\psi} \propto (f^L_V+f^L_A) d^{1}_{1,-1}(\theta) \ket{\uparrow\uparrow} + (f^L_V-f^L_A) d^{1}_{-1,-1}(\theta)  \ket{\downarrow\downarrow} ,
\label{eq:psitt}
\end{align}
\end{subequations}
where $d^{j}_{m',m}$ are the Wigner $d$-functions.
This formulation explicitly shows that the relative weight of the basis states depends on the scattering angle $\theta$ and the initial angular momentum $m$. Importantly, in the forward or backward limit ($\theta \to 0$ or $\pi$), the properties of the $d$-functions imply that one of the terms vanishes. Consequently, the state $|\psi_{RL}\rangle$ ($|\psi_{LR}\rangle$) becomes separable and the entanglement disappears regardless of the initial configuration.
As also seen in Eq.~(\ref{eq:ttab}), the maximal entanglement is achieved when the amplitudes satisfy $|a|=|b|$, corresponding to a  Bell state at the scattering angle $\theta_{\rm Max} \simeq 0.7\pi$ ($0.65\pi$) for the $RL$ ($LR$) polarization.
Shown in Fig.~\ref{fig:state_coefficient_tt} is the angular dependence of the $t\bar{t}$ state in the fully-polarized process in the high energy limit.
The figure clarifies which angular regions drive the state towards a Bell-like triplet state versus a stabilizer-like configuration, thereby anticipating the behavior observed later in the entanglement, Bell, and magic observables. The results above are equally applicable to other $s$-channel massless fermion production with chiral couplings.

\begin{figure}[tb]
  \centering
  \includegraphics[width=0.5\linewidth]{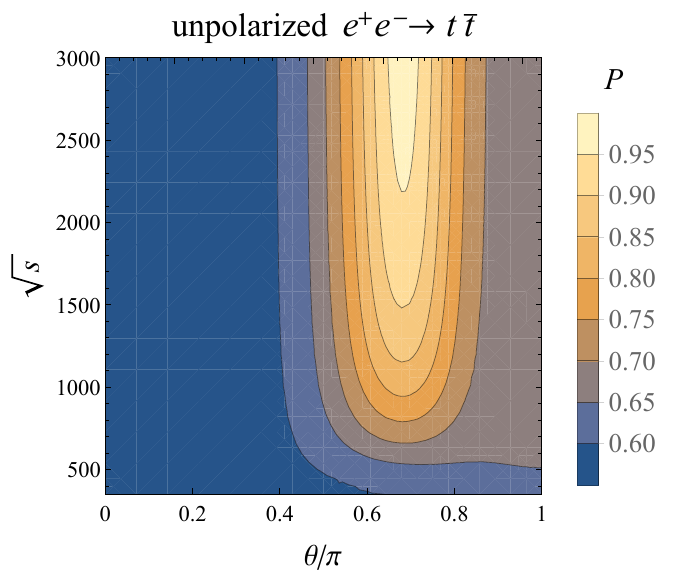}
  \caption{Contour plot for purity of the $t\bar{t}$ final state at an $e^+ e^-$ collider, shown in the kinematic plane of scattering angle $\theta$ and c.~m.~energy $\sqrt{s}$ in GeV.}
  \label{fig:ttpurity}
\end{figure}

Near the production threshold $\beta \to 0$, the $t\bar{t}$ pair is created predominantly in the $^3S_1$ state with negligible orbital angular momentum.
In the fixed-beam basis and for fully polarized initial states, the $t\bar{t}$ system is
$\ket{\downarrow\downarrow}$ ($\ket{\uparrow\uparrow}$) for $RL$ ($LR$) polarization.
In this case, the corresponding spin configuration is separable, $\rho_{t\bar t}\approx\rho_t\!\otimes\!\rho_{\bar t}$, and consequently the concurrence vanishes.
However, for unpolarized beams, the observed state is a statistical mixture of such polarized components, so the resulting final quantum state is mixed.
As shown in Fig.~\ref{fig:ttpurity}, the purity of the $t\bar{t}$ system reaches unity in the angular region around $\theta \approx 2.1$, where the final state approaches a Bell-like configuration.
These regions coincide with the maxima of concurrence and Bell nonlocality, indicating that the system is locally equivalent to a pure entangled state.
In contrast, in the scattering limits in the forward and backward direction ($\theta\to0$ and $\pi$), the purity decreases, reflecting a transition to a mixed state that corresponds to a statistical mixture of the separable configurations $LR$ and $RL$.
At the production threshold, the unpolarized final state likewise becomes a mixture of separable states arising from the two polarization modes.
%

%%%%%%%%%%%%%%%%%%%%%%%%%%%%%%%%%%%%%%%%%%%%%%%%%%%%%%%%%%%
\subsubsection{Concurrence, Bell Nonlocality, and Quantum Tomography for \texorpdfstring{$t\bar{t}$}{tt}}

\begin{figure}[tb]
  \centering
  \includegraphics[width=0.45\linewidth]{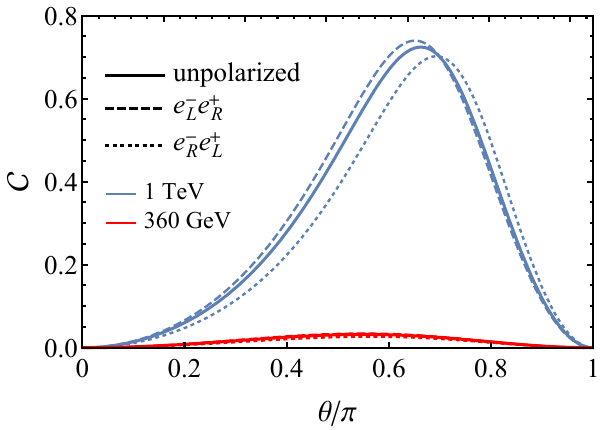}
  \qquad
  \includegraphics[width=0.45\linewidth]{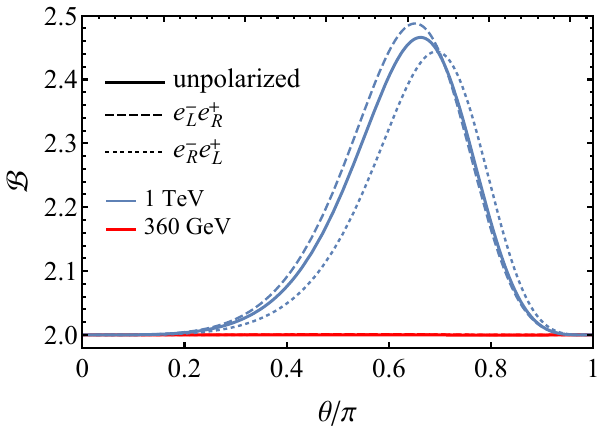}
  \caption{Concurrence $\mathcal{C}$ (left) and Bell variable $\mathcal{B}$ (right) for $e^+e^-\to t\bar{t}$ as a function of scattering angle $\theta$ at $\sqrt{s}=$ 360 GeV (red) and 1 TeV (blue), with different beam polarizations: unpolarized (solid lines), fully polarized $e^-_L e^+_R$ (dashed lines), and $e^-_R e^+_L$ (dotted lines). }
  \label{fig:tt-C&B}
\end{figure}

\begin{figure}[tb]
  \centering
  \includegraphics[width=0.42\linewidth]{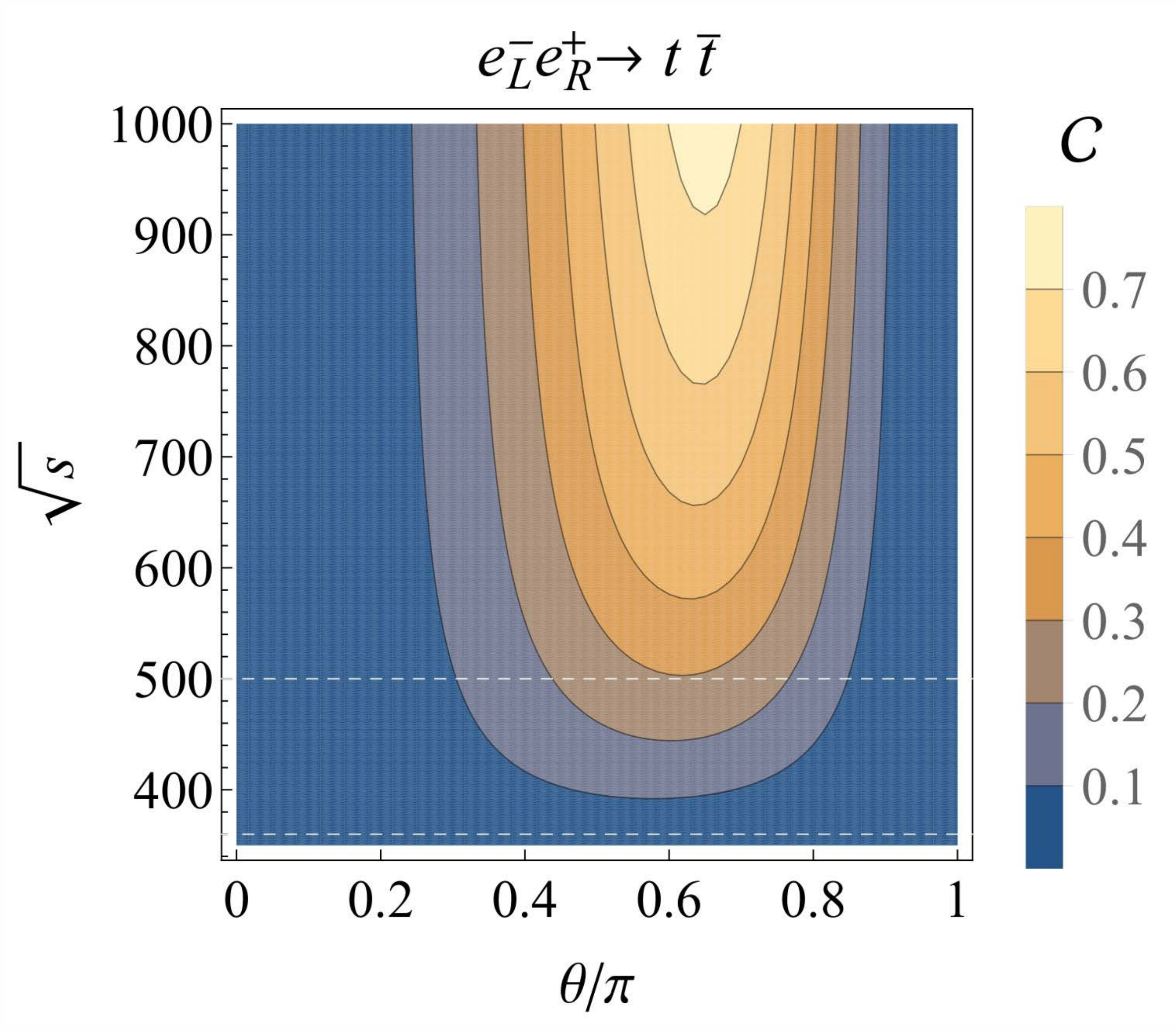}\qquad
  \includegraphics[width=0.42\linewidth]{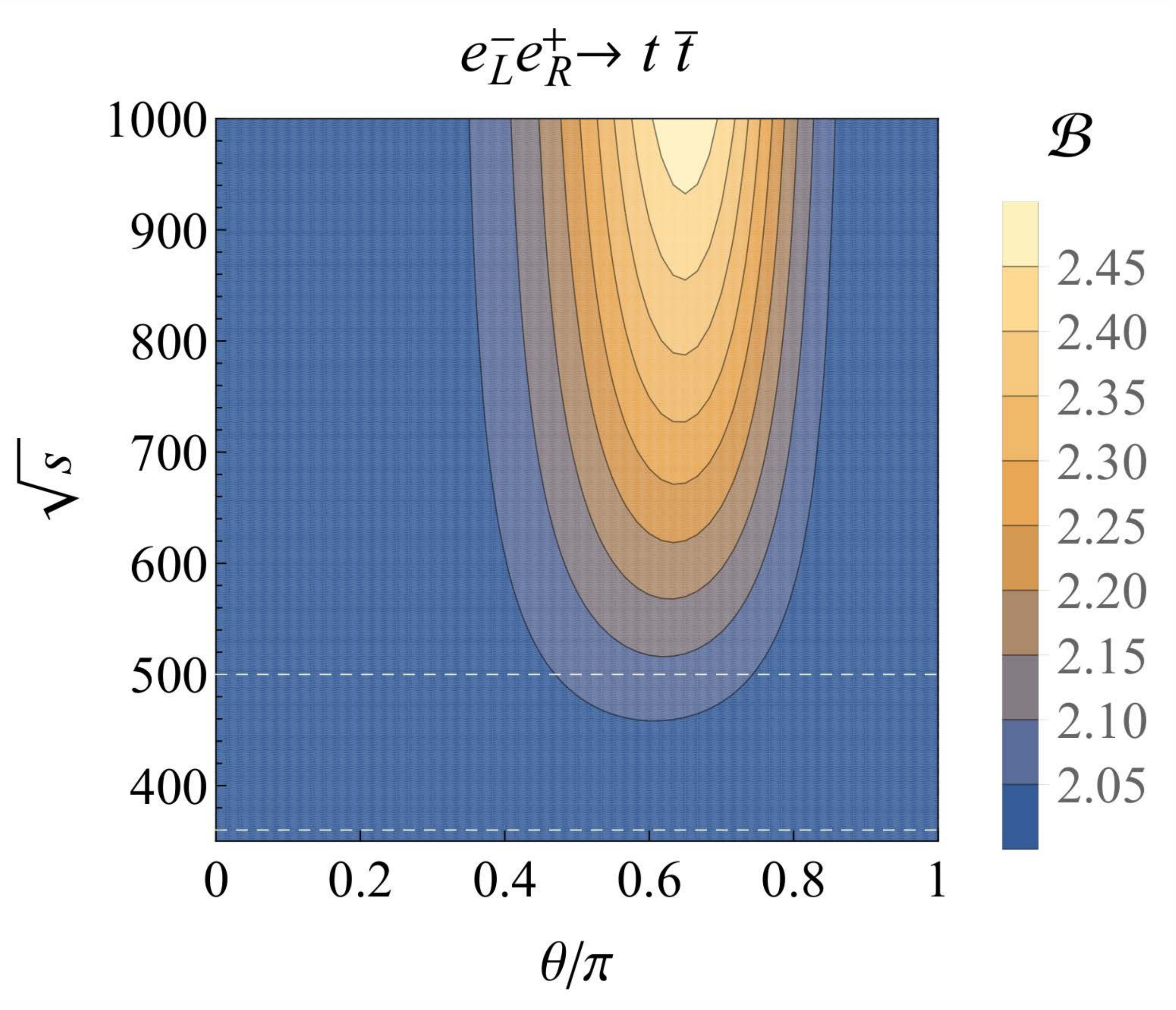}
  \caption{Contour plots for the concurrence (left) and the Bell variable (right) in the plane of $\theta-\sqrt s$ (GeV) for $e_L^-e_R^+\to t\bar{t}$.}
  \label{fig:C&B-pol}
\end{figure}

Quantum entanglement can be  quantitatively measured by concurrence $\mathcal{C}$, while the potential for Bell nonlocality is measured by the Bell variable $\mathcal{B}$. We show $\mathcal{C}$ (left panel) and $\mathcal{B}$ (right panel) in Fig.~\ref{fig:tt-C&B}  at collider energies $\sqrt s =360$ GeV near the $t\bar t$ threshold and 1 TeV, with different beam polarizations $e_L^-e_R^+$ and $e_R^-e_L^+$. We further display their energy dependence in Fig.~\ref{fig:C&B-pol} for the polarization $e_L^-e_R^+$ in the $\theta-\sqrt s$ plane. We see that the entanglement of the $t\bar{t}$ system exhibits a qualitatively different behavior in the threshold and high energy regions. Near the production threshold ($\beta\to 0$), the final spin configuration is separable and entanglement is absent in this limit. However, at higher energies ($\beta\to 1$), the $t\bar{t}$ becomes entangled and Bell nonlocal, reaching the maximum given in Eq.~(\ref{eq:CMax}) at a special scattering angle as given in Eq.~(\ref{eq:ct}).

The beam polarizations shift the peak distribution at the particular scattering angle but do not alter the qualitative features. This can be understood from the chiral structure of $C_{ij}$ as in Eqs.~(\ref{eq:fafv}) and (\ref{eq:cij4tt}). The effects of beam polarization are determined by effective couplings $f_{V,A}^{L,R}$. The similar behavior is a consequence of the approximate symmetry in the ratios of the axial and vector couplings, which vary only slightly from the threshold to 1 TeV, with $f_A^{R}/f_V^{R} \approx 0.6$ and $f_A^{L}/f_V^{L} \approx -0.5$. Therefore, in Fig.~\ref{fig:C&B-pol} and hencefore, we present one polarization $e^-_Le^+_R$ for illustration, unless otherwise specified.

\begin{figure}[tb]
  \centering
  \includegraphics[width=0.45\linewidth]{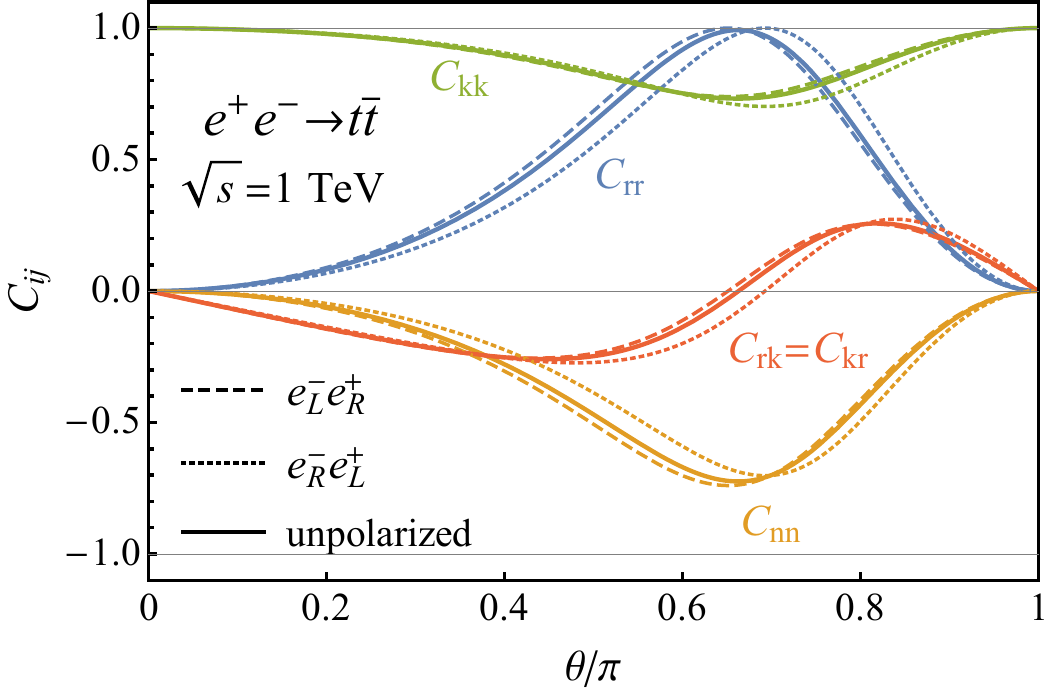}\\
  \includegraphics[width=0.45\linewidth]{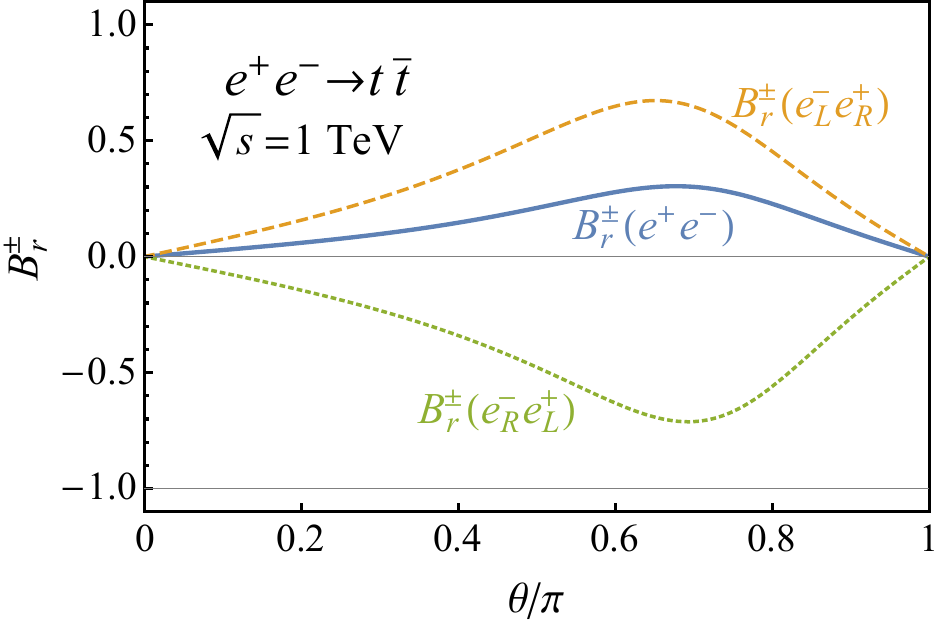}
  \qquad
  \includegraphics[width=0.45\linewidth]{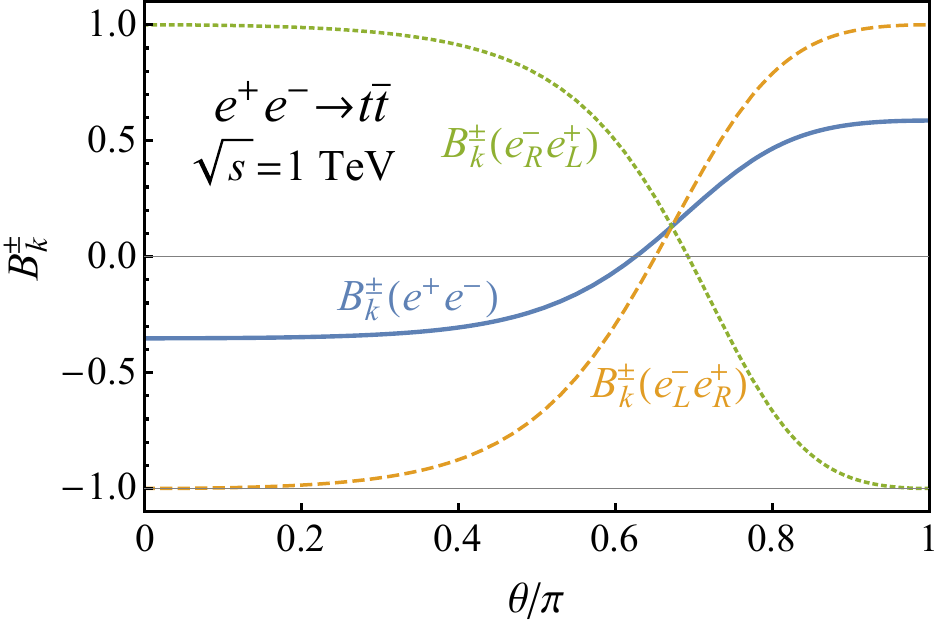}
  \caption{Fano coefficients $C_{ij}$ (top), $B_r^{\pm}$ (bottom left), and $B_k^{\pm}$ (bottom right) as functions of the scattering angle $\theta$ for $e^+e^- \to t\bar{t}$ at $\sqrt{s}=1~\mathrm{TeV}$. In the top panel, blue, green, and yellow lines correspond to $C_{nn}$, $C_{kk}$, and $C_{rr}$, respectively, while red lines denote $C_{nk}$ and $C_{kn}$. For all panels, solid, dashed, and dotted lines represent the unpolarized, $LR$-polarized, and $RL$-polarized configurations, respectively.}
  \label{fig:tt-cij&bi}
\end{figure}

\begin{figure}[tb]
  \centering
  \includegraphics[width=0.4\linewidth]{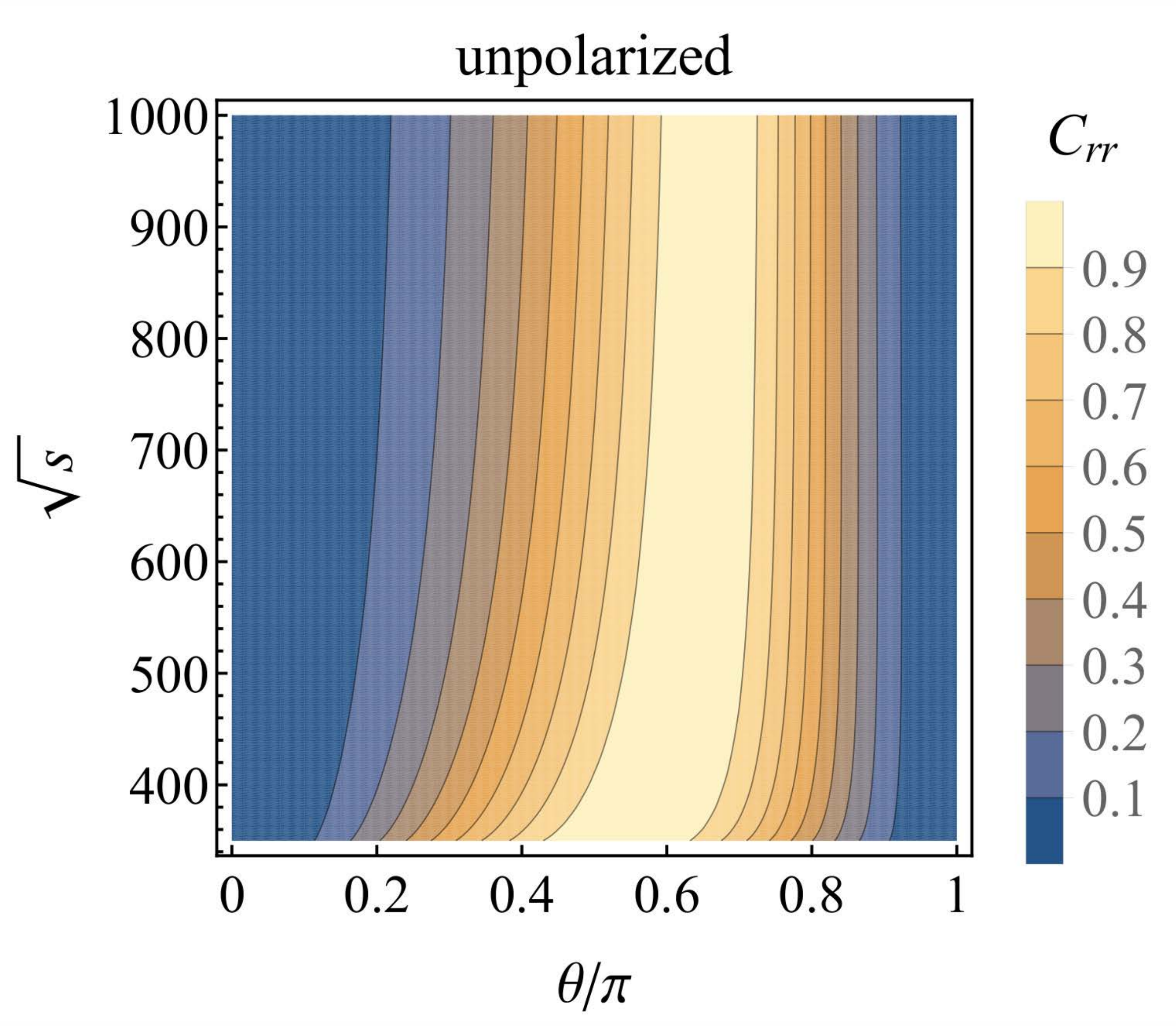}\qquad
  \includegraphics[width=0.4\linewidth]{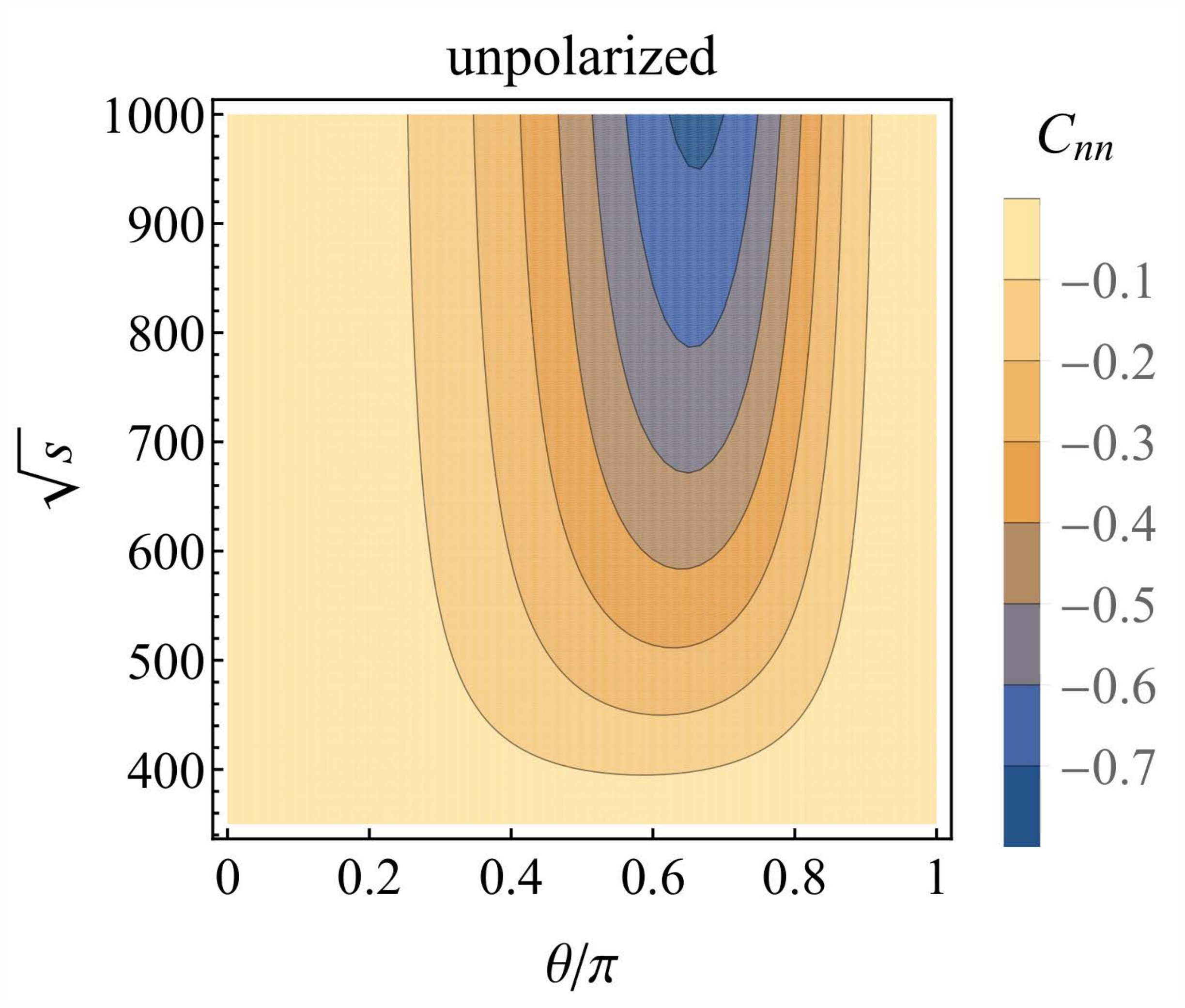}\\
  \includegraphics[width=0.4\linewidth]{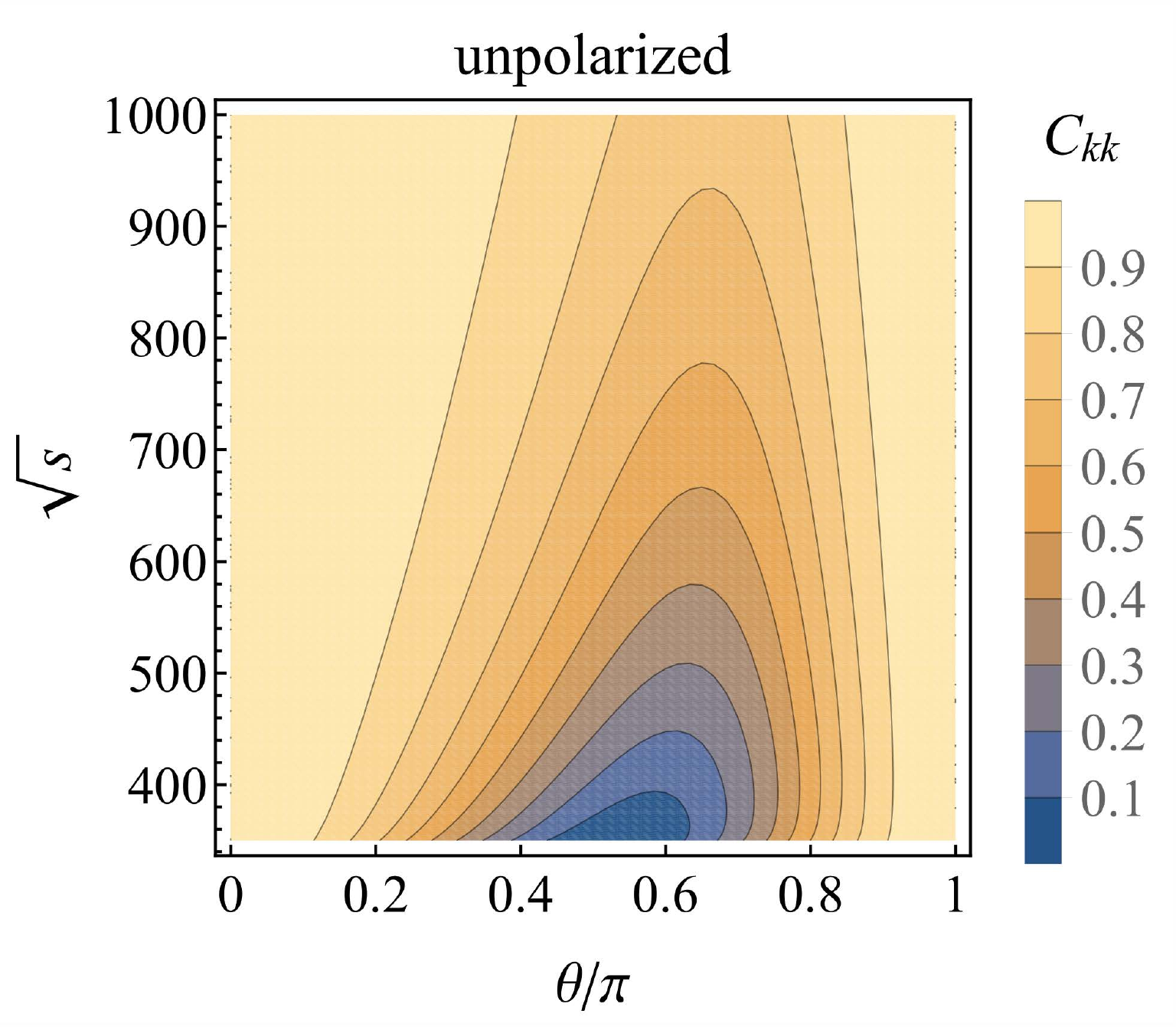}\qquad
  \includegraphics[width=0.4\linewidth]{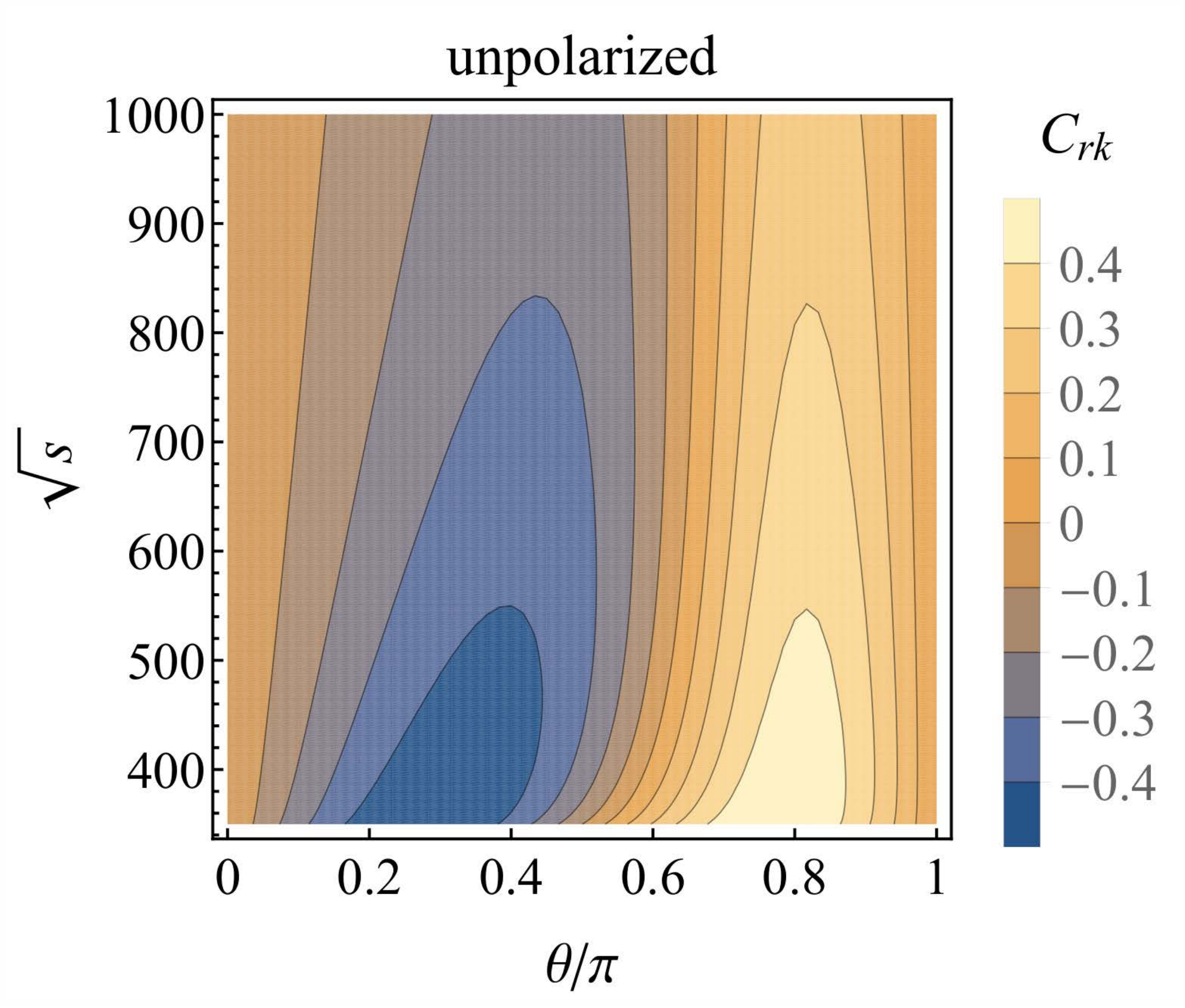}
  \caption{Contour plots for the Fano coefficients $C_{rr}$, $C_{nn}$, $C_{kk}$, and $C_{nk}$ in the plane of $\theta-\sqrt{s}$ (GeV) for unpolarized $e^+e^-\to t\bar{t}$. }
  \label{fig:tt-cij-contour}
\end{figure}
\begin{figure}[tb]
  \centering
  \includegraphics[width=0.4\linewidth]{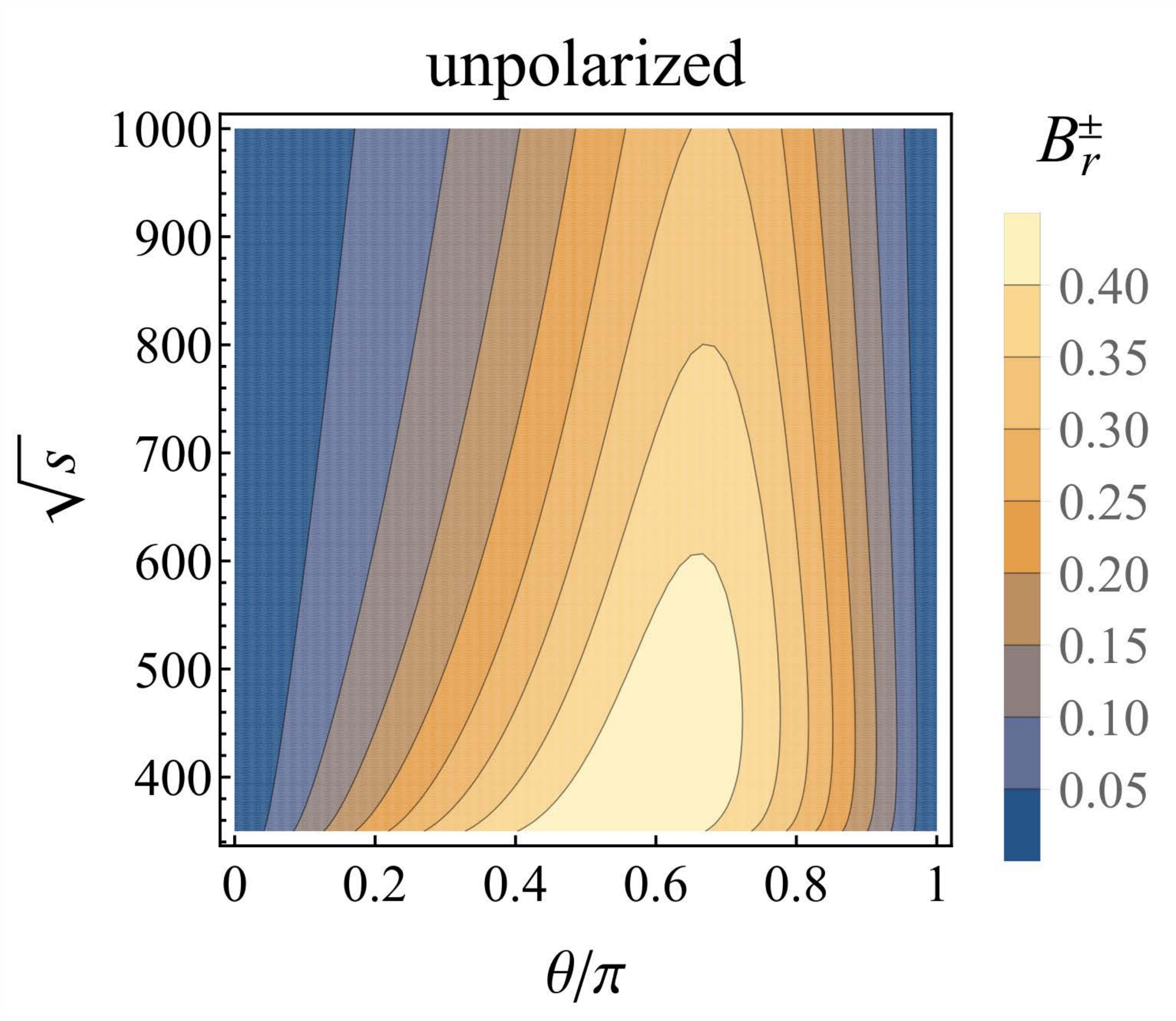}
  \includegraphics[width=0.4\linewidth]{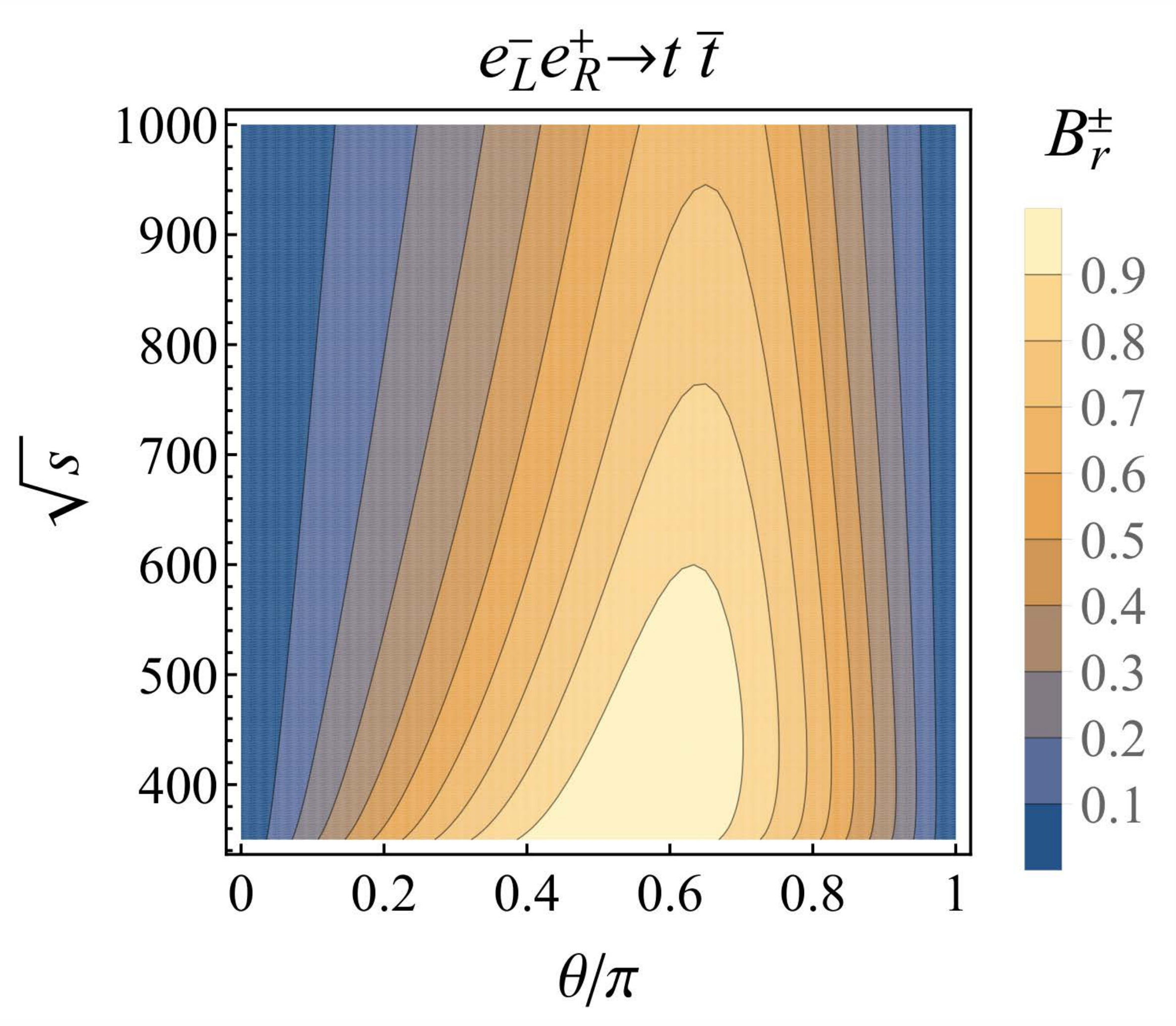}
   \includegraphics[width=0.4\linewidth]{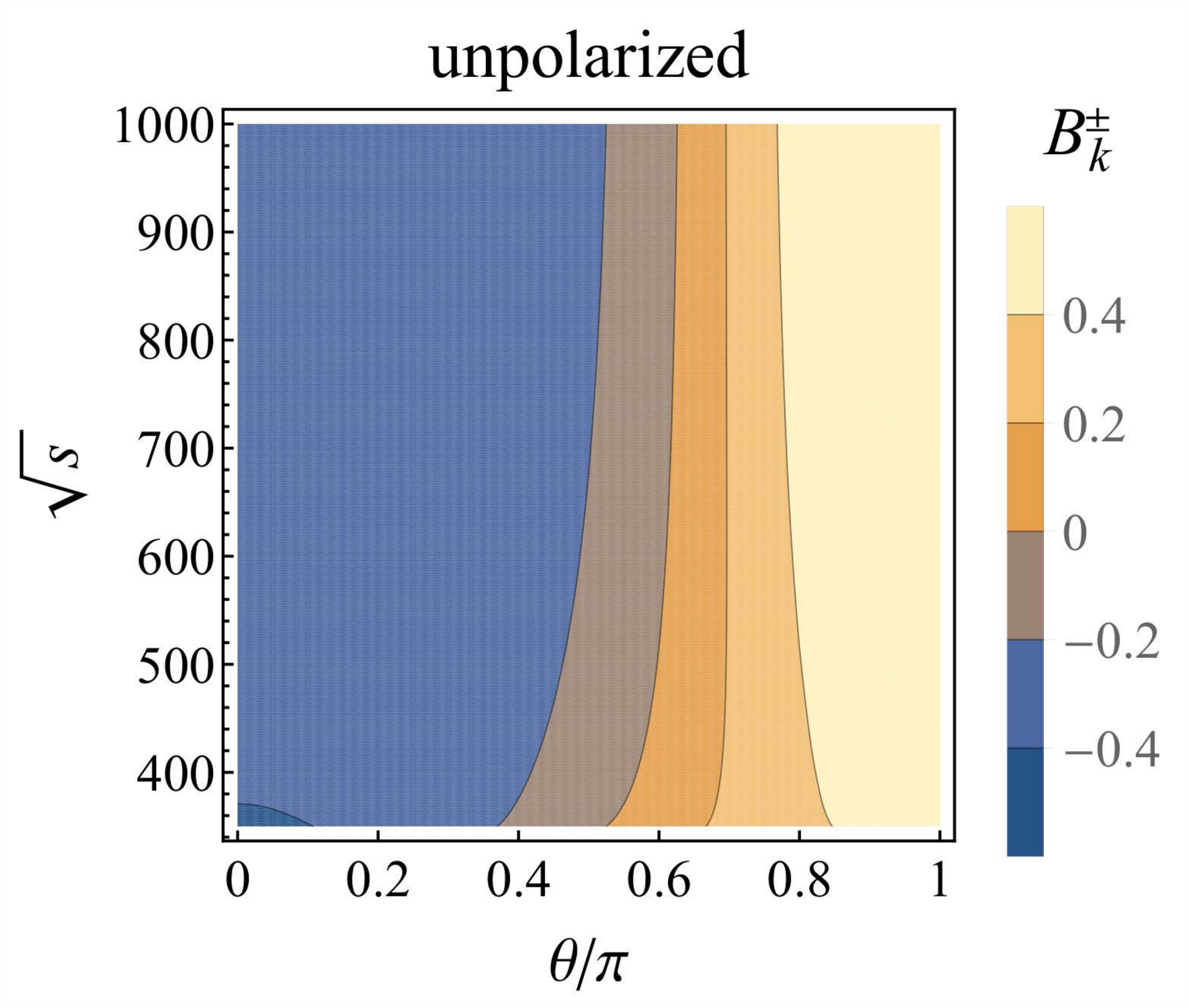}
  \includegraphics[width=0.4\linewidth]{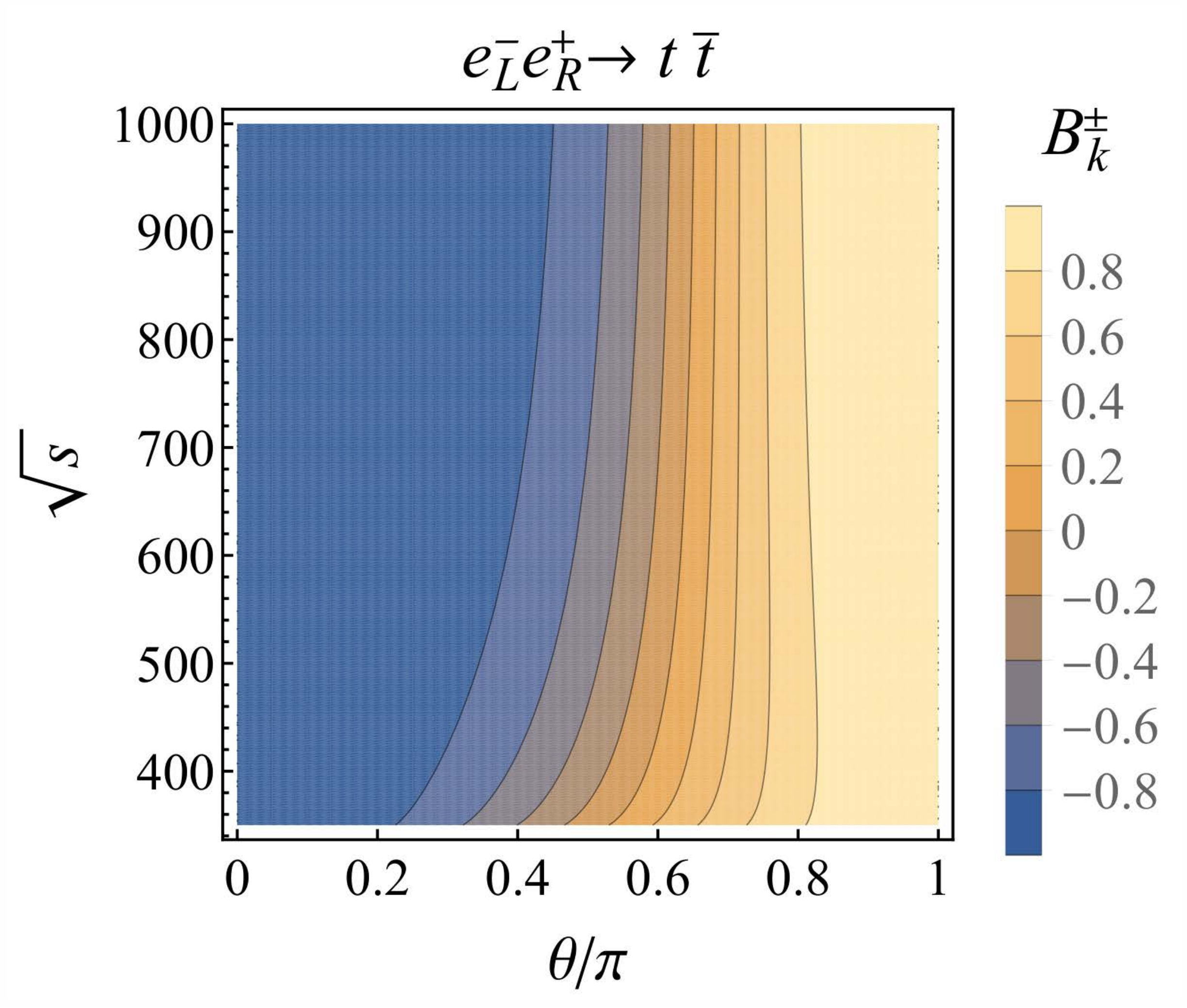}
  \caption{Contour plots of the Fano polarization coefficients, $B^\pm_{r}$ (upper row) and $B^\pm_{k}$ (bottom row), in the plane of $\theta-\sqrt{s}$ (GeV) for unpolarized (left panels) and $LR$ polarized (right panels) $e^+e^-\to t\bar{t}$.}
  \label{fig:tt-b-contour}
\end{figure}

We now perform quantum tomography for $e^+e^- \to t\bar{t}$ in terms of the Fano coefficients: the spin correlation matrix $C_{ij}$ and the polarization vectors $B_i^\pm$. In Fig.~\ref{fig:tt-cij&bi}, we show their angular dependence at $\sqrt{s}=1~{\rm TeV}$ as an example. The unpolarized curves are obtained from an incoherent average of the polarized configurations, weighted by the corresponding squared matrix elements.
We observe that the correlation matrix $C_{ij}$ depends only mildly on beam polarization, whereas the single-particle polarizations $B_i^\pm$ are strongly polarization-dependent.
We display the two-dimensional contours in the plane $\theta-\sqrt s$ for $C_{rr}$, $C_{nn}$, $C_{kk}$, and $C_{nk}$ in Fig.~\ref{fig:tt-cij-contour} with unpolarized beams. For the polarization vectors $B^\pm_{r,k}$ in Fig.~\ref{fig:tt-b-contour}, we show the results for the unpolarized and $LR$ cases, while the $RL$ configuration yields similar distribution, but with an opposite sign (see Fig.~\ref{fig:tt-cij&bi}).
As expected, the top quark longitudinal polarization $B^\pm_{k}$ in $e^-_R e^+_L$ ($e^-_L e^+_R$) collisions peaks at $\theta = 0$ ($\theta = \pi$), while the transverse polarization $B^\pm_{r}$ peaks near $\theta \approx 0.65 \pi$, while $B^\pm_{n}=0$.

\begin{figure}[htbp]
  \centering
  \includegraphics[width=0.45\linewidth]{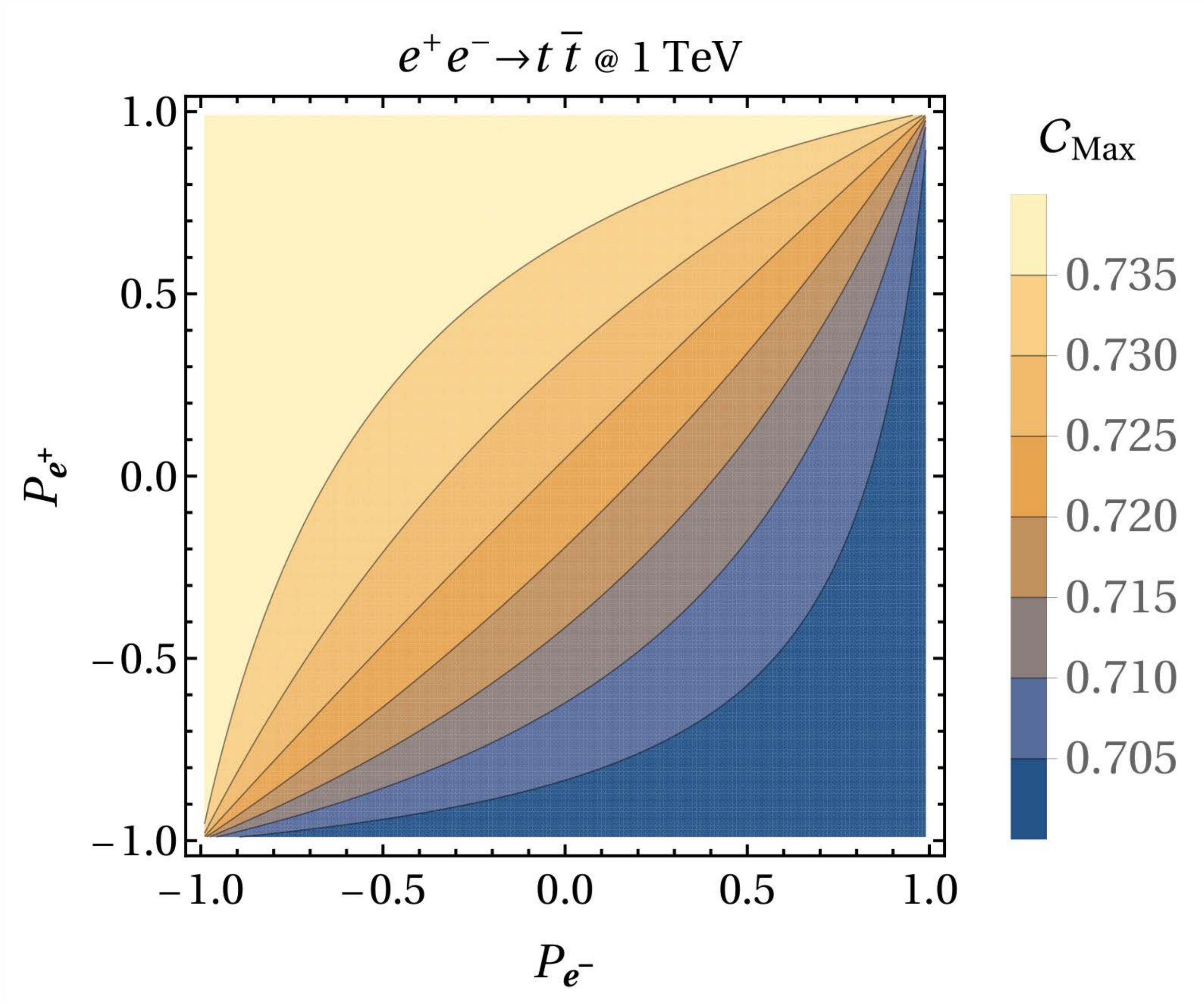}\qquad
  \includegraphics[width=0.45\linewidth]{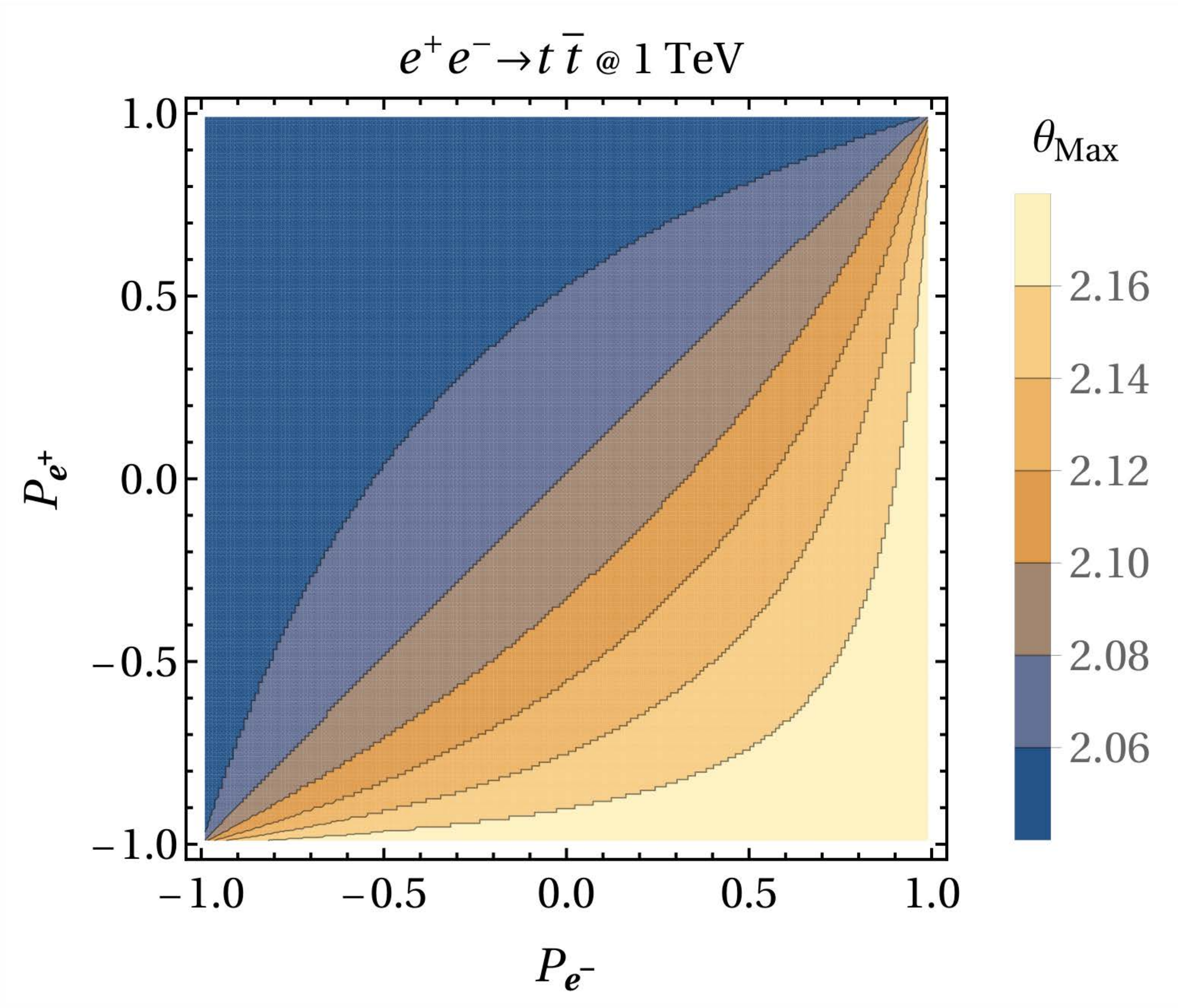}
  \caption{Contour plots for the maximum concurrence $\mathcal{C}_{\rm Max}$ (left) and the corresponding scattering angle $\theta_{\rm Max}$ (right) for $e^+e^- \to t\bar{t}$ at $\sqrt{s}=1~\text{TeV}$, evaluated over the polarizations of the $e^-$ and $e^+$ beams.}
  \label{fig:max_C}
\end{figure}

We further scan the polarization plane $(P_{e^-},P_{e^+})$ at $\sqrt{s}=1~{\rm TeV}$ in Fig.~\ref{fig:max_C}, and extract the maximal concurrence $\mathcal{C}_{\rm Max}$ (left panel) and the corresponding optimal angle $\theta_{\rm Max}$ (right panel). Over the same kinematic range, the Bell variable $\mathcal{B}$ exhibits a similar dependence, while the optimal angles $\theta_{\rm Max}$ for $LR$ and $RL$ differ due to parity-violating electroweak couplings.
Specifically, $|f_A^R/f_V^R|$ is slightly larger than $|f_A^L/f_V^L|$ indicating that the electroweak interaction pushes the result farther away from the QED limit in the $e^-_R e^+_L \!\to\! t\bar{t}$ process.

This also explains why $LR$ is more inclined towards the $\pi/2$ symmetric distribution characteristic of QED than $RL$, as seen earlier in Fig.~\ref{fig:tt-cij&bi}.
This angular asymmetry further explains the difference in the maximum values of $\mathcal{B}$ between the two polarization configurations.
At the angles $\theta_{\rm Max}^{LR}$ and $\theta_{\rm Max}^{RL}$, where the Bell nonlocality reaches its peak, the quantum state of the $t\bar{t}$ system can be expressed as
\begin{equation}
\ket{\psi} \propto
\frac{1}{\sqrt{2}}\left(\ket{\uparrow\uparrow}+\ket{\downarrow\downarrow}\right)
+ \epsilon\,\frac{1}{\sqrt{2}}\left(\ket{\uparrow\downarrow}+\ket{\downarrow\uparrow}\right),
\quad \rm with \ \  \epsilon \propto \frac{\sqrt{1-\beta^2}}{\sin\theta_{\mathrm{Max}}}.
\end{equation}
In the high-energy limit $\epsilon \to 0$, corresponding to the massless approximation, the $t\bar{t}$ pair forms a Bell triplet state and the entanglement reaches its theoretical maximum.
As $|\epsilon|$ increases, quantum entanglement and Bell nonlocality both decrease.
Since $|\sin\theta_{\rm Max}^{LR}|>|\sin\theta_{\rm Max}^{RL}|$, one finds $|\epsilon_{LR}|<|\epsilon_{RL}|$, which leads to a larger maximum value of $\mathcal{B}$ for the $LR$ polarization. The same trend is also reflected in the concurrence.

%%%%%%%%%%%%%%%%%%%%%%%%%%%%%%%%%%%%%%%%%%%%%%%%%%%%%%%%%%%
\subsubsection{Second Stabilizer R\'{e}nyi Entropy for \texorpdfstring{$t\bar{t}$}{tt}} \label{sec:magic-tt}
%%%%%%%%%%%%%%%%%%%%%%%%%%%%%%%%%%%%%%%%%%%%%%%%%%%%%%%%%%%

\begin{figure}[tb]
  \centering
  \includegraphics[width=0.45\linewidth]{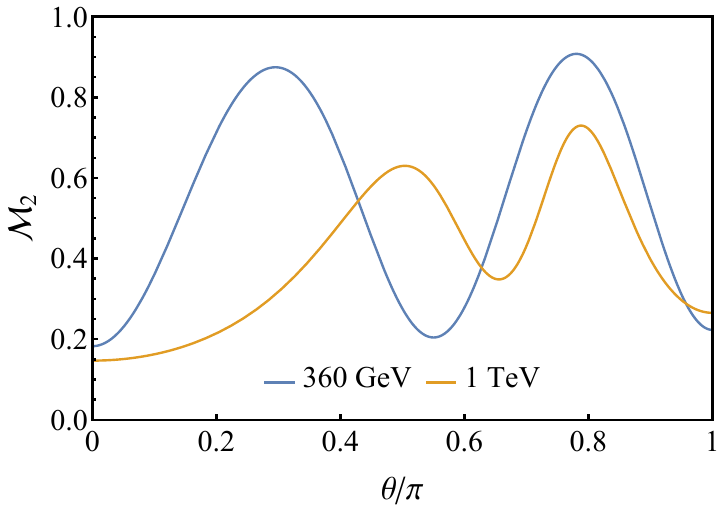}
  \qquad
  \includegraphics[width=0.45\linewidth]{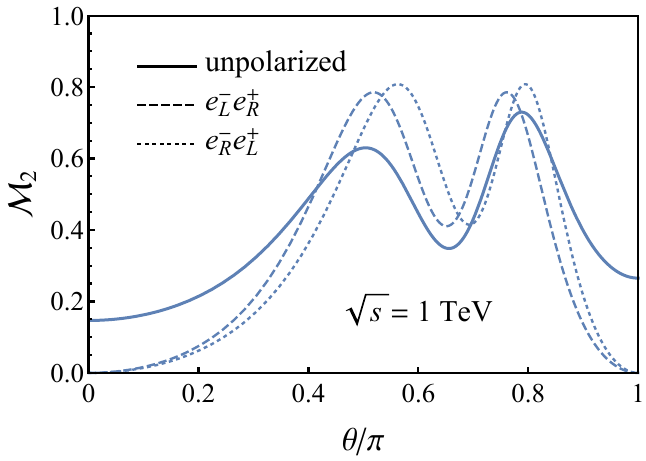}
  \caption{The SSRE $\mathcal{M}_2$ for $e^+e^-\to t\bar{t}$ as a function of scattering angle $\theta$. Left: Unpolarized results at $\sqrt{s}=360$ GeV (blue) and 1 TeV (orange). Right: Results at $\sqrt{s}=1$ TeV for unpolarized (solid), $LR$ (dashed), and $RL$ (dotted) beam polarizations.}
  \label{fig:Magic-tt}
\end{figure}

\begin{figure}[tb]
  \centering
  \includegraphics[width=0.4\linewidth]{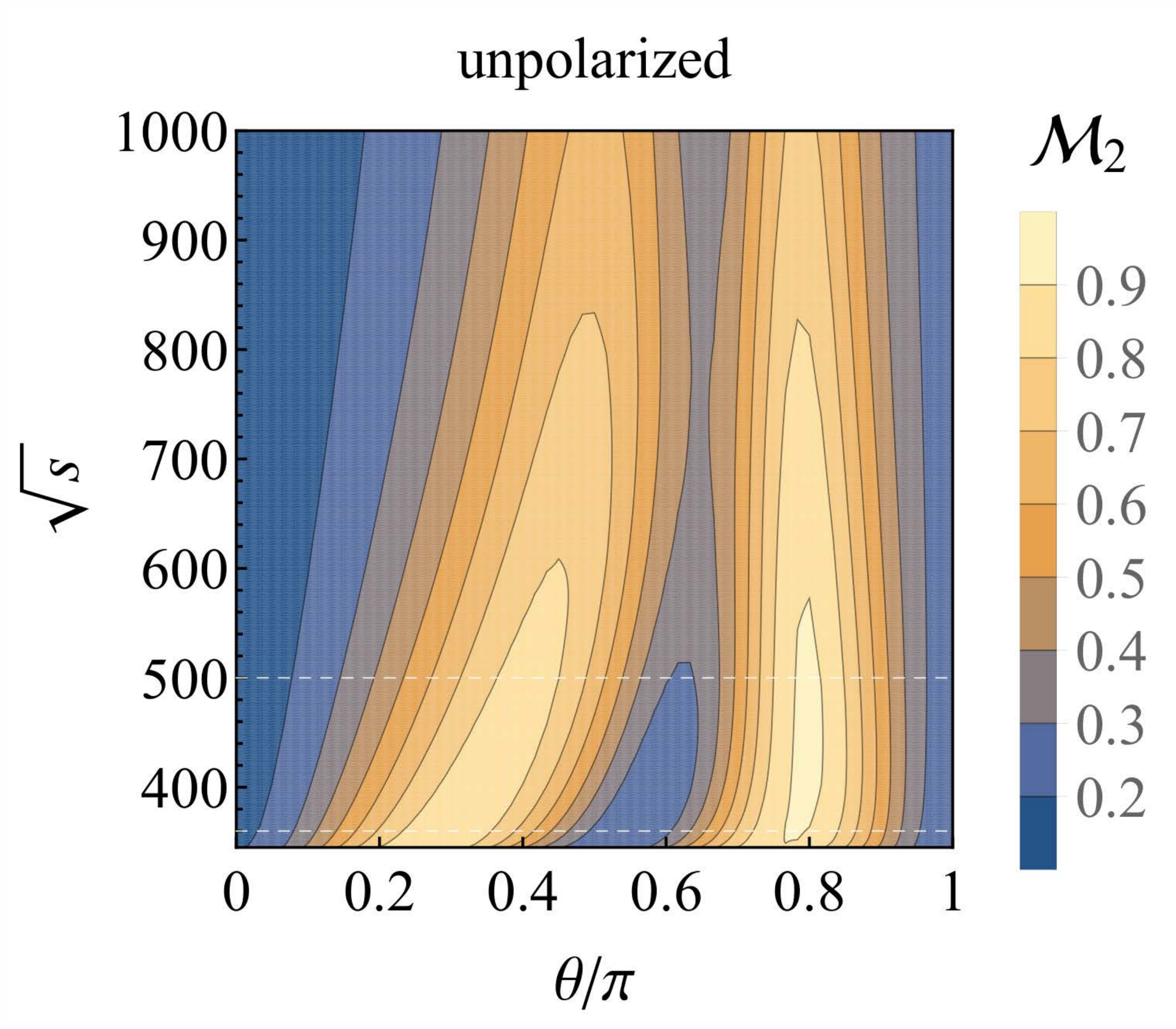}  \qquad
  \includegraphics[width=0.4\linewidth]{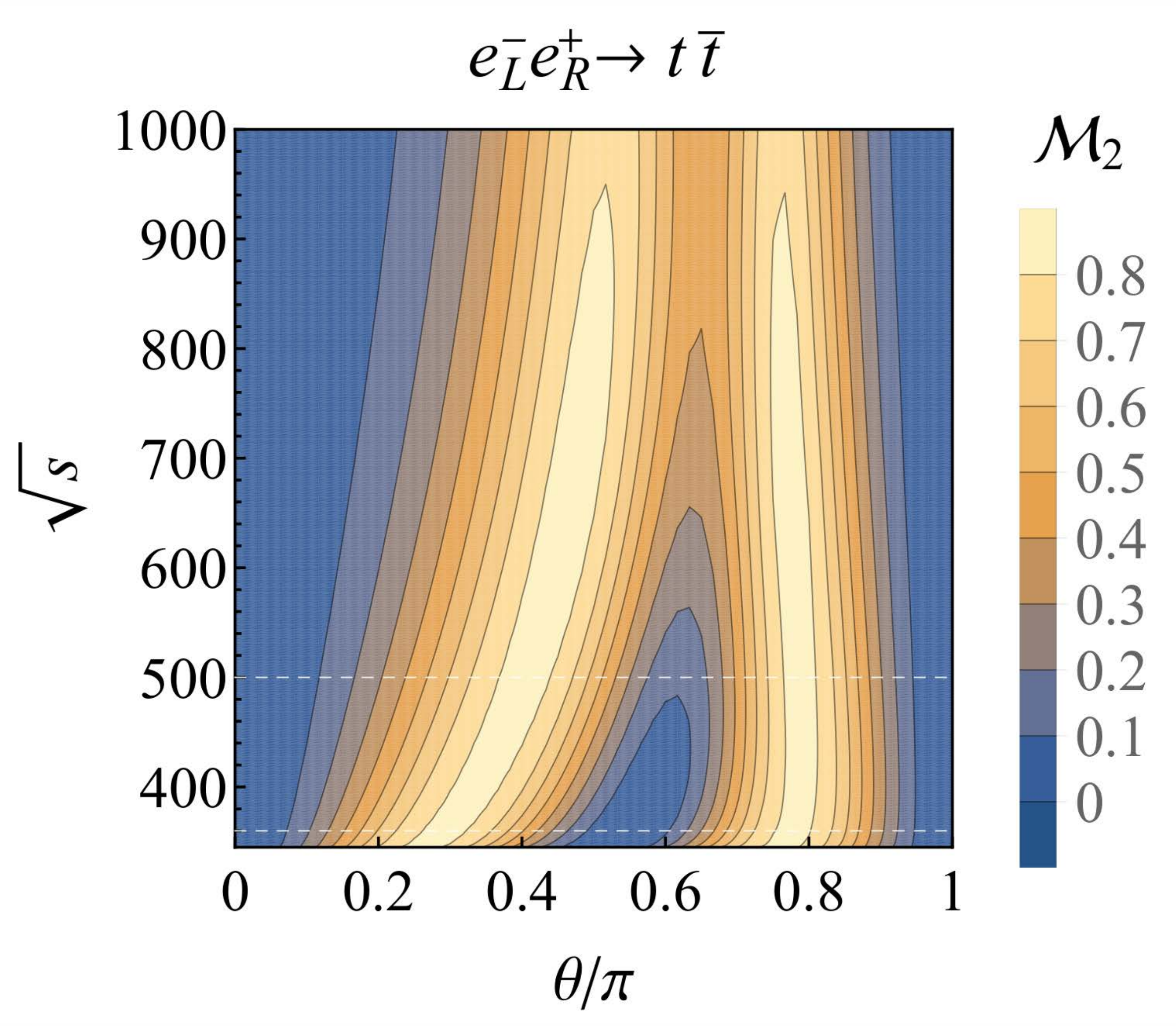}
  \caption{Contour plots for the SSRE $\mathcal{M}_2$ in the plane of $\theta-\sqrt{s}$ (GeV) for the unpolarized (left) and $LR$ (right) polarized $e^+e^-\to t\bar{t}$ process.}
  \label{fig:magic-contour-tt}
\end{figure}

The SSRE magic ($\mathcal{M}_2$) quantifies the deviation of a quantum state from the set of stabilizer states, serving as a measure of the resource required for non-Clifford quantum simulation.
We present the angular dependence of $\mathcal{M}_2$ on the $t\bar{t}$ system in Fig.~\ref{fig:Magic-tt} near the threshold ($360~\text{GeV}$) and at high energy ($1~\text{TeV}$) in the left panel, and the impact of beam polarizations for  $\sqrt{s}=1~\text{TeV}$ in the right panel.
The full angular and energy dependence is mapped in the $\theta-\sqrt{s}$ plane in Fig.~\ref{fig:magic-contour-tt}. We display only the $LR$ polarization, as the $RL$ case exhibits a qualitatively similar distribution.

A distinctive feature of the magic distribution is its ``twin-peak'' structure in the angular distribution, flanking a central valley. Near the $t\bar t$ threshold, the distribution is symmetric about $\theta=\pi/2$, reflecting the dominance of the parity-conserving QED contribution.
As $\sqrt{s}$ increases, the interference from the $Z$ boson exchange introduces parity violation, breaking this symmetry and shifting the regions of large magic. Comparing the unpolarized and fully polarized cases reveals that beam polarization significantly enhances the magnitude of magic. Notably, significant magic persists near the threshold in the forward/backward scattering, where entanglement is negligible. In contrast, magic vanishes in the forward and backward limits. A detailed physical interpretation of these regimes—specifically why separable states can possess magic,  while maximally entangled states may not—is provided in Sec.~\ref{subsec:SecIII-summary}.

\begin{figure}[tb]
  \centering
  \includegraphics[width=0.45\linewidth]{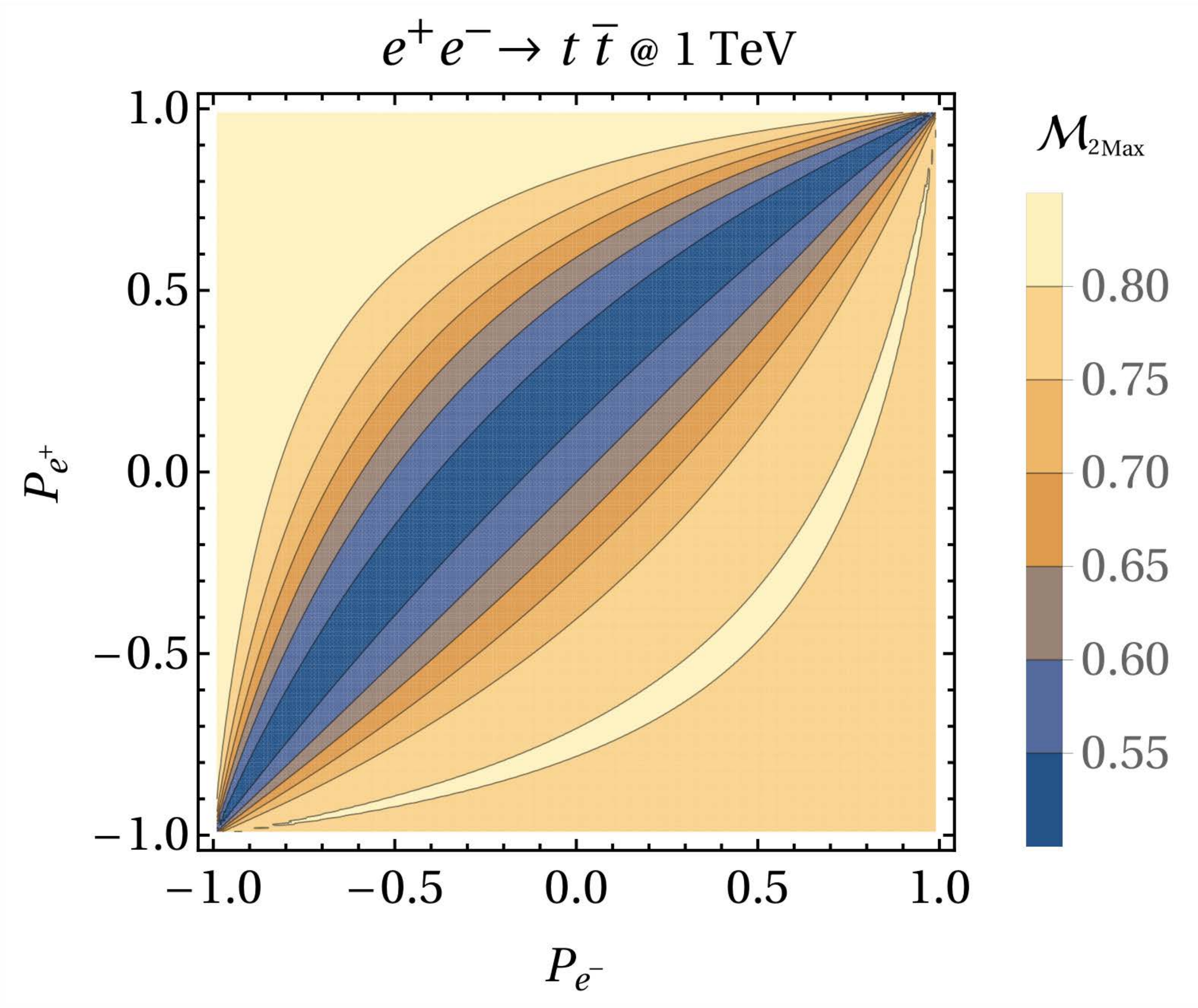} \qquad
  \includegraphics[width=0.45\linewidth]{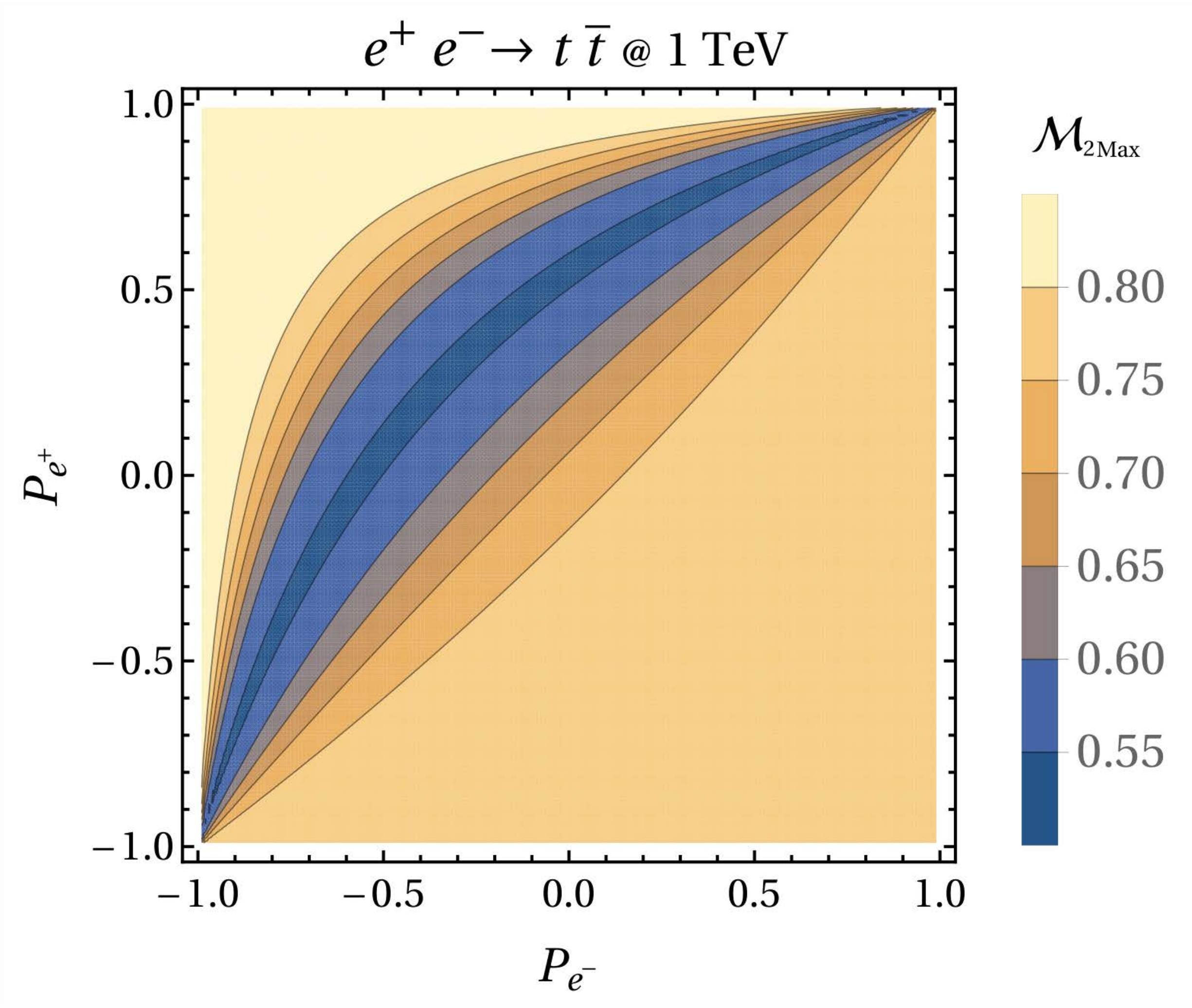}\\
  \includegraphics[width=0.45\linewidth]{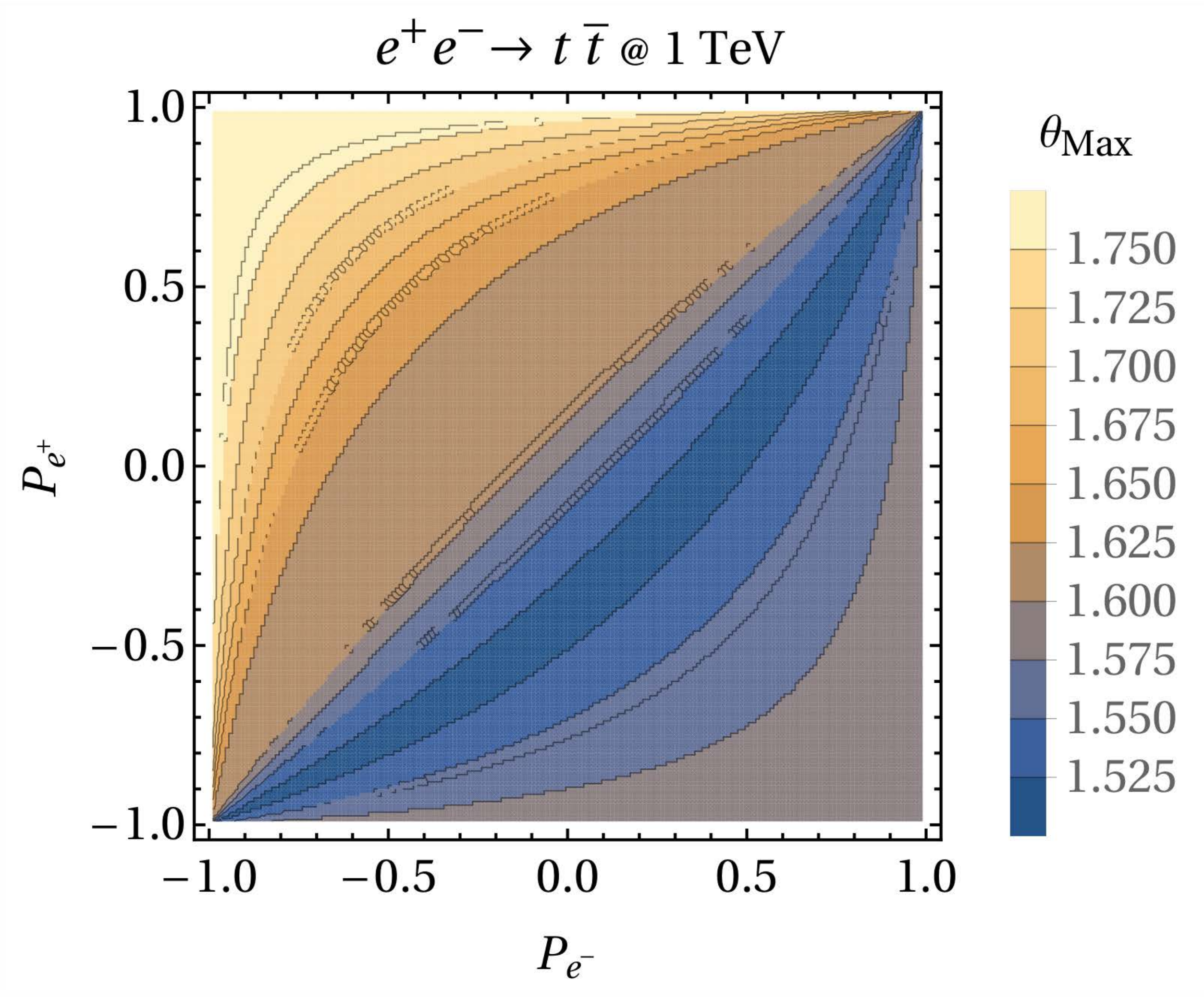} \qquad
  \includegraphics[width=0.45\linewidth]{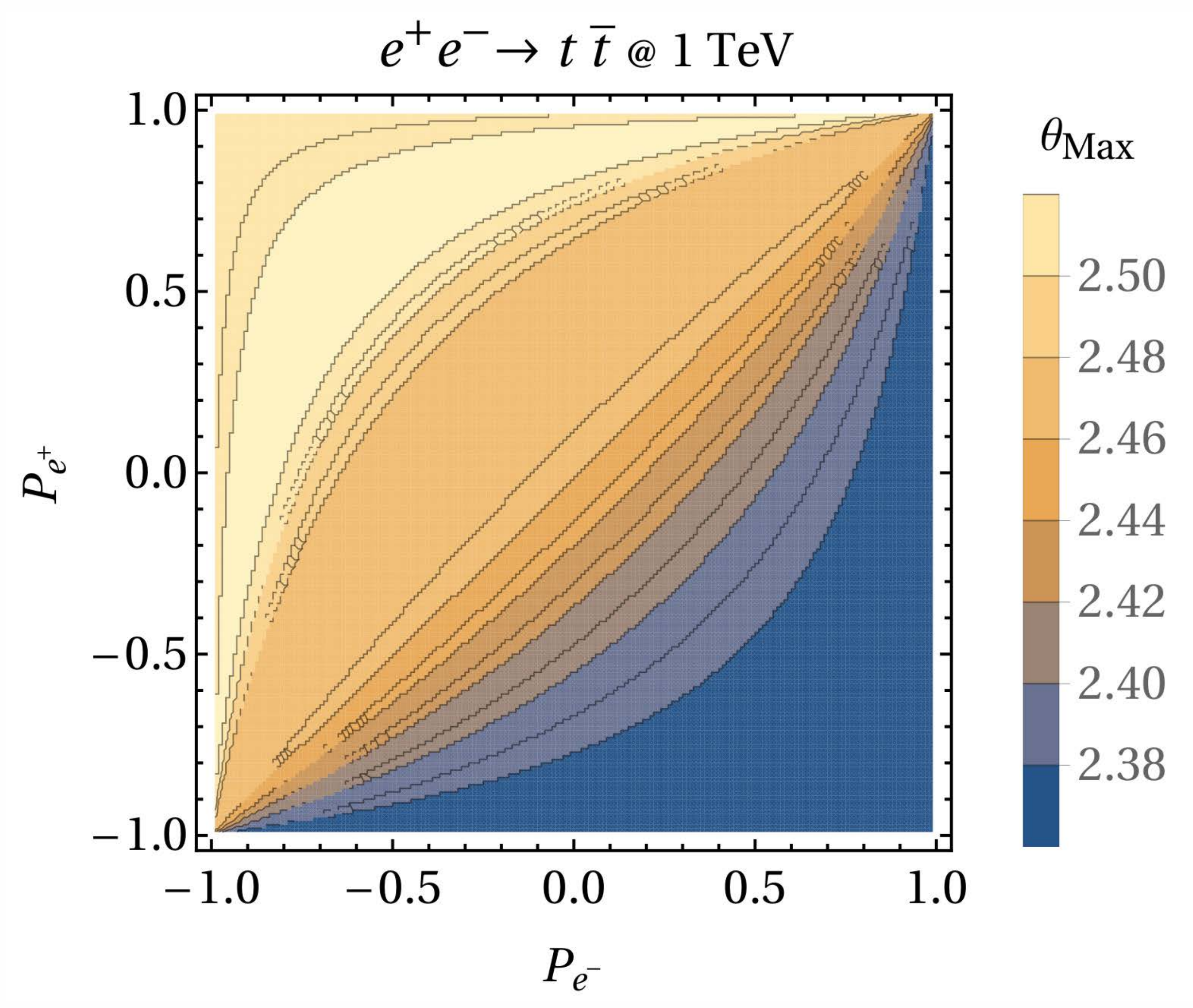}
  \caption{The maximum SSRE $\mathcal{M}_2$ (top) and the corresponding scattering angle $\theta$ (bottom) for $e^+e^- \to t\bar{t}$ at $\sqrt{s}=1~\text{TeV}$, evaluated over the polarizations of the $e^-$ and $e^+$ beams. The left (right) panels correspond to the left (right) peaks of the twin-peak structure in Figs.~\ref{fig:Magic-tt} and \ref{fig:magic-contour-tt}.}
  \label{fig:tt-maxM&theta}
\end{figure}

Finally, Fig.~\ref{fig:tt-maxM&theta} explores the optimal conditions for generating magic resources by plotting the maximum values of $\mathcal{M}_2$ over the $(P_{e^-},P_{e^+})$ polarization plane at $\sqrt{s}=1~\text{TeV}$, for the two left-right peaks separately (upper panels), and their corresponding angles (bottom panels).
The maximal $\mathcal{M}_2$ increases monotonically as the beams approach full polarization.
This behavior contrasts with entanglement and Bell nonlocality, which are largely robust against polarization due to their dependence on $C_{ij}$.
Instead, magic shows a strong dependence on the local polarization vectors $B^\pm_i$, making it a highly polarization-tunable quantum resource in high-energy collisions.

%%%%%%%%%%%%%%%%%%%%%%%%%%%%%%%%%%%%%%%%%%%%%%%%%%%%%%%%%%%
\subsection{Lepton Pair Production}
\label{subsec:mumu}
%%%%%%%%%%%%%%%%%%%%%%%%%%%%%%%%%%%%%%%%%%%%%%%%%%%%%%%%%%%

The quantum states of the $\mu^+\mu^-$ and $\tau^+\tau^-$ systems from $e^+e^-$ collisions exhibit qualitative similarities, governed by lepton universality. The primary differences arise from kinematic threshold effects because of their distinct masses. The $\tau^+\tau^-$ process  has been discussed in detail in Ref.~\cite{Han:2025ewp}. In this work, we focus on the polarization effects in the $\mu^+\mu^-$ system, treating it as an effectively massless two-particle quantum system in the high-energy regime.

%%%%%%%%%%%%%%%%%%%%%%%%%%%%%%%%%%%%%%%%%%%%%%%%%%%%%%%%%%%
\subsubsection{The Quantum State for \texorpdfstring{$\mu^+\mu^-$}{mm}}
%%%%%%%%%%%%%%%%%%%%%%%%%%%%%%%%%%%%%%%%%%%%%%%%%%%%%%%%%%%

In the massless limit $m_\mu=0$, chirality coincides with helicity. Consequently, the coefficients $c$ and $d$ in Eq.~\eqref{eq:psi} vanish, and $a$ and $b$ are given in Eq.~\eqref{eq:ttab}  with $\beta=1$.

\begin{figure}[tb]
  \centering
  \includegraphics[width=0.45\linewidth]{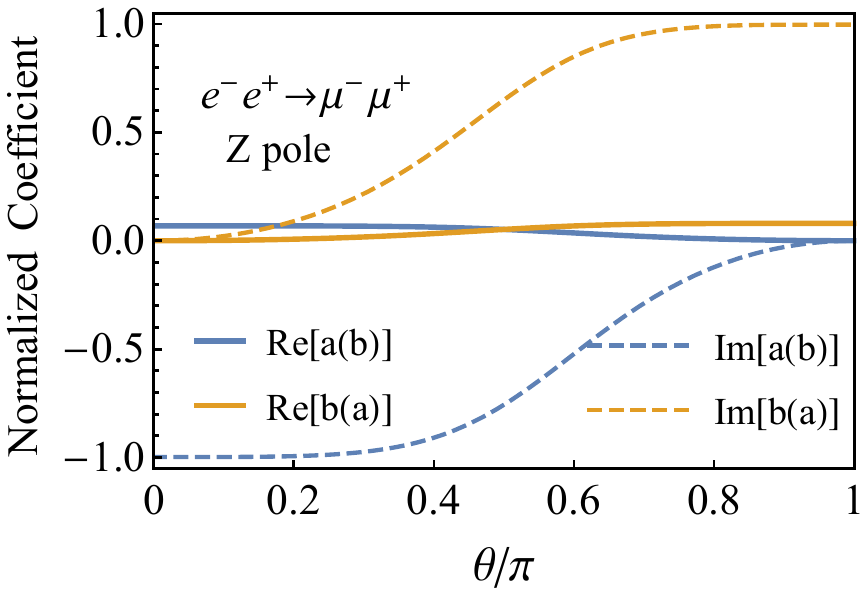}  \qquad
  \includegraphics[width=0.43\linewidth]{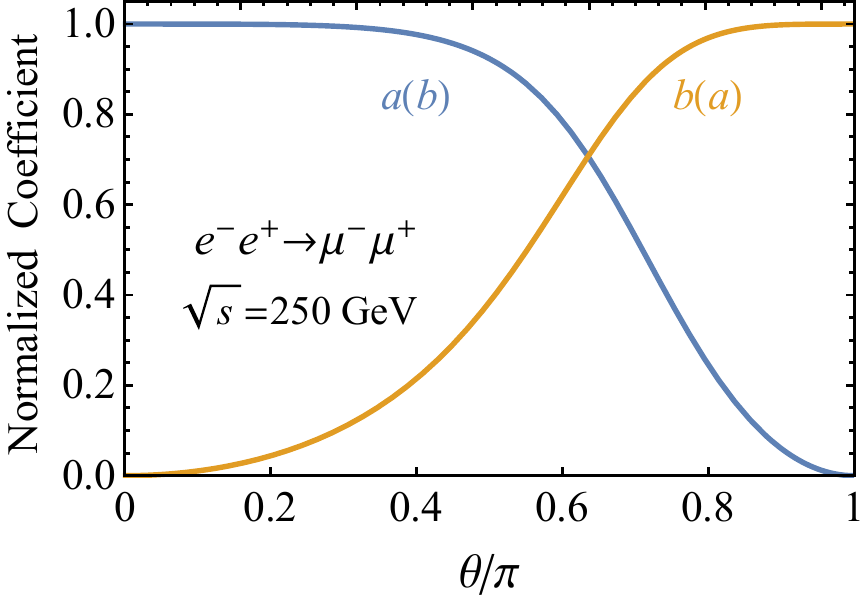}
  \caption{Coefficients $a$ and $b$ in Eq.~\eqref{eq:psi} for the quantum state of the $\mu^+ \mu^-$ system for $LR$ ($RL$) beam polarizations as functions of the scattering angle $\theta$ at $\sqrt{s}=m_Z$ (left) and $250$ GeV (right). }
  \label{fig:state_coefficient_mm}
\end{figure}
Figure \ref{fig:state_coefficient_mm} illustrates the angular dependence of the coefficients $a$ and $b$ for the $LR$ and $RL$ configurations at the $Z$ pole ($\sqrt{s}=m_Z=91.2~\rm{GeV}$) and at $\sqrt{s}=250~\rm{GeV}$.
Near the $Z$ resonance, the process is dominated by $s$-channel $Z$ exchange, and the propagator is largely controlled by its imaginary part $i s\Gamma_Z/m_Z$.
In this regime, the photon contribution is subleading, so the $\gamma$--$Z$ interference induces only a small relative phase in the normalized coefficients, while the chiral structure is entirely governed by the $Z$ couplings $g_L$ and $g_R$.
Since the magnitude of the resonance term overwhelms the real part of the interference, the relative phase between the $LR$ and $RL$ configurations is suppressed.
As shown in Fig.~\ref{fig:state_coefficient_mm} (left), the coefficients $a$ and $b$ are comparable in magnitude for both polarization channels, showing only a dependence on $\theta$ governed by angular factors $(1 \pm \cos\theta)$. Entanglement thus arises from the coherent superposition of these spin configurations.

In contrast, at energies far above the $Z$ pole ($\sqrt{s} \gg m_Z$), the interference between the $\gamma$ and $Z$ amplitudes becomes significant. The coefficients $a$ and $b$ for the beam polarization configuration $RL$ ($LR$) simplify to
\begin{equation}
a \propto \,\big(1 + \frac{f_A^{R/L}}{f_V^{R/L}} \big)\, (1\pm \cos\theta ), \quad
b \propto \,\big(1 - \frac{f_A^{R/L}}{f_V^{R/L}} \big)\, ( 1\mp\cos\theta ),
\end{equation}
with numerical values ${f_A^{R}}/{f_V^{R}}\approx 0.36$ and ${f_A^{L}}/{f_V^{L}}\approx -0.32$ at high energies. Unlike the resonance region, here the normalized coefficients depend sensitively on $\cos\theta$. Since $f_A/f_V\neq 0$, the coefficients $a$ and $b$ show a significant forward–backward asymmetry. Consequently, the condition $|a|=|b|$, corresponding to maximally entangled Bell states, is displaced from the symmetric point $\theta=\pi/2$, reflecting the underlying parity violation.

\begin{figure}[tb]
  \centering
  \includegraphics[width=0.5\linewidth]{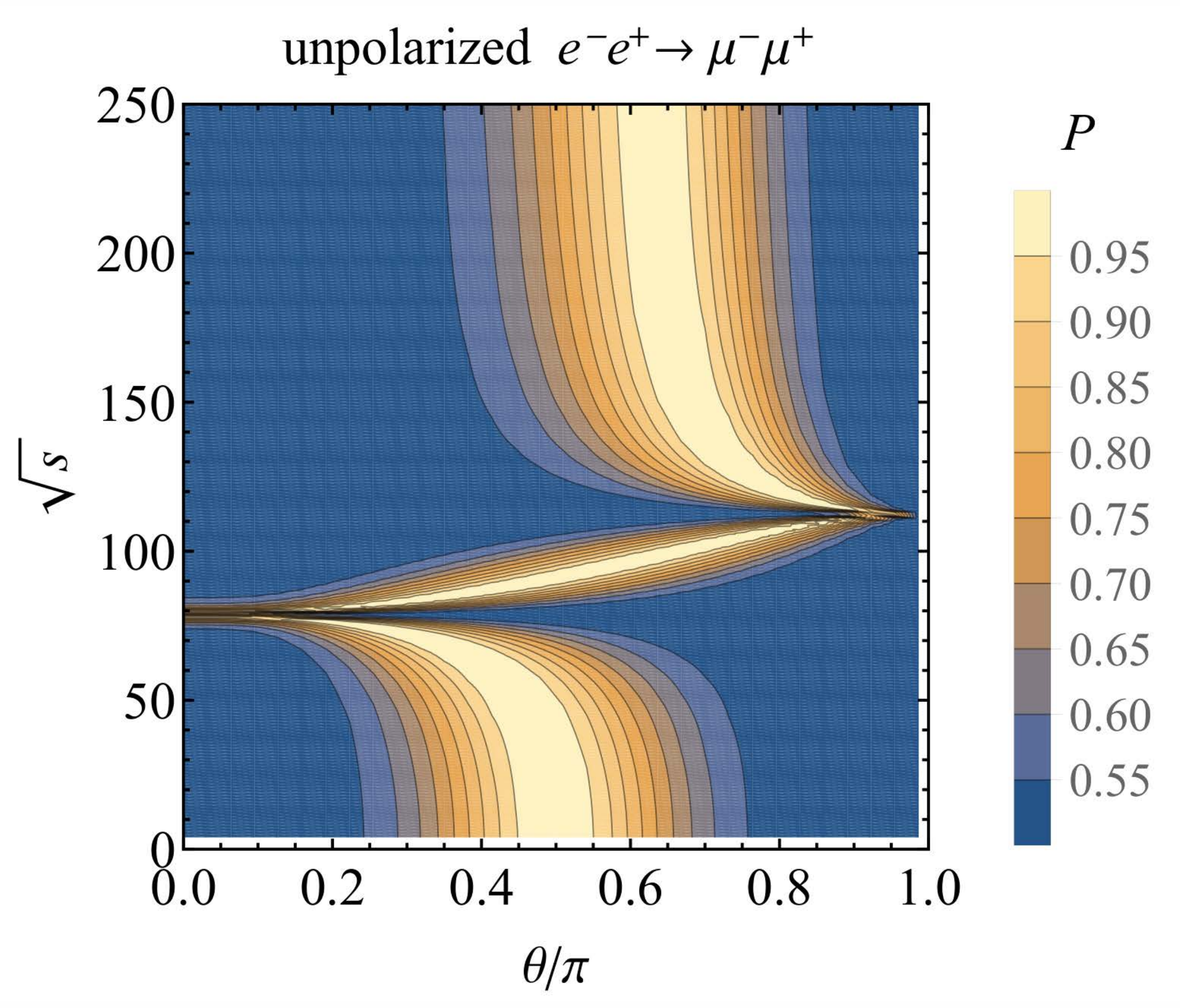}
  \label{fig:rho-mm}
  \caption{Contour plot for the purity of the $\mu^+\mu^-$ final state at an $e^+e^-$ collider, shown in the kinematic plane of scattering angle $\theta$ and c.m.~energy $\sqrt{s}$ in GeV. }
  \label{fig:llpurity}
\end{figure}

For unpolarized or partially polarized beams, the system is described by a density matrix. In the massless limit, the diagonal basis is preserved, and the mixed state takes the form
\begin{equation}\label{eq:Cdiagonal}
\rho_{\rm mixed}=\begin{pmatrix}      \rho_{\uparrow\uparrow,\uparrow\uparrow} &0&0 & \rho_{\uparrow\uparrow,\downarrow\downarrow}\\
        0&0 &0 & 0\\
        0&0&0 & 0 \\
                \rho_{\downarrow\downarrow,\uparrow\uparrow}&0&0 & \rho_{\downarrow\downarrow,\downarrow\downarrow} \\
    \end{pmatrix}.
\end{equation}
where $ \rho_{\uparrow\uparrow,\uparrow\uparrow}$, $\rho_{\uparrow\uparrow,\downarrow\downarrow}$, $\rho_{\downarrow\downarrow,\uparrow\uparrow}$, and
$\rho_{\downarrow\downarrow,\downarrow\downarrow}$ denote the non-zero entries of the density matrix.
While fully polarized collisions yield a pure state, the unpolarized case results in a mixed state with purity $P < 1$. We calculate the purity according to Eq.~(\ref{eq:P}), as shown in Fig.~\ref{fig:llpurity}. Here and henceforth, the contour plots are displayed up to $\sqrt{s} = 250~\text{GeV}$, as the structure of the quantum state remains unchanged at higher scales.
Compared with the $t\bar t$ state in Fig.~\ref{fig:ttpurity}, a key difference emerges from the $Z$ pole resonance, as well as the mass effect $\beta_t<1$.

%%%%%%%%%%%%%%%%%%%%%%%%%%%%%%%%%%%%%%%%%%%%%%%%%%%%%%%%%%%
\subsubsection{Concurrence, Bell Nonlocality and Quantum Tomography for \texorpdfstring{$\mu^+\mu^-$}{mm}}

\begin{figure}[tb]
  \centering
  \includegraphics[width=0.43\linewidth]{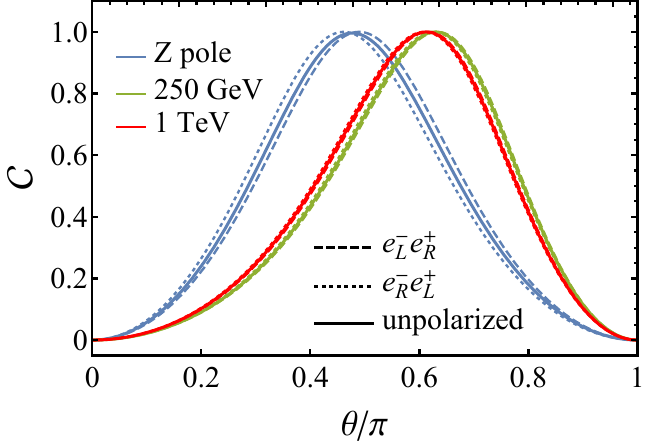}
  \qquad
  \includegraphics[width=0.43\linewidth]{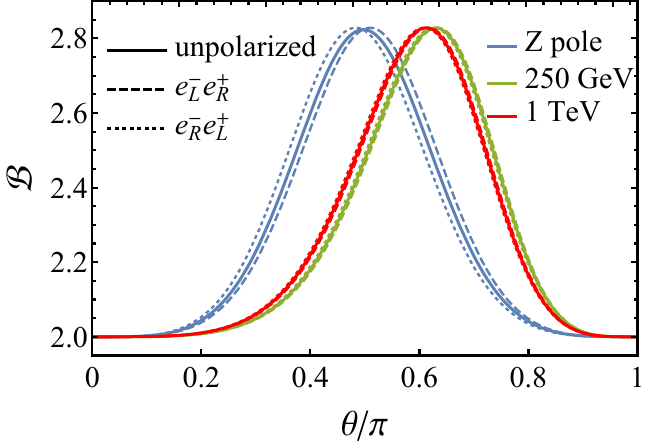}
  \caption{Concurrence $\mathcal{C}$ (left) and Bell nonlocality $\mathcal{B}$ (right) of the $\mu^+\mu^-$ state as a function of scattering angle $\theta$ at three representative energies: $m_Z$ (blue), 250 GeV (green) and 1 TeV (red). Solid, dashed, and dotted lines represent unpolarized, fully polarized $e^-_L e^+_R$ and $e^-_R e^+_L$ configurations, respectively.}
  \label{fig:mm-C&B}
\end{figure}

We evaluate the concurrence $\mathcal{C}$ and Bell nonlocality $\mathcal{B}$ for the $\mu^+\mu^-$ system at three representative energies: the $Z$ pole ($\sqrt{s}=m_Z = 91.2$ GeV), the energy of a Higgs factory ($250~\text{GeV}$), and the higher energy frontier ($1~\text{TeV}$), as shown in Fig.~\ref{fig:mm-C&B}.
For fully-polarized collisions, the final state is pure, and the concurrence reaches unity at specific scattering angles governed by the coupling ratio $f^{L,R}_A/f^{L,R}_V$.
At the $Z$ pole, maximal entanglement occurs near the symmetric point $\theta \approx 0.5\pi$. In contrast, at higher energies well above the $Z$ pole, the optimal angles shift to $\theta \approx 0.63\pi$. While switching between the $LR$ and $RL$ polarization configurations induces a slight shift in these peaks, the overall features of the distributions remain similar.

\begin{figure}[tb]
  \centering
  \includegraphics[width=0.45\linewidth]{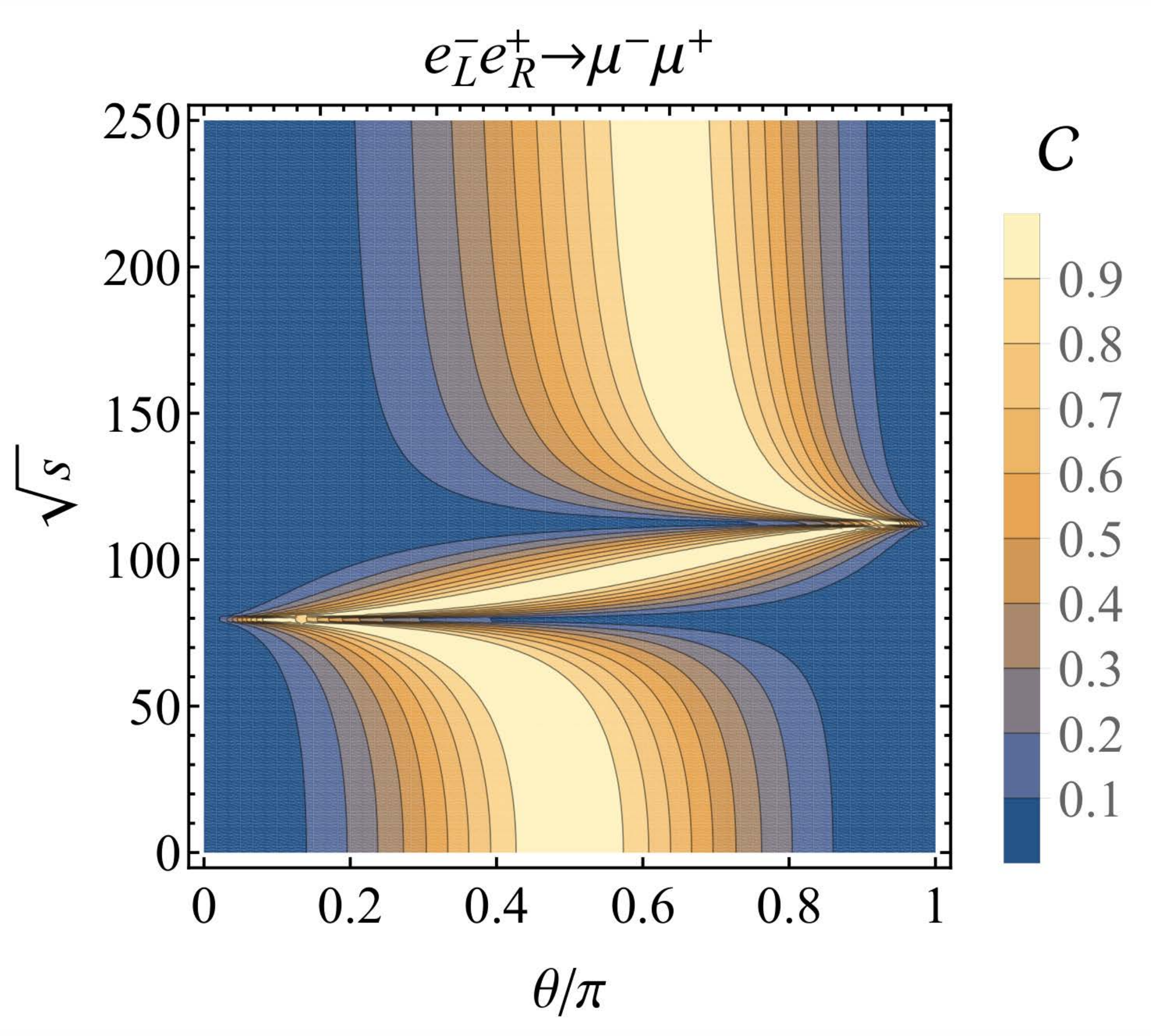} \qquad
  \includegraphics[width=0.45\linewidth]{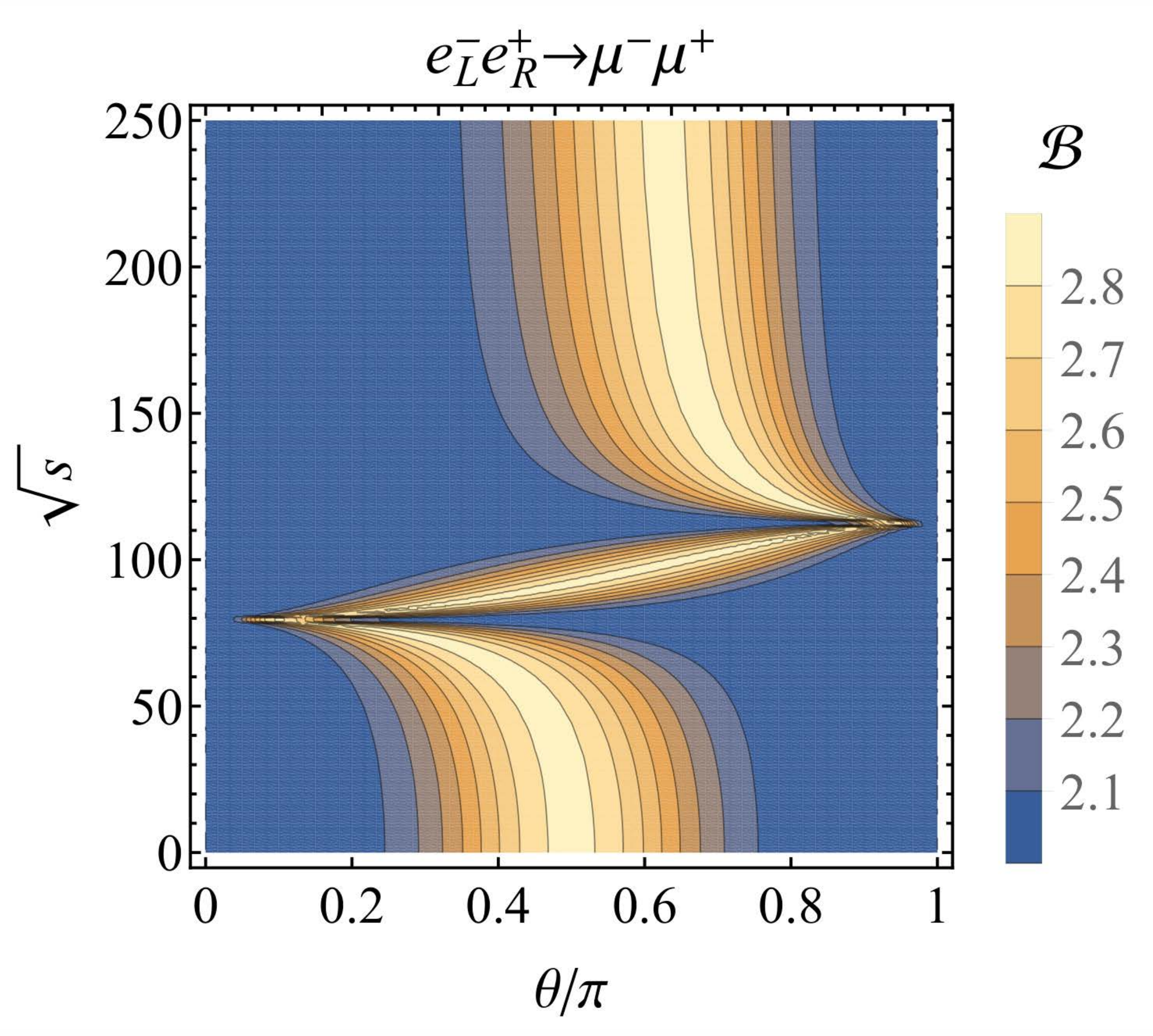}
  \caption{Contour plots for the concurrence (left) and Bell nonlocality (right) in the plane  $\theta-\sqrt{s}$ (GeV) for the fully polarized process $e^-_Le^+_R \to \mu^-\mu^+$.}
  \label{fig:mm_contour_C&B}
\end{figure}

To elucidate the kinematic dependence, we present the contours of $\mathcal{C}$ and $\mathcal{B}$ in the $\theta$--$\sqrt{s}$ plane for a polarized configuration $e^-_Le^+_R\to \mu^-\mu^+$ in Fig.~\ref{fig:mm_contour_C&B}.
The behavior can be categorized into three distinct kinematic regions:
\begin{itemize}
    \item \textbf{Low-energy Region $\sqrt{s} \ll m_Z$:} The process is dominated by QED $s$-channel photon exchange. Parity is approximately conserved as $f_A\sim s/m_Z^2$, resulting in distributions symmetric about $\theta=\pi/2$, with the $\mu^-\mu^+$ state forming a spin triplet.
    \item \textbf{$Z$-pole Region $\sqrt{s} \approx m_Z$:} %Electroweak effects are maximized.
    The resonant enhancement leads to maximal values for $\mathcal{C}$ and $\mathcal{B}$, driving the $\mu^+\mu^-$ system close to a Bell state across a wide angular range.
    %Need to show $f_A/f_V$
    \item \textbf{High-energy Region $\sqrt{s} \gg m_Z$:} Interference between QED and electroweak amplitudes becomes significant. The system approaches a stable asymptotic configuration where quantum correlations become less sensitive to energy variations. However, enhanced parity-violating effects shift the angle of maximal entanglement $\theta_{\mathrm{Max}}$ towards the backward region ($\cos\theta < 0$).
\end{itemize}

\begin{figure}[tb]
  \centering
  \includegraphics[width=0.45\linewidth]{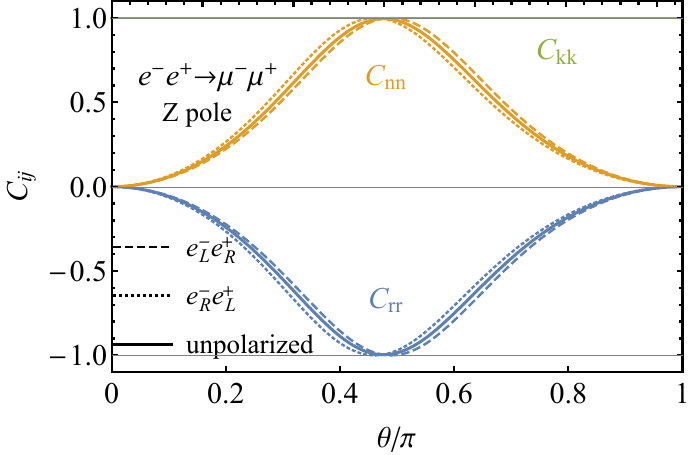}\qquad
  \includegraphics[width=0.45\linewidth]{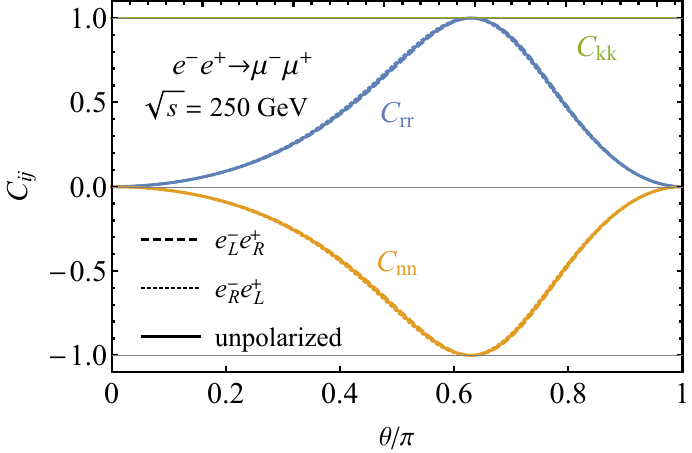}\\
  \includegraphics[width=0.45\linewidth]{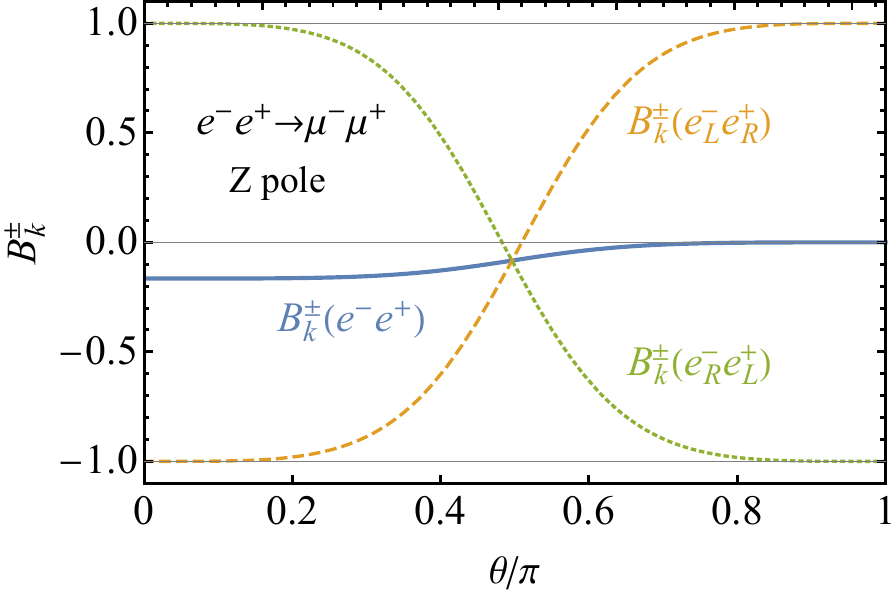}\qquad
  \includegraphics[width=0.45\linewidth]{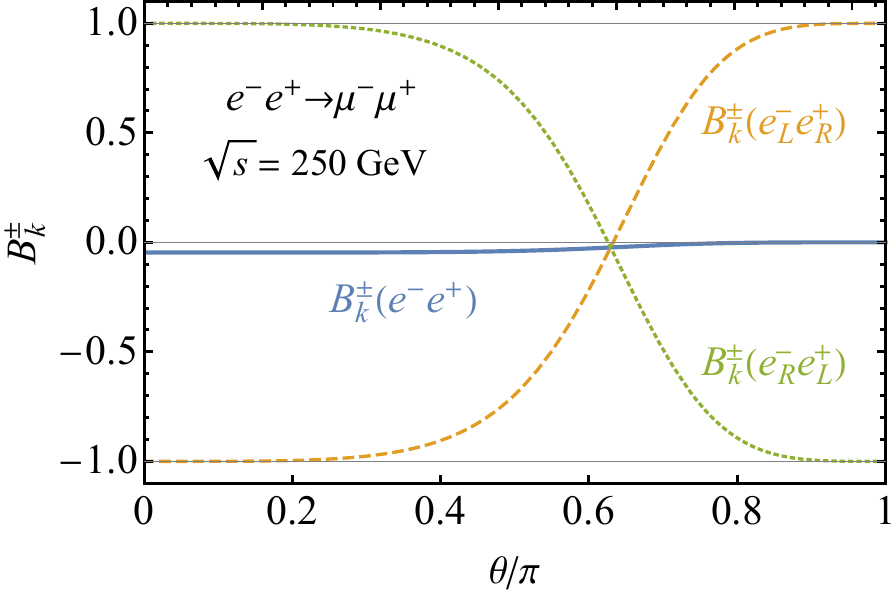}
  \caption{Fano coefficients $C_{ij}$ (top) and $B_k^\pm$ (bottom) for $e^+e^- \to \mu^+\mu^-$ at $\sqrt{s}=m_Z$ (left) and $250$ GeV (right). In the $C_{ij}$ panel, blue, green, and yellow lines correspond to $C_{nn}$, $C_{kk}$, and $C_{rr}$, respectively. For all panels, solid, dashed, and dotted lines represent the unpolarized, $LR$-, and $RL$-polarized configurations, respectively.}
  \label{fig:mm-cij&bi}
\end{figure}
From the perspective of quantum tomography, $\mathcal{C}$ and $\mathcal{B}$ depend on the spin correlation matrix $C_{ij}$. Figure \ref{fig:mm-cij&bi} illustrates the impact of beam polarization on the Fano coefficients $C_{ij}$ and $B_k^\pm$, as a function of the scattering angle $\theta$ at the $Z$ pole and 250 GeV. A striking feature is that while the polarization vector $B_k^\pm$ flips sign or vanishes depending on the $LR/RL$ beam configuration, the spin correlation matrix $C_{ij}$ remains remarkably stable. A similar feature was seen for the $t\bar t$ case in Fig.~\ref{fig:tt-cij&bi}. Again, the ratio of vector to axial-vector currents satisfies the approximate symmetry $\bigl|f_{A}^{L}/f_{V}^{L}\bigr|\simeq \bigl|f_{A}^{R}/f_{V}^{R}\bigr|$. This mechanism ensures that the entanglement structure, governed by $C_{ij}$, is largely independent of the initial beam polarization, making $\mathcal{C}$ and $\mathcal{B}$ polarization-insensitive.
\begin{figure}
  \centering
  \includegraphics[width=0.33\linewidth]{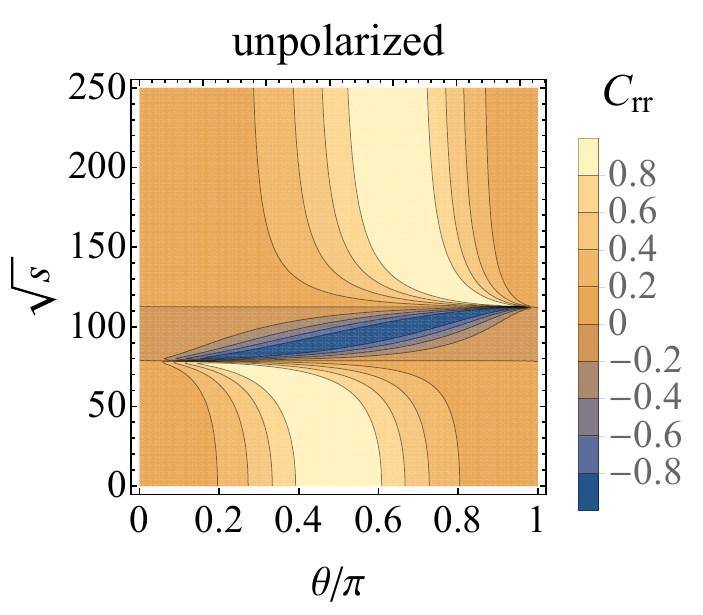}
  \includegraphics[width=0.33\linewidth]{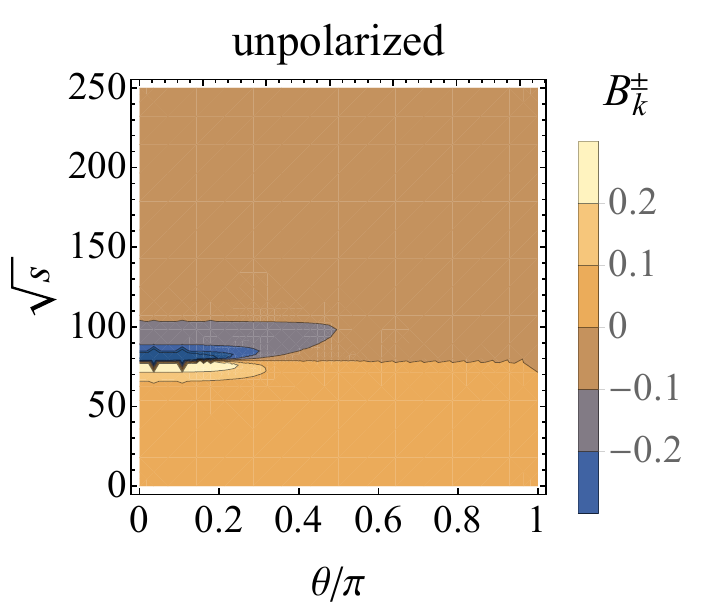}
  \includegraphics[width=0.32\linewidth]{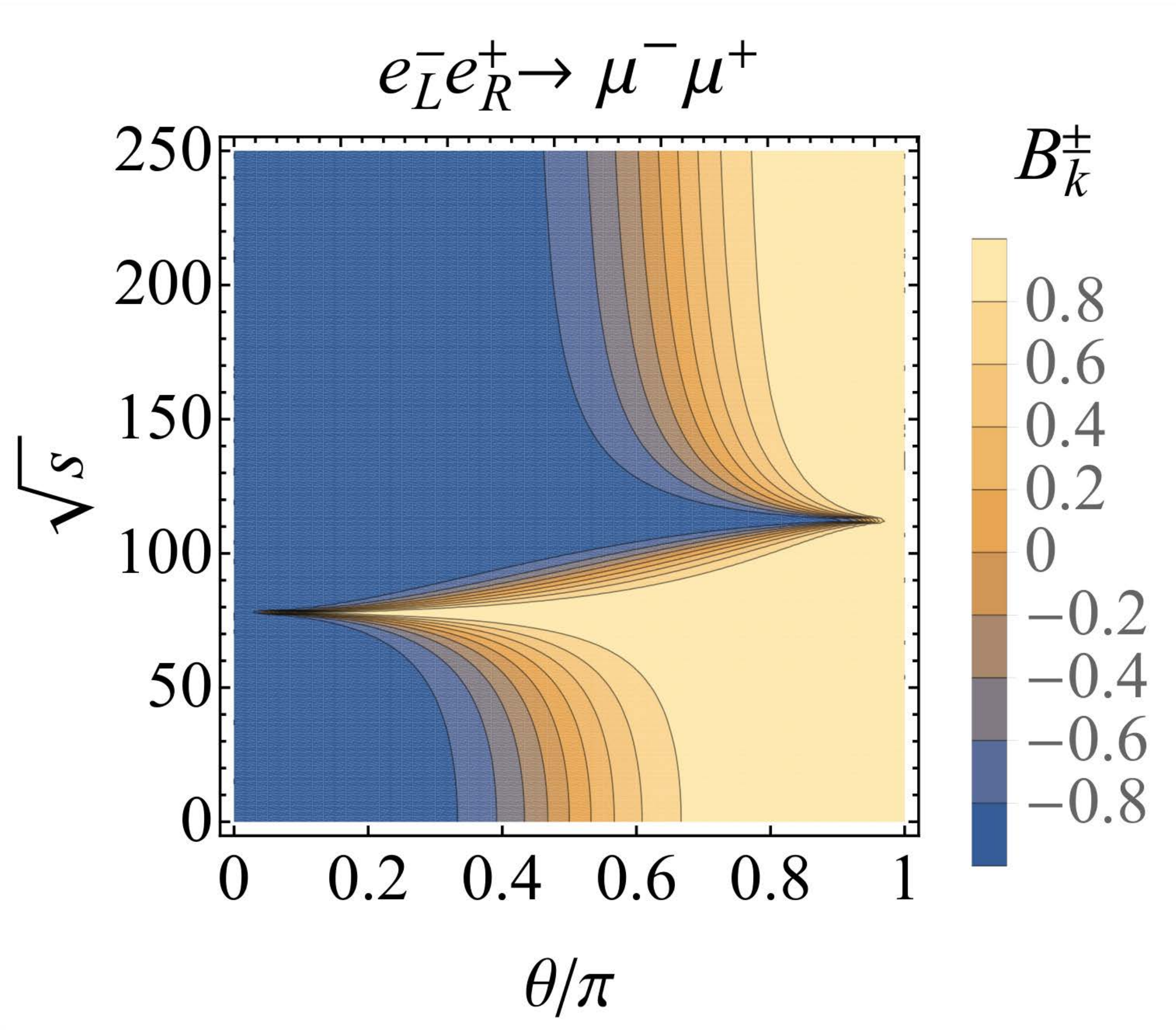}
  \caption{Contour plots of Fano coefficients for $e^+e^- \to \mu^+\mu^-$ in the $\theta-\sqrt{s}$ plane. The panels display unpolarized $C_{rr}$ (left) and $B_k^\pm$ (middle), and $LR$-polarized $B_k^\pm$ (right).}
  \label{fig:mm-cij-contour}
\end{figure}

In the massless $\mu^+\mu^-$ system, kinematic constraints enforce $C_{kk}=1$ and $B_{r,n}^\pm=0$. The remaining non-trivial dynamics are captured by $C_{rr}$, $C_{nn}\ (=-C_{rr})$, and $B^\pm_k$.  Since the $C_{ij}$ components are largely insensitive to beam polarization, we present the contour of $C_{rr}$ solely for the unpolarized case in Fig.~\ref{fig:mm-cij-contour} (left panel). In contrast, the  longitudinal polarization vector $B^\pm_k$ is highly sensitive to the beam polarization. The $LR$ and $RL$ configurations yield qualitatively similar distributions but with an opposite sign. Thus, we display the $B^\pm_k$ contours for the unpolarized and $LR$ cases in the middle and right panels of Fig.~\ref{fig:mm-cij-contour}, respectively.

Specifically, at the phase space points where $f_A/f_V = \pm 1$, the spin configuration becomes separable (purely $\ket{\uparrow\uparrow}$ or $\ket{\downarrow\downarrow}$), implying $\mathcal{C}=0$ and a saturated minimal Bell variable $\mathcal{B}=2$.
Conversely, maximal entanglement occurs in the intermediate regions where the state approaches the superposition $(\ket{\uparrow\uparrow}+\ket{\downarrow\downarrow})/\sqrt{2}$.

\begin{figure}[tb]
  \centering
  \includegraphics[width=0.45\linewidth]{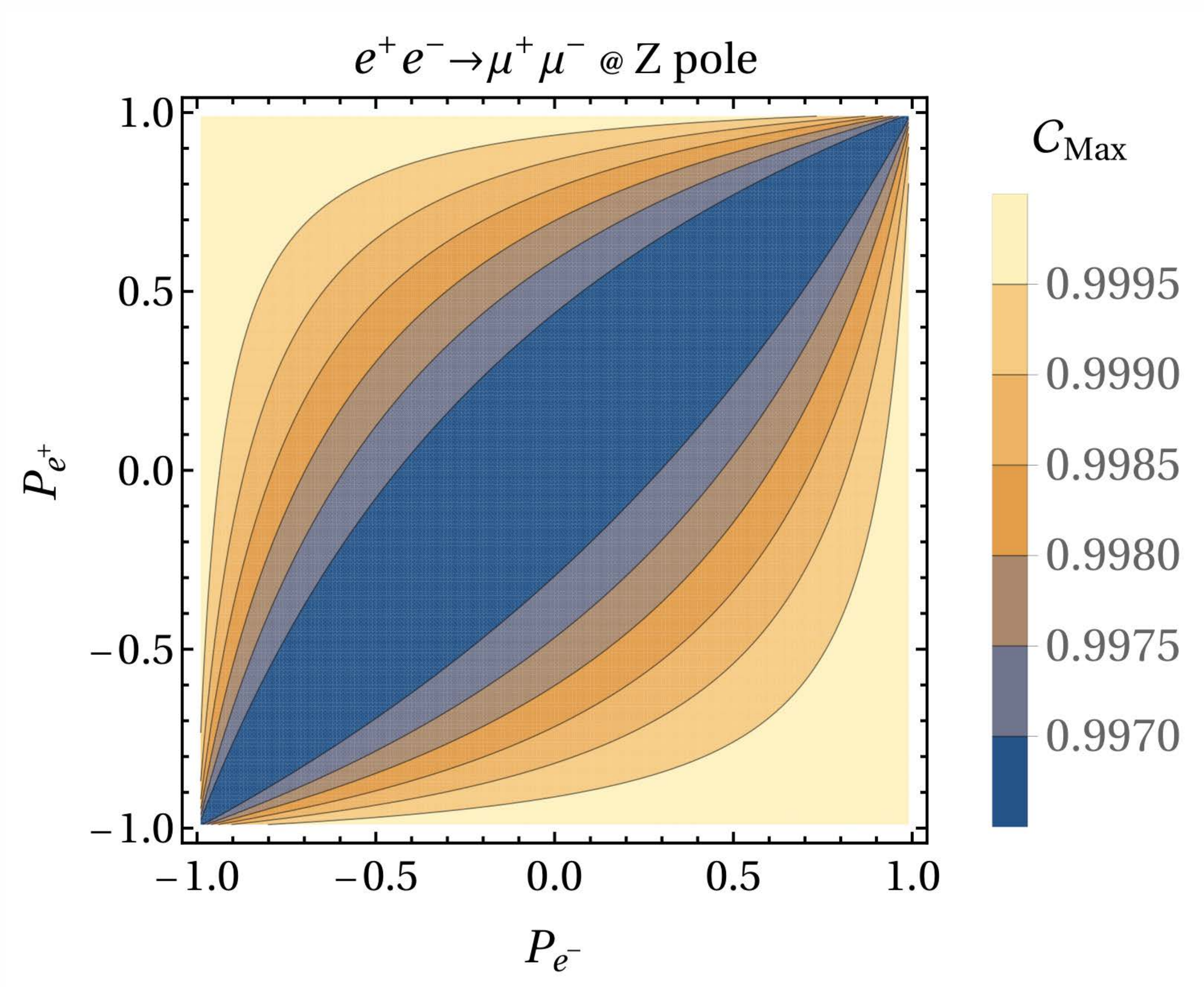} \qquad
  \includegraphics[width=0.42\linewidth]{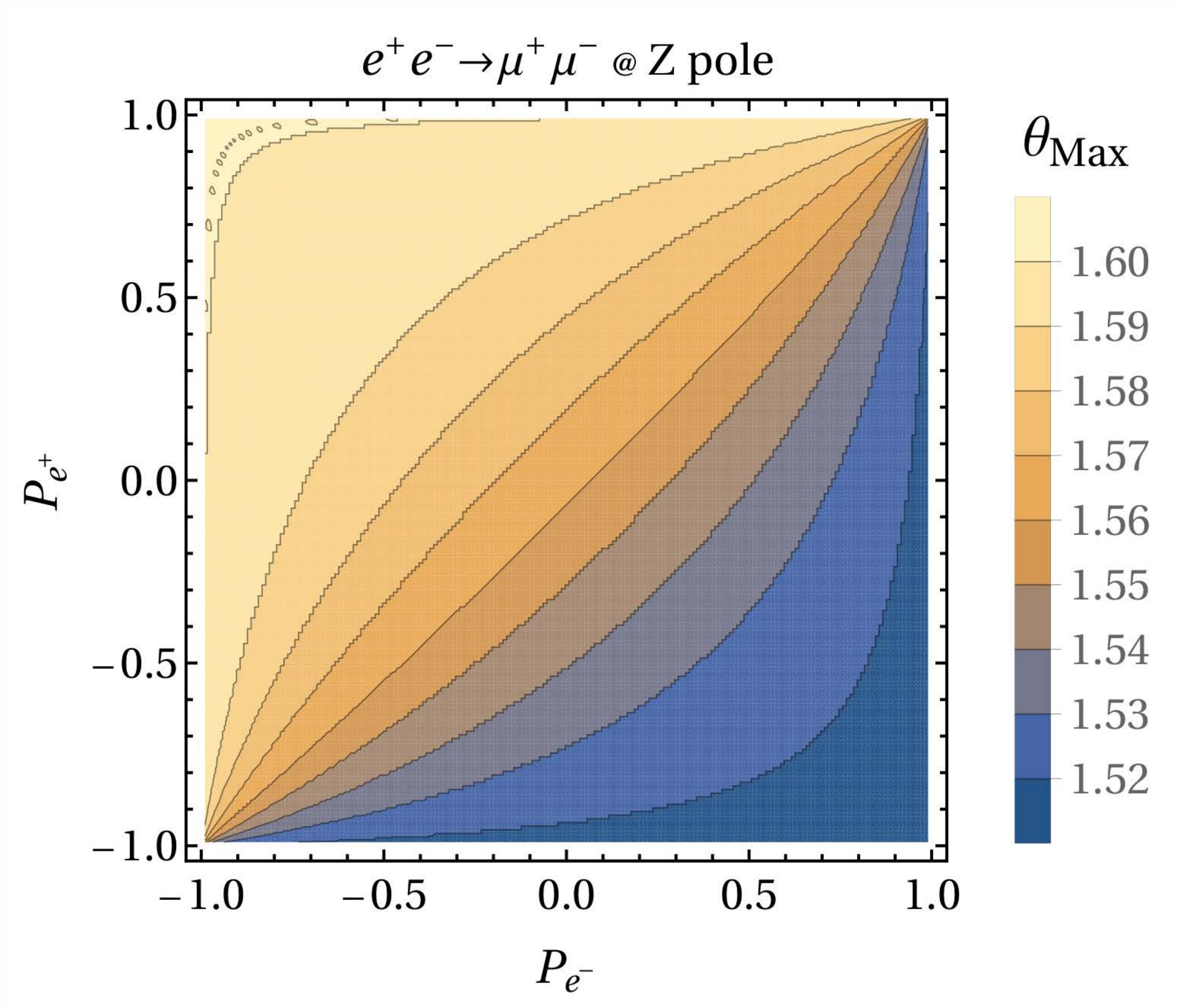}
  \caption{Contour plots for the maximum concurrence $\mathcal{C}_{\rm Max}$ (left) and corresponding $\theta_{\rm Max}$ (right) for the $\mu^+\mu^-$ system at $Z$ pole evaluated over the polarization of the $e^-$ and $e^+$  beams. }
  \label{fig:mmCmax}
\end{figure}

The maximum values of concurrence $\mathcal{C}_{\rm Max}$ for $e^+e^-\to\mu^+\mu^-$ at $Z$ pole in the  ($P_{e^-},P_{e^+}$) polarization plane, along with their corresponding optimal angles $\theta_{\rm Max}$, are detailed in Fig.~\ref{fig:mmCmax}.
At energies off the $Z$ pole, the distribution of the maximum value of concurrence largely remains the same as in Fig.~\ref{fig:mmCmax}, although the $\theta_{\rm Max}$ is slightly shifted. Similar behavior is exhibited for Bell nonlocality $\mathcal{B}$.

%%%%%%%%%%%%%%%%%%%%%%%%%%%%%%%%%%%%%%%%%%%%%%%%%%%%%%%%%%%
\subsubsection{Second Stabilizer R\'{e}nyi Entropy for \texorpdfstring{$\mu^+\mu^-$}{mm}}
\label{sec:magic-mm}
%%%%%%%%%%%%%%%%%%%%%%%%%%%%%%%%%%%%%%%%%%%%%%%%%%%%%%%%%%%
\begin{figure}[tb]
  \centering
  \includegraphics[width=0.5\linewidth]{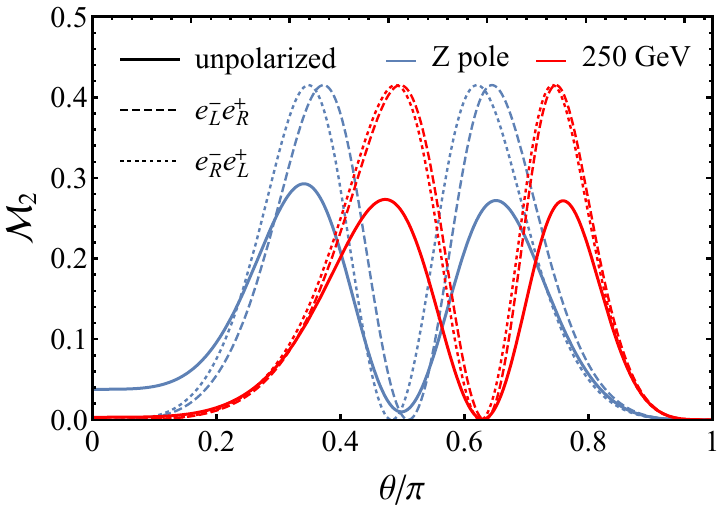}
  \caption{The SSRE $\mathcal{M}_2$ as a function of scattering angle $\theta$ for the $\mu^+\mu^-$ system at $\sqrt{s}=m_Z$ (blue) and 250 GeV (red) with unpolarized (solid), $LR$ (dashed), and $RL$ (dotted) beam polarization configurations.}
  \label{fig:magic-mm}
\end{figure}

\begin{figure}[tb]
  \centering
  \includegraphics[width=0.4\linewidth]{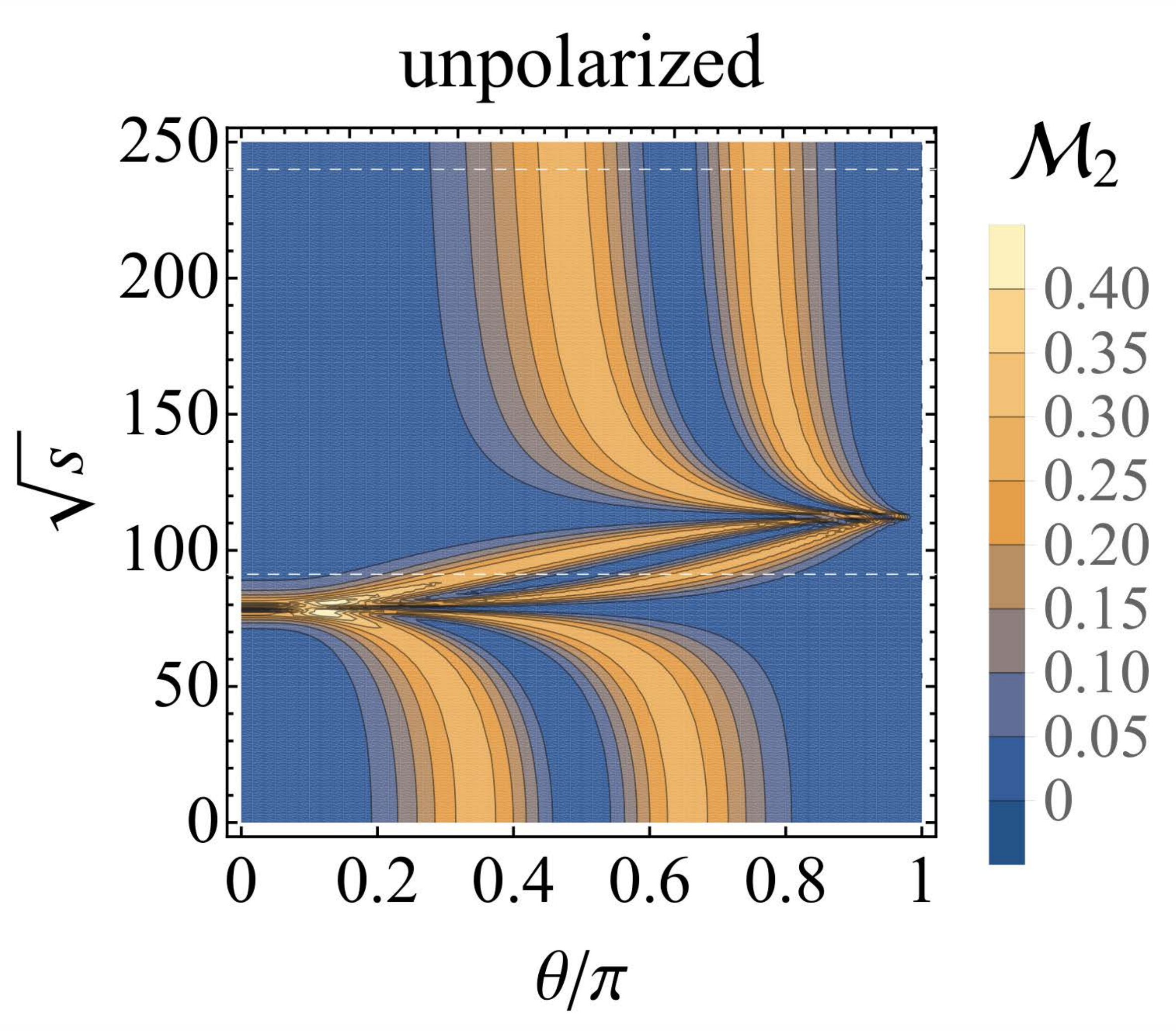}
  \includegraphics[width=0.4\linewidth]{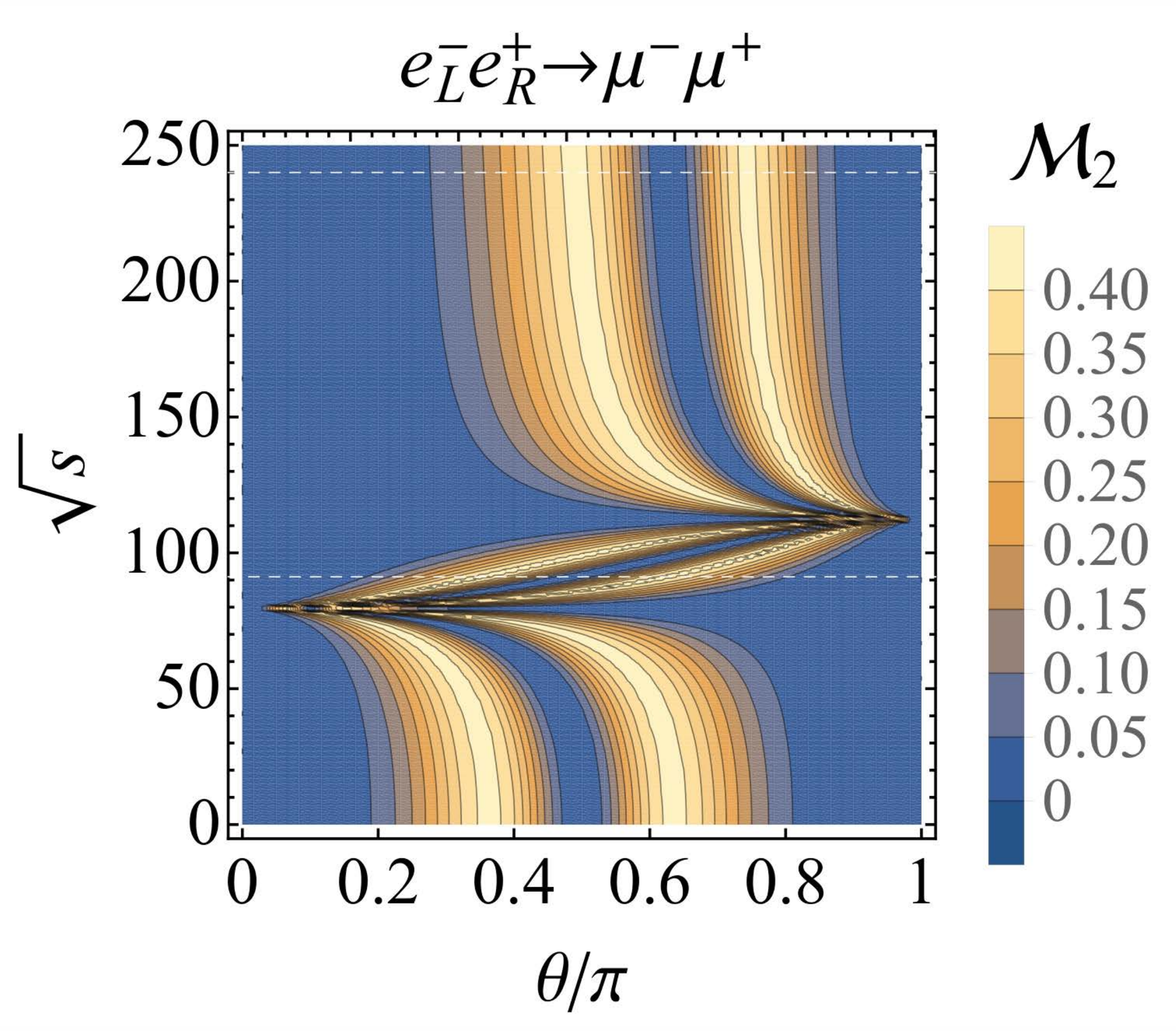}
  \caption{Contour plots for the SSRE $\mathcal{M}_2$ in the plane of $\theta-\sqrt{s}$ (GeV) for the unpolarized (left), and $LR$ (right) polarized $e^+e^-\to \mu^+\mu^-$ process.}
  \label{fig:magic-contour-mm}
\end{figure}

Following the example of $t\bar t$, we examine the ``non-stabilizerness'' of the $\mu^+\mu^-$ quantum state by evaluating the Second Stabilizer R\'{e}nyi Entropy or magic ($\mathcal{M}_2$). The results of the magic of unpolarized and $LR$, $RL$ fully polarized $e^+e^-\to\mu^+\mu^-$ process at $Z$ pole and $\sqrt{s}=250$ GeV are shown in Fig.~\ref{fig:magic-mm}. The contour plots of magic for the $\mu^+\mu^-$ system are shown in Fig.~\ref{fig:magic-contour-mm}. We display only the $LR$ polarization, as the $RL$ case exhibits a qualitatively similar distribution.
The patterns resemble those of entanglement or Bell nonlocality, except that the regions of maximal entanglement correspond to stabilizer states with vanishing magic.
In the $\mu^+\mu^-$ process, the system approaches a product state in the forward/backward limits ($\theta \to 0, \pi$), as well as near the points where $f_V \approx \pm f_A$.
Simultaneously, the system approaches a Bell state at the scattering angle of maximal entanglement $\theta_{\rm Max}$.
We found $\mathcal{M}_2 \to 0$ in both the regions of zero entanglement and maximal entanglement.
Non-zero magic emerge in the \textit{transition regions} between these two regions, naturally giving rise to the ``twin peaks'' structure as observed in Figs.~\ref{fig:magic-mm} and
Fig.~\ref{fig:magic-contour-mm}. We show  that the ``valleys'' of magic coincide precisely with the ``ridges'' of maximal entanglement and Bell nonlocality in the $\theta-\sqrt{s}$ plane. We reiterate our findings in Sec.~\ref{subsec:SecIII-summary}.

\begin{figure}[htbp]
  \centering
    \includegraphics[width=0.45\linewidth]{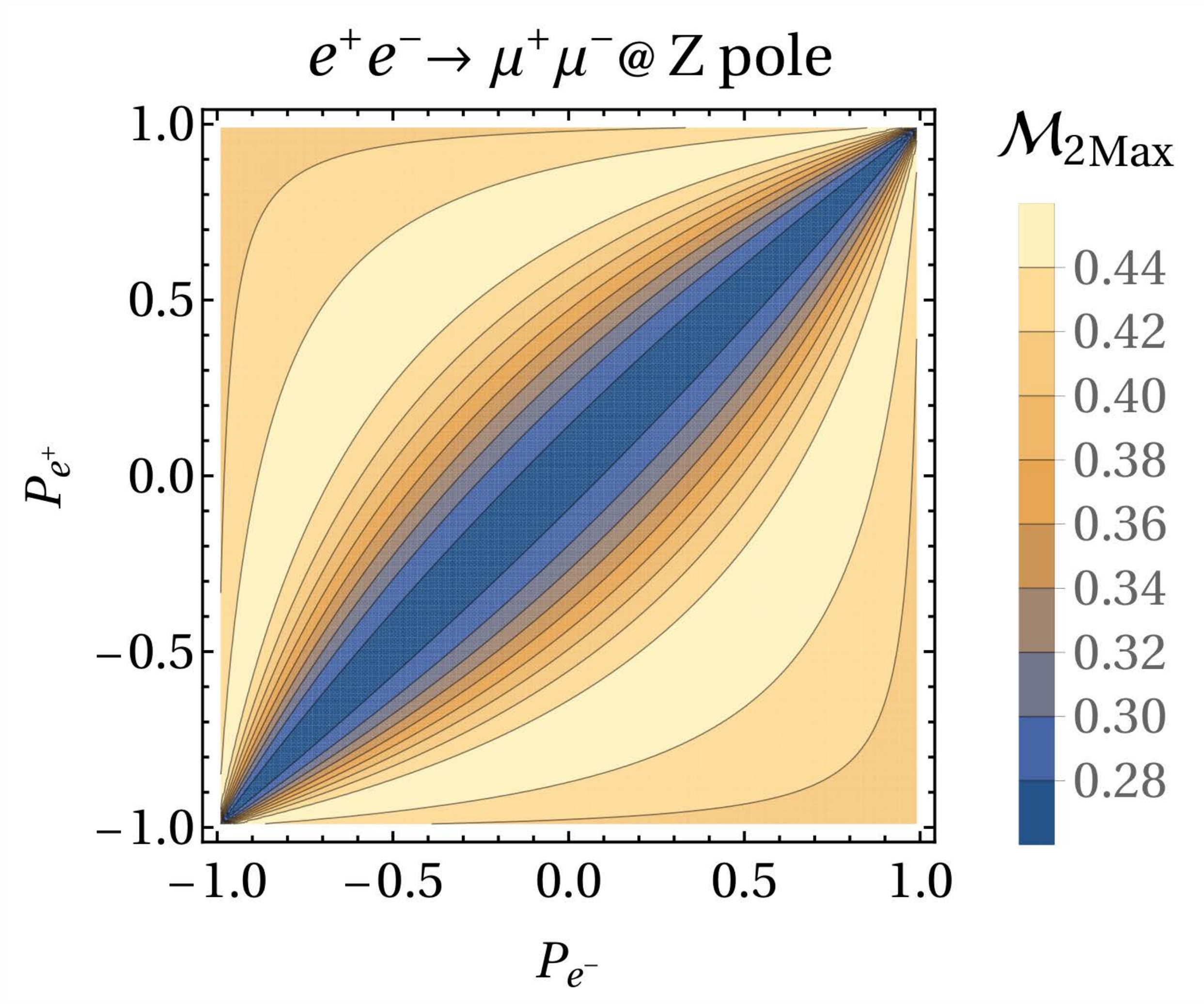}\qquad
    \includegraphics[width=0.45\linewidth]{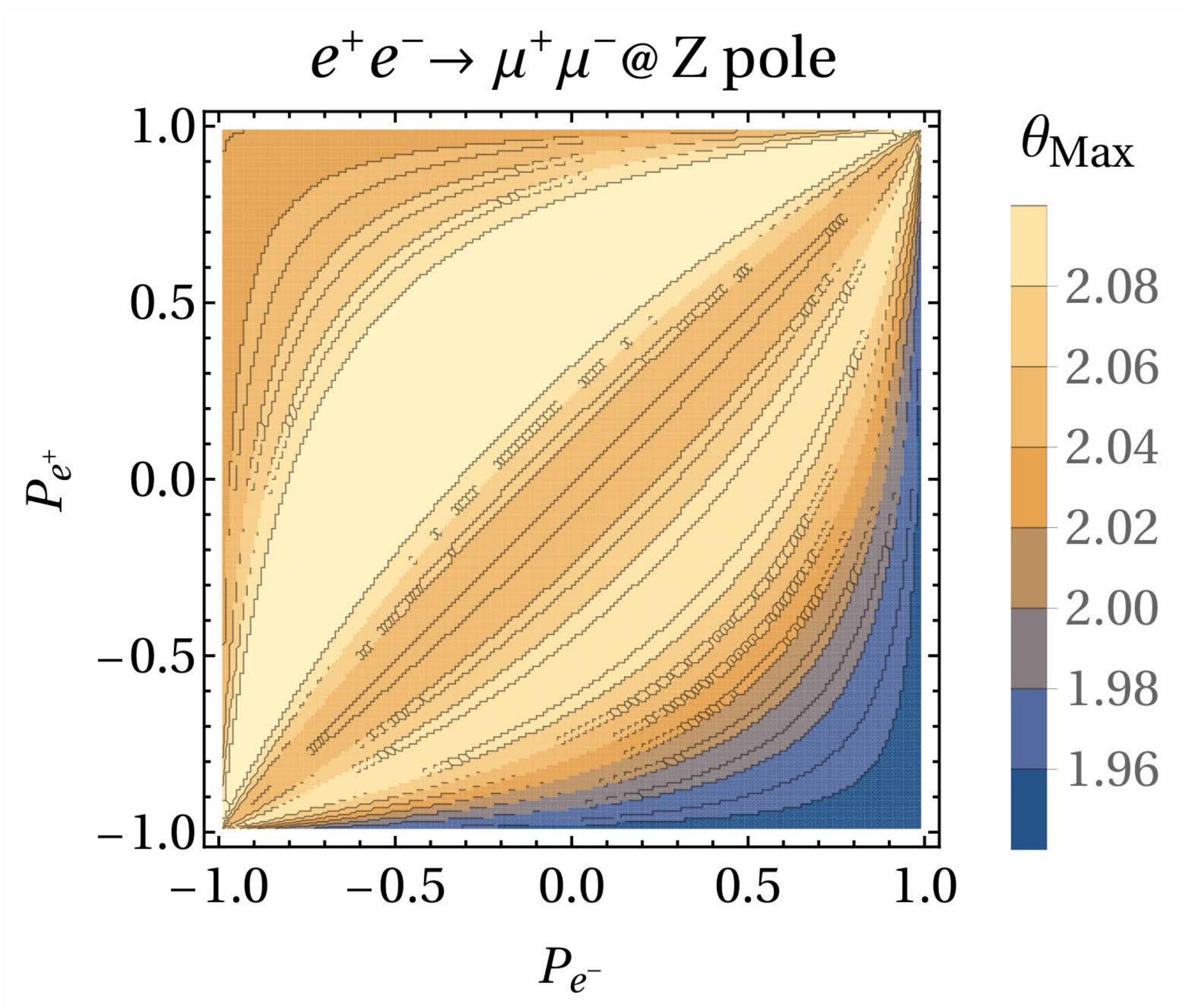}
  \caption{Contour plots for the  maximum $\mathcal{M}_2$ (left) and corresponding $\theta_{\rm Max}$ (right) for the $e^+e^-\to \mu^+\mu^-$ evaluated over the polarization space at the $Z$ pole. The plots correspond to one of the twin peaks (with a larger angle) in Figs. \ref{fig:magic-mm} and \ref{fig:magic-contour-mm}.}
  \label{fig:fig:Mmax91}
\end{figure}

We analyze the optimal conditions for generating magic resources. Figure \ref{fig:fig:Mmax91} shows the maximum value of $\mathcal{M}_{\mathrm{2Max}}$ in the plane of the beam polarizations ($P_{e^-}, P_{e^+}$) at the $Z$ pole. Given that the twin peaks in Figs.~\ref{fig:magic-contour-mm} and \ref{fig:fig:Mmax91} share a similar distribution for the maximum $\mathcal{M}_{\mathrm{2Max}}$, we restrict our display in Fig.~\ref{fig:fig:Mmax91} to the peak of the right side one. There is no significant difference at other energies. It is interesting to note that the global maximum of magic ($\mathcal{M}_{\mathrm{2Max}}\simeq 0.43$) is \textit{not} achieved at the fully polarized limits ($P=\pm 1$). Instead, the optimal operating point lies within the partially polarized region, extended around $(0.4, -0.5)$ and $(-0.5, 0.4)$.
This suggests that for quantum information tasks requiring magic,  moderate beam polarization may be superior to full polarization.
This finding emphasises that magic and entanglement are distinct resources. While entanglement is maximized by purity and full polarization, magic thrives on the complex superposition structure that differs from both product and Bell states.

%%%%%%%%%%%%%%%%%%%%%%%%%%%%%%%%%%%%%%%%%%%%%%%%%%%%%%%%%%%
\subsection{Bhabha Scattering}
\label{subsec:ee}
%%%%%%%%%%%%%%%%%%%%%%%%%%%%%%%%%%%%%%%%%%%%%%%%%%%%%%%%%%%

Bhabha scattering of $e^+e^-\!\to e^+e^-$ has a distinct quantum structure different from that of $\mu^+\mu^-$ and $\tau^+\tau^-$ production due to the additional contributions from $t$-channel exchanges. The $s$-channel exchange, identical to that in $\mu^+\mu^-$ and $\tau^+\tau^-$ production, can generate quantum entanglement, whereas the $t$-channel contribution tends to suppress entanglement by dominating the forward region and diluting the spin correlations. As a result, the entanglement structure in Bhabha scattering arises from the interplay between the $s$- and $t$-channel amplitudes and is substantially more intricate than in purely $s$-channel processes.
Beam polarization provides an experimentally controllable handle to modify the relative weights of the two channels, enabling targeted enhancement of entanglement in the $e^+e^-$ final state.
In the following, we provide a detailed analysis of how entanglement emerges and evolves under the combined influence of the $s$- and $t$-channel dynamics.

%%%%%%%%%%%%%%%%%%%%%%%%%%%%%%%%%%%%%%%%%%%%%%%%%%%%%%%%%%%
\subsubsection{The Quantum State for \texorpdfstring{$e^+e^-$}{ee}}
%%%%%%%%%%%%%%%%%%%%%%%%%%%%%%%%%%%%%%%%%%%%%%%%%%%%%%%%%%%
%Helicity amplitudes
In the massless limit, the process $e^+e^-\!\to e^+e^-$ contains six nonvanishing helicity configurations. Their amplitudes naturally separate into three categories: pure $s$-channel contributions, pure $t$-channel contributions, and configurations receiving both $s$- and $t$-channel amplitudes.  The spin-helicity factors $(1\pm\cos\theta)/2$ arise from angular-momentum conservation, while the $1/t$ behavior originates from the well-known collinear enhancement of the $t$-channel photon propagator. It is useful to express the helicity amplitudes for $\mathcal{M}(e^-e^+ \to  e^-e^+)$ in terms of the Mandelstam variables $s$, $t$, and $u$,
\begin{subequations}
\label{eq:eeAM1}
\begin{align}
\mathcal{M}(RL,\uparrow\uparrow) &=
 2u\!\left[\frac{1+f_{1}(s)}{s}+\frac{1+f_{1}(t)}{t}\right],\
\mathcal{M}(LR,\downarrow\downarrow) =
 2u\!\left[\frac{1+f_{2}(s)}{s}+\frac{1+f_{2}(t)}{t}\right],
 \nonumber \\
\mathcal{M}(RL,\downarrow\downarrow) &= \mathcal{M}(LR,\uparrow\uparrow) = -\frac{2t}{s}\!\left[1+f_{3}(s)\right],  \\
\mathcal{M}(LL,\downarrow\uparrow) &=
\mathcal{M}(RR,\uparrow\downarrow) = -\frac{2s}{t}\!\left[1+f_{3}(t)\right],
\nonumber
\end{align}
\end{subequations}
where the electroweak form factors are given by
\begin{subequations}
\label{eq:eeAM2}
\begin{align}
f_{1}(x)&=\frac{x}{x-m_Z^2+is\Gamma_Z/ m_Z}\ \frac{g_R^e g_R^e}
          {\sin^2\theta_W\cos^2\theta_W}, \\
f_{2}(x)&=\frac{x}{x-m_Z^2+is\Gamma_Z/m_Z}\
          \frac{g_L^e g_L^e}
          {\sin^2\theta_W\cos^2\theta_W}, \\
f_{3}(x)&=\frac{x}{x-m_Z^2+is\Gamma_Z/m_Z}\
          \frac{g_L^e g_R^e}
          {\sin^2\theta_W\cos^2\theta_W}.
\end{align}
\end{subequations}
where $g_L^e=I^3_e-Q_e s_W^2=-1/2+s_W^2$ and $g_R^e=-Q_e s_W^2=s_W^2$.

% Channel structure
From these amplitudes one immediately identifies the channel structure:
\begin{itemize}
\item $\mathcal{M}(LR,\uparrow\uparrow)$ and $\mathcal{M}(RL,\downarrow\downarrow)$ receive contributions exclusively from the $s$-channel, identical to $e^+e^-\!\to\mu^+\mu^-$ (Sec.~\ref{subsec:mumu}).
\item $\mathcal{M}(LL,\downarrow\uparrow)$ and $\mathcal{M}(RR,\uparrow\downarrow)$ originate purely from the $t$-channel; they produce only a single helicity configuration leading to a separable state, and therefore cannot generate entanglement.
\item $\mathcal{M}(LR,\downarrow\downarrow)$ and $\mathcal{M}(RL,\uparrow\uparrow)$ receive contributions from both the $s$- and $t$-channels and encode the interference between the two.
\end{itemize}
Such $t$-channel-only structures and $s$--$t$ interference patterns do not appear in $\mu^-\mu^+$ or $t\bar{t}$ production, making Bhabha scattering a qualitatively different case in the entanglement study of lepton–pair production.

\begin{figure}[tb]
  \centering
  \includegraphics[width=0.43\linewidth]{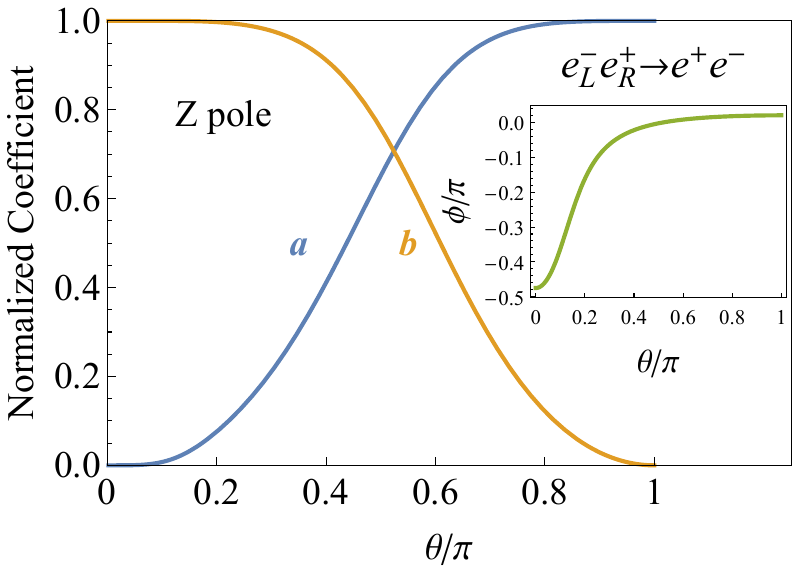}  \qquad
  \includegraphics[width=0.40\linewidth]{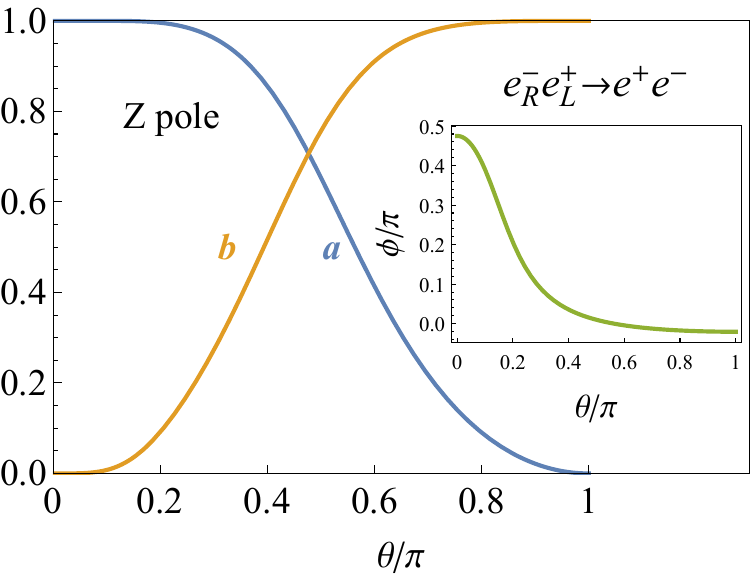}  \\
  \includegraphics[width=0.45\linewidth]{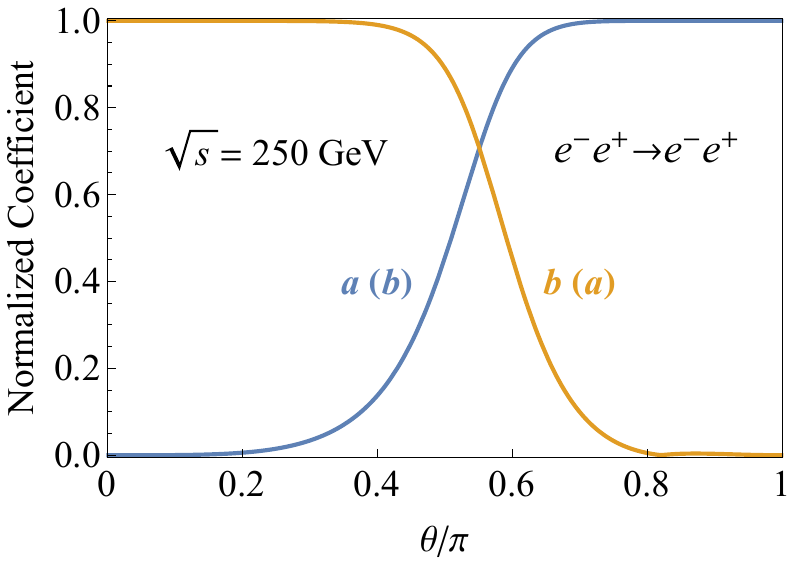}
  \caption{Coefficients $a$ and $b$ in Eq.~(\ref{eq:psi}) for Bhabha scattering as a function of scattering angle $\theta$. Upper panels: Results at the $Z$ pole for the $LR$ (left) and $RL$ (right) polarizations. The insets display the phase difference induced by the imaginary part of the $Z$ boson propagator. Lower panels: Corresponding coefficients at $\sqrt{s}=250~\text{GeV}$ for the $LR$ ($RL$) configurations.}
  \label{fig:state_coefficient_ee}
\end{figure}

\begin{figure}[tb]
  \centering
  \includegraphics[width=0.42\linewidth]{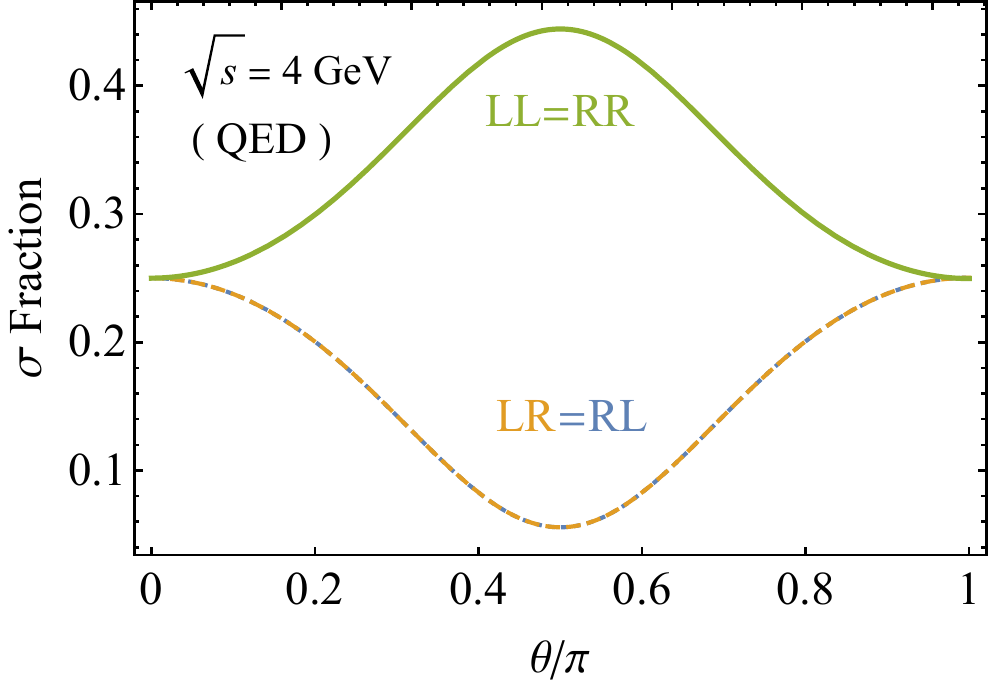}
  \qquad
  \includegraphics[width=0.42\linewidth]{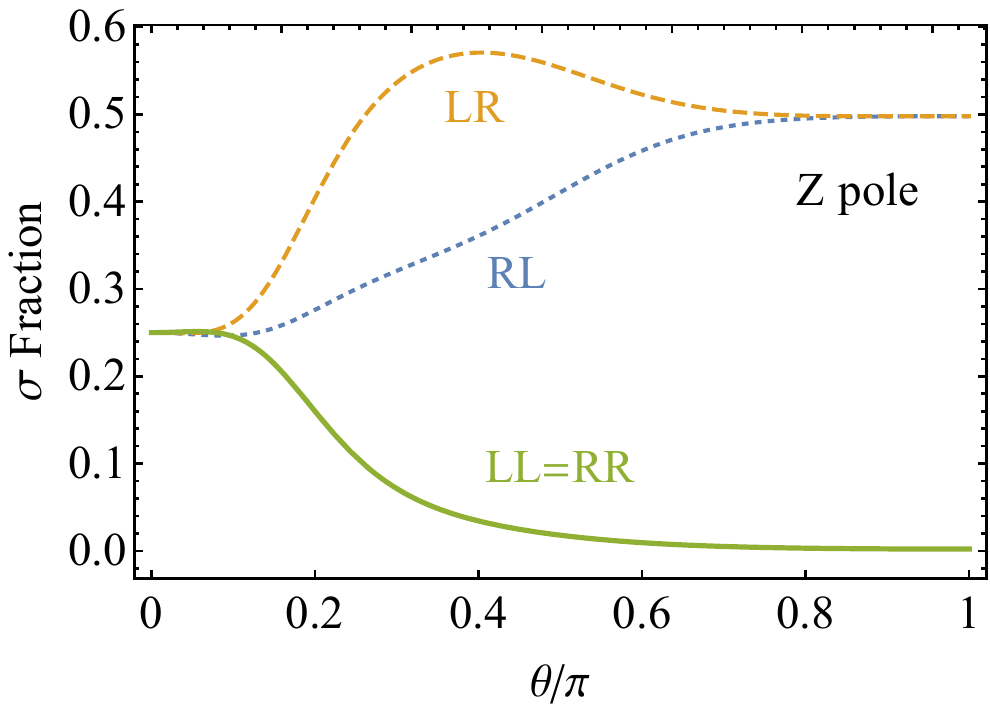}\\
  \includegraphics[width=0.42\linewidth]{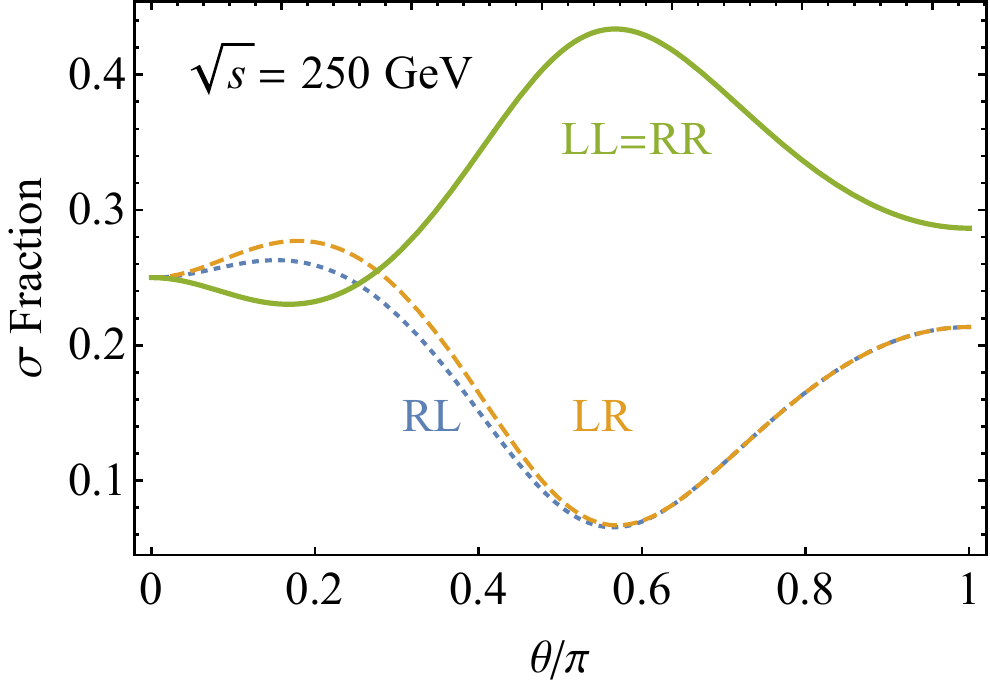}
  \qquad
  \includegraphics[width=0.42\linewidth]{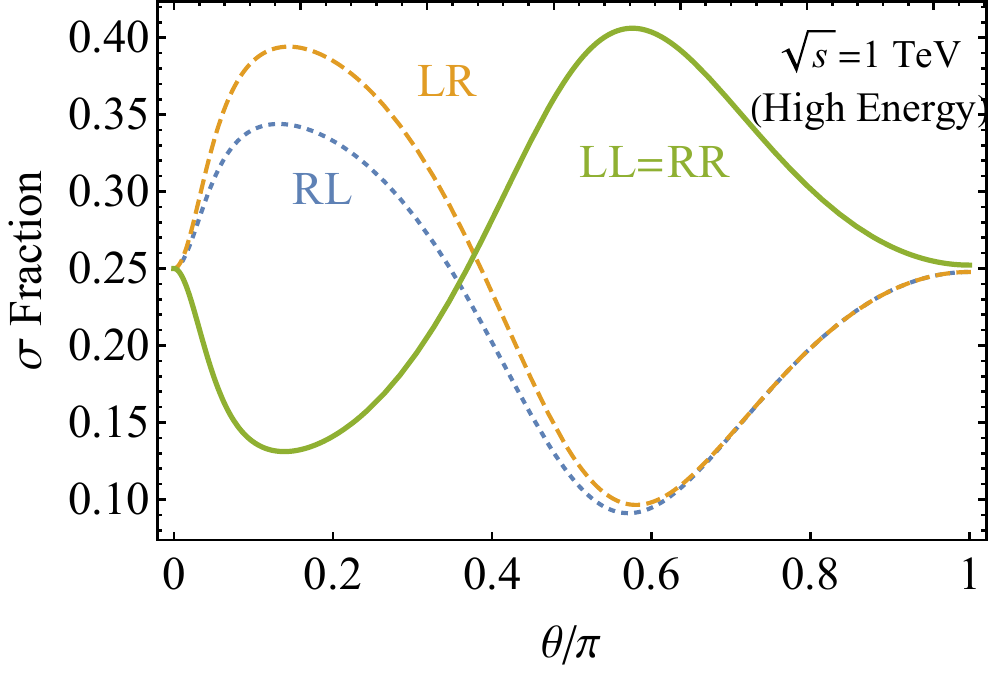}
  \caption{Fractions of the helicity components in the unpolarized $e^-e^+\!\to e^-e^+$ process as functions of the scattering angle~$\theta$ for four representative energies: the low-energy QED regime ($\sqrt{s}=4~\text{GeV}$), the $Z$ pole, $\sqrt{s}=250$ GeV and 1 TeV. }
  \label{fig:ee-FRrho}
\end{figure}

Based on the general two qubit state in Eq.~\eqref{eq:psi}, the final pure state in the helicity-basis for $e^- e^+ \to e^-e^+$ can be written as
\begin{subequations}\label{eq:eepsi}
\begin{align}
\ket{\psi_{RL}} &= a_{RL}\,\ket{\uparrow\uparrow} + b_{RL}\,\ket{\downarrow\downarrow}, \\
\ket{\psi_{LR}} & = a_{LR}\,\ket{\uparrow\uparrow} + b_{LR}\,\ket{\downarrow\downarrow}, \\
\ket{\psi_{RR}} &= c_{RR}\,\ket{\uparrow\downarrow}, \quad \ket{\psi_{LL}} = d_{LL}\,\ket{\downarrow\uparrow},
\end{align}
\end{subequations}
with
\begin{subequations}
\label{eq:coef4eepsi}
\begin{align}
a_{RL} &= \mathcal{M}(RL,\uparrow\uparrow), \qquad b_{RL} = \mathcal{M}(RL,\downarrow\downarrow), \\
a_{LR} &= \mathcal{M}(LR,\uparrow\uparrow), \qquad b_{LR} = \mathcal{M}(LR,\downarrow\downarrow),\\
c_{RR} &= \mathcal{M}(RR,\uparrow\downarrow), \qquad
d_{LL} = \mathcal{M}(LL,\downarrow\uparrow).
\end{align}
\end{subequations}
Only the  beam polarization configurations $LR$ and $RL$ contain coherent superpositions of the final states  $\ket{\uparrow\uparrow}$ and $\ket{\downarrow\downarrow}$, therefore capable of producing entanglement.
The $LL$ and $RR$ amplitudes generate only separable states.
The coefficients $a$ and $b$ of the entangled component are shown in Fig.~\ref{fig:state_coefficient_ee} for $\sqrt{s}=m_Z$ and 250 GeV.
The range of angles for which $a$ and $b$ are both nonzero is significantly broader at the $Z$ pole, making entanglement more experimentally accessible in this region.
Near the $Z$ pole, the imaginary part of the propagator introduces a phase difference between the two helicity components, yielding the state $$\ket{\psi}=a\ket{\uparrow\uparrow}+b\,e^{i\Delta\phi}\ket{\downarrow\downarrow}.$$
This effect, absent in $\mu^+\mu^-$ production, is driven by the interference between the imaginary $s$-channel and real $t$-channel amplitudes, allowing $\Delta\phi$ to reach values
as large as $\pi/2$.\footnote{The Fano coefficient $C_{12}$ will be induced by $\Delta\phi$, but it becomes prominent only for $a \approx b$ and $\Delta\phi \approx \pi/2$. Figure~\ref{fig:state_coefficient_ee} indicates that the effect is negligible in the $e^+e^-$ system.}
At higher energies, the propagator becomes real and the phase difference rapidly vanishes.
It should be noted that, according to
Eqs.~\eqref{eq:eeAM1} and \eqref{eq:eeAM2}, the coefficients satisfy the relations $a_{LR} = b_{RL}$ and $b_{LR} \approx a_{RL}$.

\begin{figure}[tb]
  \centering
  \includegraphics[width=0.45\linewidth]{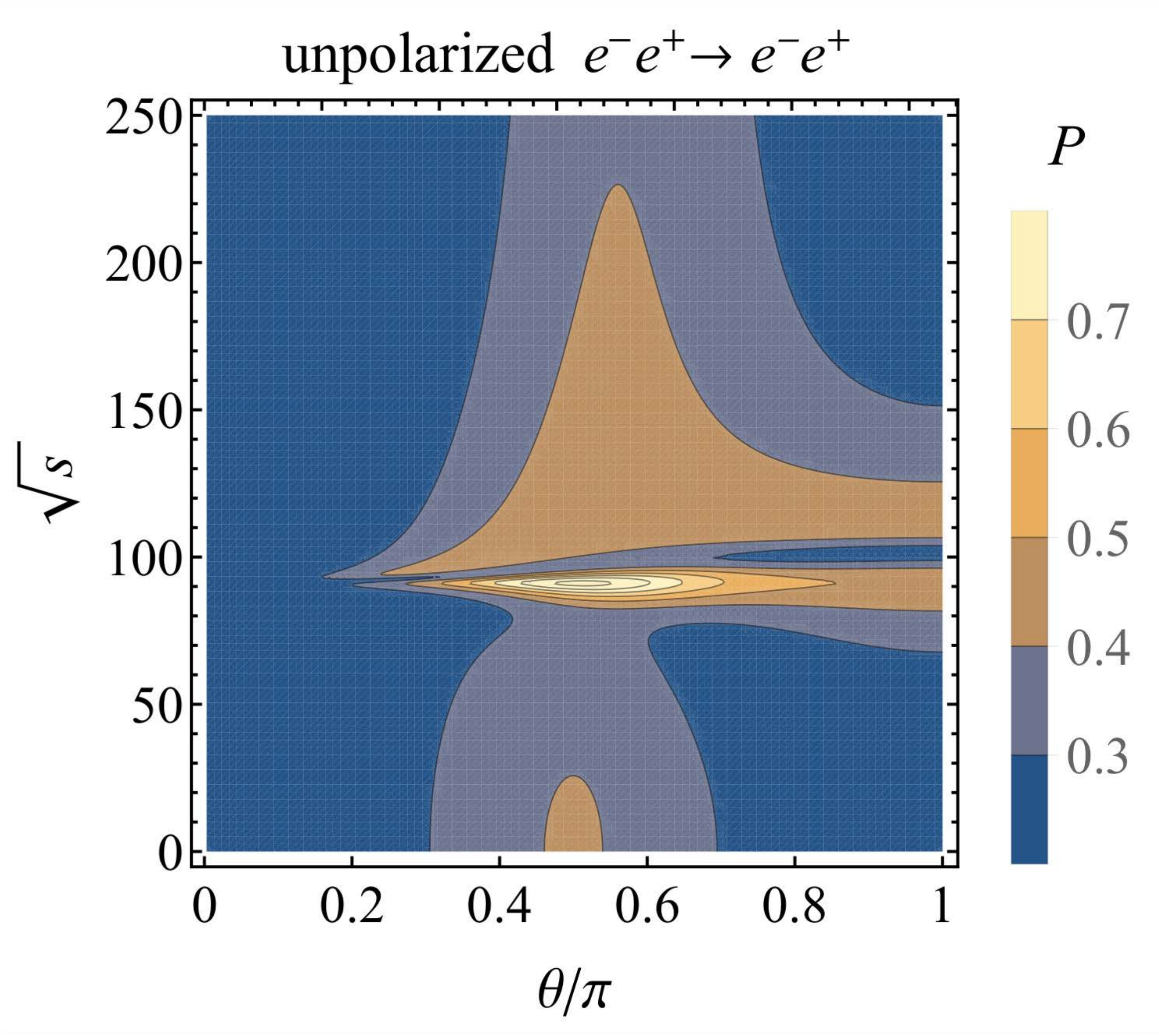}
  \label{fig:rho-ee}
  \caption{Contour plot for the purity $P$ of the process $e^-e^+\to e^-e^+$, shown in the kinematic plane of scattering angle $\theta$ and $\sqrt{s}$ in GeV.}
  \label{fig:eepurity}
\end{figure}

The unpolarized process corresponds to an incoherent mixture of the four initial-helicity sectors as given in Eq.~(\ref{eq:rho_in}).
In the massless limit, the general density matrix of the unpolarized $e^+ e^-$ system is
\begin{equation}
\rho_{\rm mixed}=
\begin{pmatrix}
 \rho_{\uparrow\uparrow,\uparrow\uparrow} & 0 & 0 & \rho_{\uparrow\uparrow,\downarrow\downarrow} \\
 0 & \rho_{\uparrow\downarrow,\uparrow\downarrow} & 0 & 0 \\
 0 & 0 &  \rho_{\downarrow\uparrow,\downarrow\uparrow} & 0 \\
 \rho_{\downarrow\downarrow,\uparrow\uparrow} & 0 & 0 & \rho
_{\downarrow\downarrow,\downarrow\downarrow} \\
\end{pmatrix},
\end{equation}
where the symbols $ \rho_{\uparrow\uparrow,\uparrow\uparrow}$, $\rho_{\uparrow\uparrow,\downarrow\downarrow}$, $\rho_{\uparrow\downarrow,\uparrow\downarrow}$, $\rho_{\downarrow\uparrow,\downarrow\uparrow}$, $\rho_{\downarrow\downarrow,\uparrow\uparrow}$, and $\rho
_{\downarrow\downarrow,\downarrow\downarrow}$ denote non-zero entries of the density matrix.

Figure~\ref{fig:ee-FRrho} shows the angular dependence of contributions from different initial polarizations.
The curves show the contributions from the entanglement-generating channels $\rho_{LR}$ and $\rho_{RL}$, as well as the separable $t$-channel-dominated components $\rho_{LL}=\rho_{RR}$.
At low energies, the $t$-channel photon exchange  dominates over almost the entire angular range.
At $\theta=\pi/2$ we find $\sigma_{LL}\simeq \sigma_{RR}\simeq 8\,\sigma_{LR}\simeq 8\,\sigma_{RL}$, indicating that the unpolarized ensemble is dominated by the separable $t$-channel components. Near the $Z$ pole the $s$-channel contribution enhances the $\rho_{LR}$ and $\rho_{RL}$ fractions primarily in the intermediate-angle region, where the $t$-channel forward enhancement is absent and the backward suppression from interference is reduced. The fraction of the $s$-channel contribution within $\rho_{RL/LR}$ is maximized at $\cos\theta=\pm f_A/f_V$, consistent with the pattern observed in $\mu^+\mu^-$ production.
At higher energies, the $t$-channel again dominates for $\theta \to \pi$, while the entangled $LR/RL$  contributions become increasingly  important in the forward region ($\theta \lesssim \pi/3$).

Finally, the purity of the unpolarized state, shown in Fig.~\ref{fig:eepurity}, reflects the relative strength of the entangled $LR$/$RL$ components and the separable $LL$/$RR$ components on the $\theta-\sqrt{s}$ plane.
The high purity observed near the $Z$ pole is crucial. This region corresponds to a dominant $s$-channel contribution via $Z/\gamma$ exchange, which enhances the entangled $LR/RL$ channels while suppressing the separable $t$-channel component. The purity approaching unity ($P \to 1$) in this region  confirms that the state is locally equivalent to a nearly pure Bell state. This identifies the optimal domain for maximal entanglement generation.

%%%%%%%%%%%%%%%%%%%%%%%%%%%%%%%%%%%%%%%%%%%%%%%%%%%%%%%%%%%
\subsubsection{Concurrence, Bell Nonlocality, and Quantum Tomography for \texorpdfstring{$e^+e^-$}{ee}}
%%%%%%%%%%%%%%%%%%%%%%%%%%%%%%%%%%%%%%%%%%%%%%%%%%%%%%%%%%%

For Bhabha scattering
$e^- e^+ \to e^- e^+$, the Fano coefficients exhibit a simple structure, following Eqs.~(\ref{eq:Bk}) and (\ref{eq:cij4mm})
\begin{equation}
    C_{ij}=\left (\begin{array}{ccc}
        C_{rr} & 0 & 0 \\
        0 & C_{nn} & 0 \\
        0 & 0 & C_{kk}
    \end{array}\right ),
    \qquad
    B^\pm_i=\left (\begin{array}{c}
        0 \\
        0 \\
        B_k
    \end{array}\right ),
\end{equation}
where $C_{nn}=-C_{rr}$. The non-vanishing Fano coefficients  $C_{rr}$, $C_{kk}$, and $B_k$ are defined in terms of the coefficients of the qubit state in Eq.~\eqref{eq:coef4eepsi}
\begin{subequations}
\label{eq:ee-cji&bi}
    \begin{align}
        C_{rr} &=\frac{2|a_{LR}b_{LR}^*+a_{RL}b_{RL}^*|}{\tilde{A}}, \\
        C_{kk} &=\frac{a_{LR}^2+b_{LR}^2+a_{RL}^2+b_{RL}^2-(c_{LL}^2+d_{RR}^2)}{\tilde{A}}, \\
        B_k &=\frac{(a_{RL}^2-b_{RL}^2)+(a_{LR}^2-b_{LR}^2)}{\tilde{A}},
    \end{align}
\end{subequations}
where $\tilde{A}=a_{LR}^2+b_{LR}^2+a_{RL}^2+b_{RL}^2+c_{LL}^2+d_{RR}^2$ is the normalization factor.

Given the diagonal form of the spin-correlation matrix $C_{ij}$ and the polarization vectors, the concurrence $\mathcal{C}$ of the final state simplifies to:
\begin{equation}
    \mathcal{C}=\max\left\{0,C_{rr}+\frac{1}{2}C_{kk}-\frac{1}{2}\right\}
\end{equation}
The Bell nonlocality is
\begin{equation}
\mathcal{B}=\left\{
\begin{array}{ll}
2\sqrt{C_{rr}^2+C_{kk}^2},
& \qquad\text{if } |C_{kk}|>|C_{rr}|, \\
2\sqrt{2}|C_{rr}|,
& \qquad\text{if } |C_{kk}|<|C_{rr}|.
\end{array}
\right.
\end{equation}

\begin{figure}[tb]
  \centering
  \includegraphics[width=0.45\linewidth]{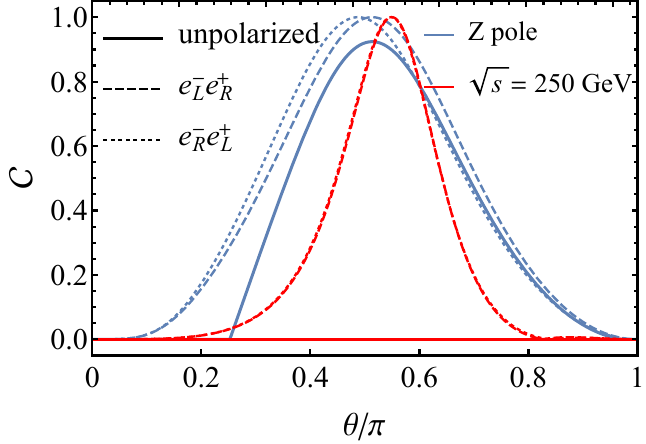}
  \qquad
  \includegraphics[width=0.45\linewidth]{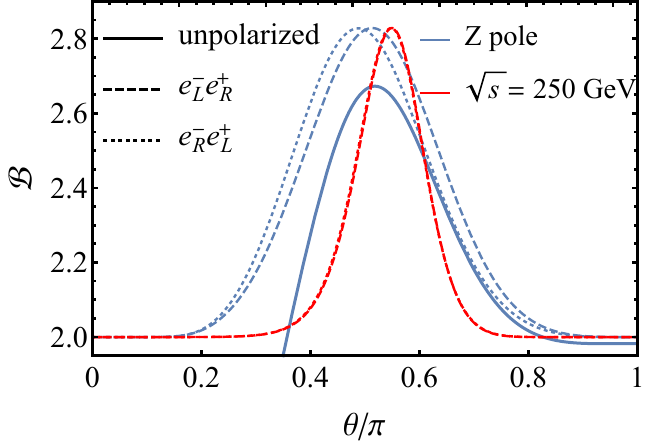}
  \caption{Concurrence $\mathcal{C}$ (left) and the Bell nonlocality $\mathcal{B}$ (right) of the $e^-e^+\to e^-e^+$ state at two operating points of future colliders: $Z$ pole (blue) and 250 GeV (red), with three beam polarization configurations: unpolarized (solid lines), fully polarized $e^-_L e^+_R$ (dashed lines) and $e^-_R e^+_L$ (dotted lines).}
  \label{fig:ee-C&B}
\end{figure}

\begin{figure}[tb]
  \centering
  \includegraphics[width=0.45\linewidth]{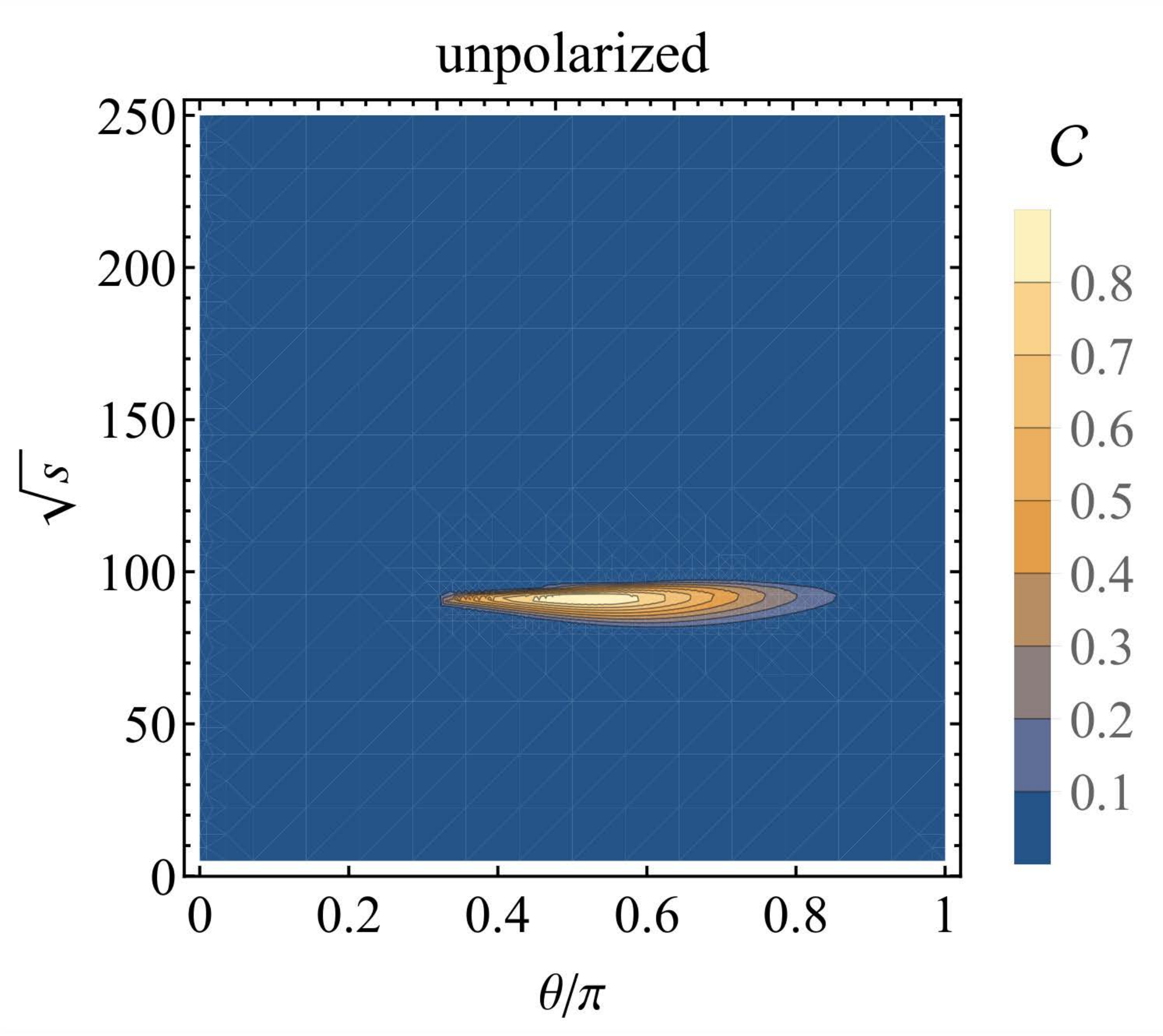}
  \qquad
  \includegraphics[width=0.45\linewidth]{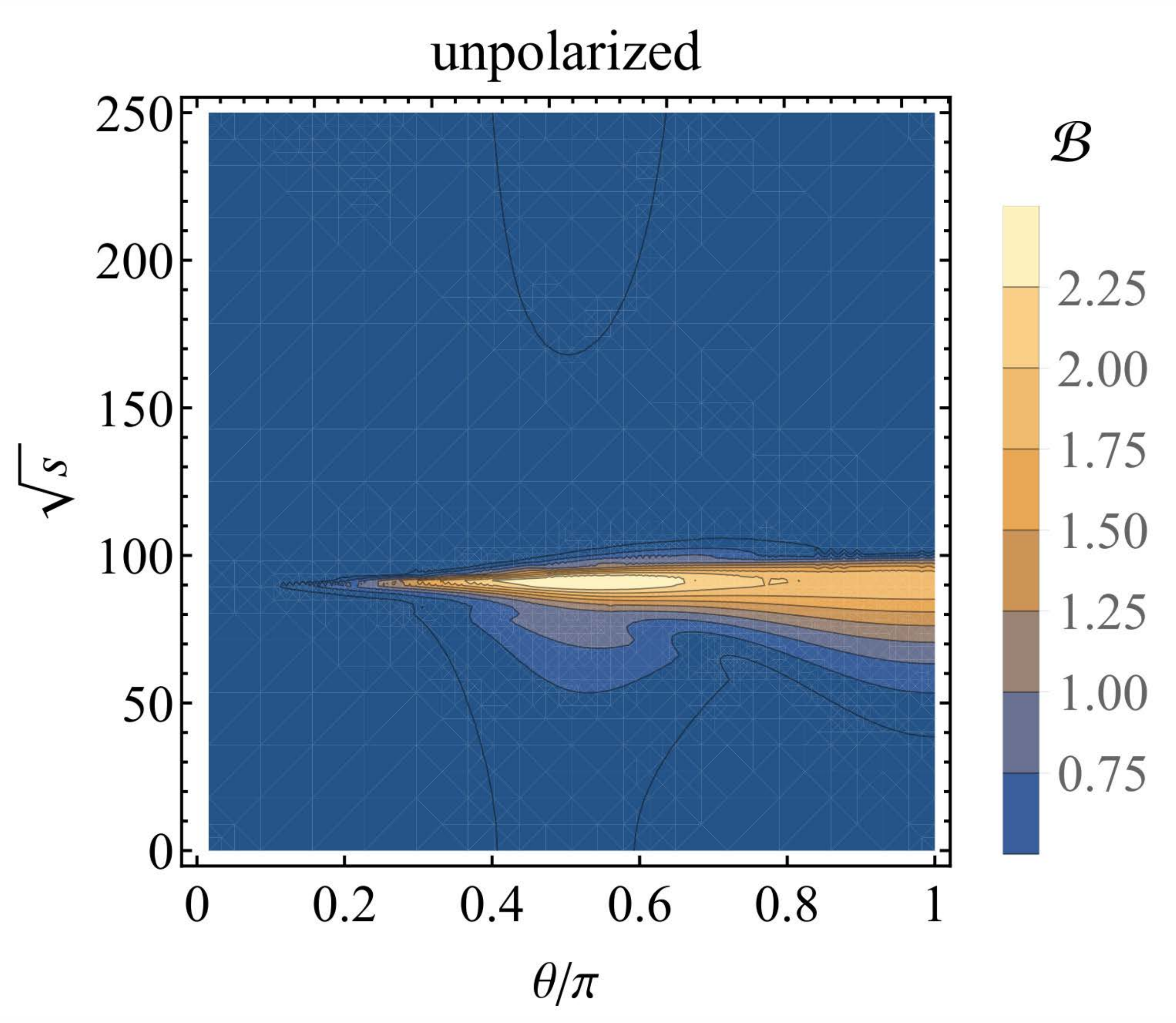}
  \centering
  \includegraphics[width=0.45\linewidth]{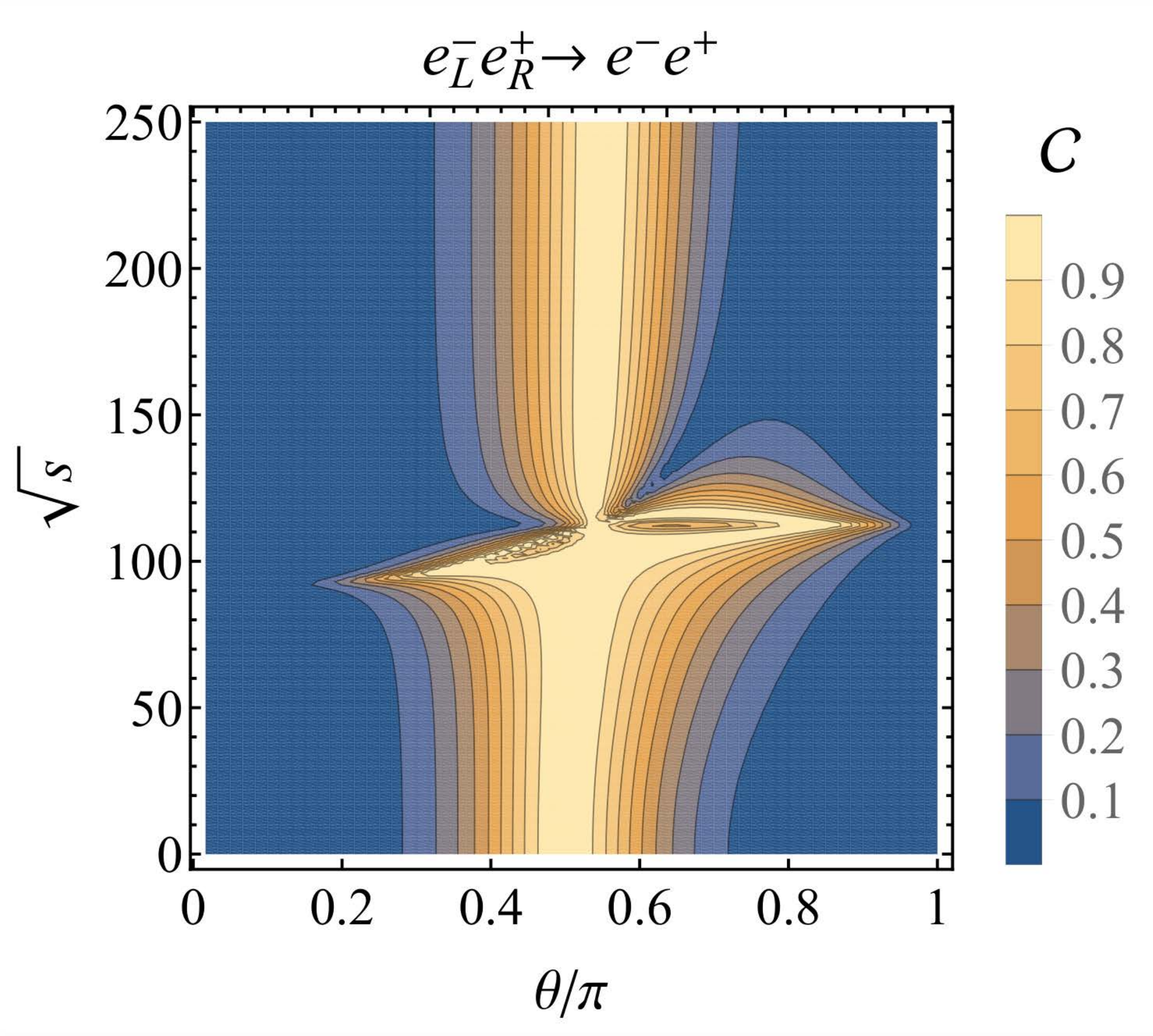} \qquad
  \includegraphics[width=0.45\linewidth]{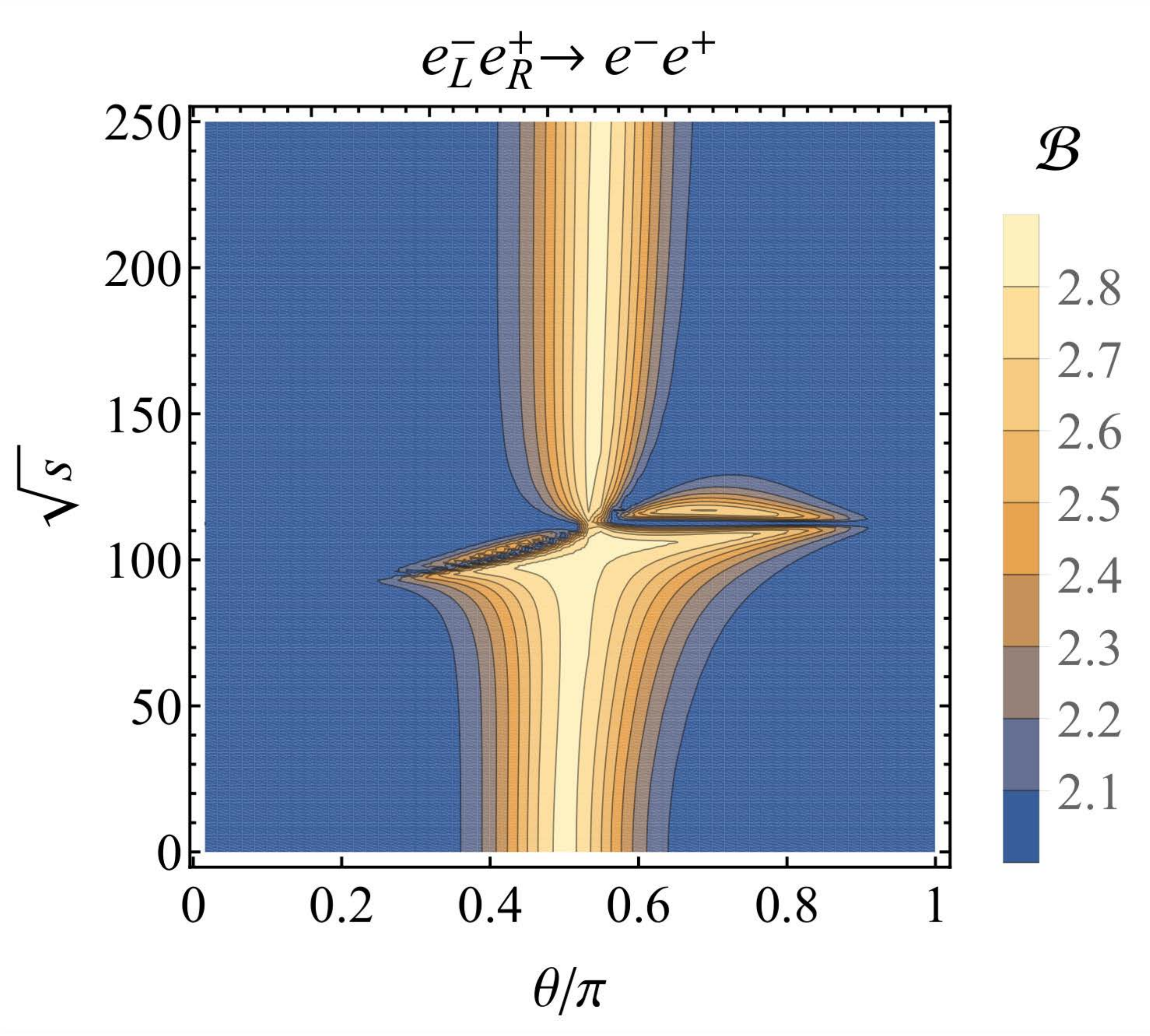}
  \caption{Contour plots for the concurrence $\mathcal{C}$ (left) and Bell variable $\mathcal{B}$ (right) in the plane of $\theta-\sqrt{s}$ (GeV) for the unpolarized process $e^- e^+ \to e^- e^+$ (upper panels) and fully polarized process $e^-_Le^+_R \to e^-e^+$ (lower panels).}
  \label{fig:ee_contour_C&B}
\end{figure}

We evaluate the concurrence $\mathcal{C}$ and Bell nonlocality $\mathcal{B}$ for the $e^+e^-$ system at the $Z$ pole and at $\sqrt{s}=1~\text{TeV}$, including the effects of beam polarization, as shown in Fig.~\ref{fig:ee-C&B}.
To illustrate energy dependence, we present their distributions in the $\theta-\sqrt{s}$ plane for unpolarized $e^+e^-$ in Fig.~\ref{fig:ee_contour_C&B}. The entanglement properties of the unpolarized process at $Z$ pole differ notably from those at higher energies. We can see for the unpolarized situation, the systems are only entangled near the $Z$ resonance and there is no entanglement across the entire angular range at high energy, as seen in the upper panels of Fig.~\ref{fig:ee_contour_C&B}.
However, for fully polarized configurations ($LR$ or $RL$), both the concurrence and Bell nonlocality recover their rich physics pattern as seen in the lower panels of  Fig.~\ref{fig:ee_contour_C&B}. The overall features of the distributions between the $LR$ and $RL$ polarization remain similar. Therefore,  we will present one polarization $e^-_L e^+_R$ for illustration. The theoretical maximal values take place by the condition $|a| = |b|$. Specifically, at the $Z$ pole of $\sqrt s = m_Z$ ($\sqrt s=250$ GeV), the maxima are reached at the angles $\theta_{\rm Max} \approx 0.513\pi\ (0.549\pi)$ for $LR$, and $0.487\pi\ (0.548\pi)$ for $RL$.

\begin{figure}[tb]
  \centering
  \includegraphics[width=0.45\linewidth]{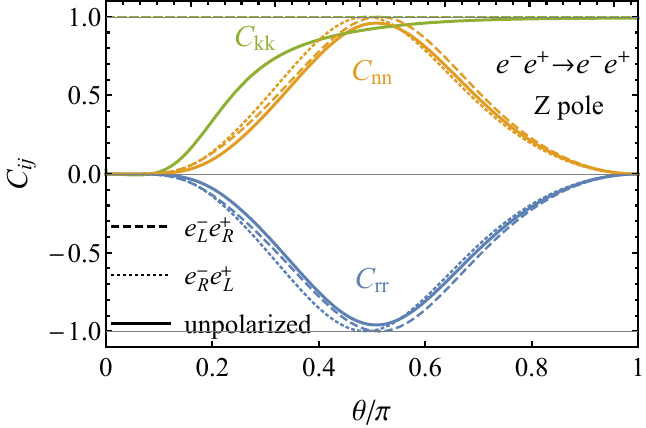}
  \qquad
  \includegraphics[width=0.45\linewidth]{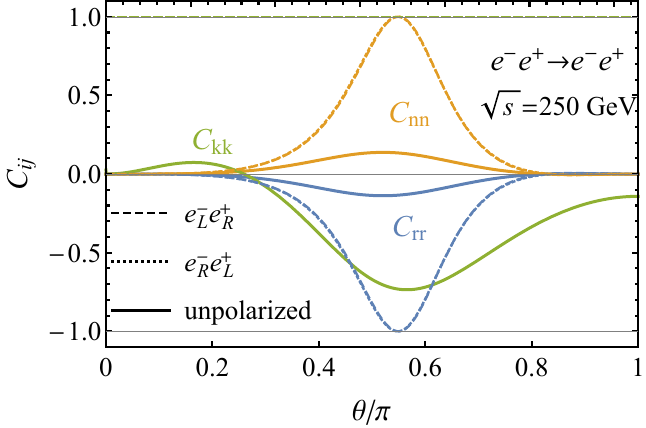}\\
  \includegraphics[width=0.45\linewidth]{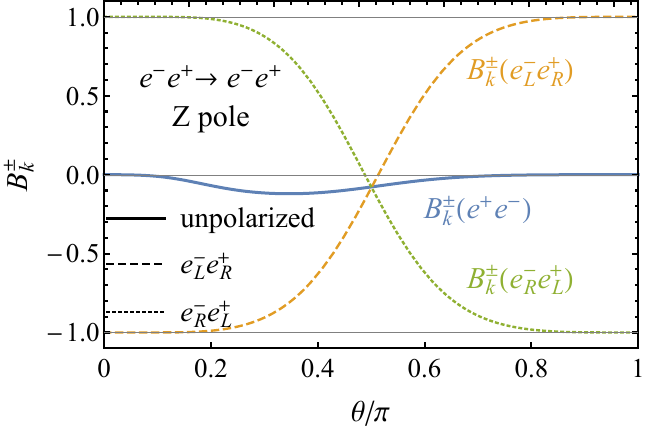}
  \qquad
  \includegraphics[width=0.45\linewidth]{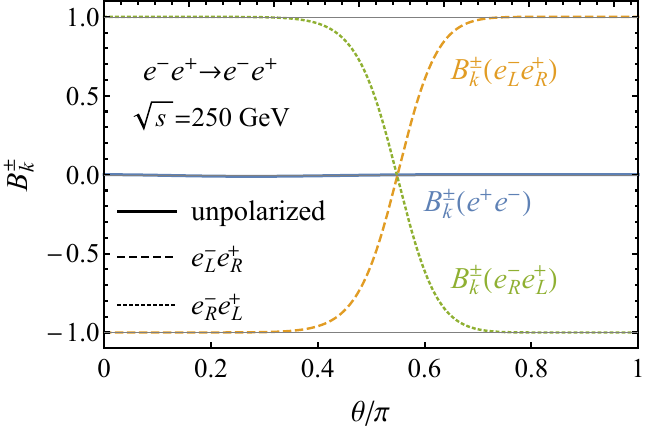}
  \caption{
  Fano coefficients $C_{ij}$ (top) and $B_k^\pm$ (bottom) for $e^-e^+ \to e^-e^+$ at $Z$ pole (left) and $\sqrt{s}=250~\text{GeV}$ (right). In the $C_{ij}$ panel, blue, yellow, and green lines correspond to  $C_{rr}$, $C_{nn}$, and $C_{kk}$, respectively. For all panels, solid, dashed, and dotted lines represent the unpolarized, $LR$, and $RL$ polarized configurations, respectively.}
  \label{fig:ee-cij&bi}
\end{figure}

Figure \ref{fig:ee-cij&bi} compares the Fano coefficients $C_{ij}$ (top panels) and $B_k^\pm$ (bottom panels) for unpolarized and fully polarized ($LR$ and $RL$) configurations at the $Z$ pole (left panels) and $\sqrt{s}=250~\rm{GeV}$ (right panels).
As seen earlier in $e^+e^-\!\to t\bar{t}$ (Fig.~\ref{fig:tt-cij&bi}) and $e^+e^-\!\to\mu^+\mu^-$ (Fig.~\ref{fig:mm-cij&bi}), the beam polarization has a pronounced effect on the final state  polarization parameters $B_k^\pm$, but not much on $C_{ij}$.
Unlike those purely $s$-channel dominated processes, however, Bhabha scattering shows a substantial polarization dependence in the spin-correlation matrix $C_{ij}$.
This occurs because beam polarization modifies the relative weights of the $s/t$-channel interference in $LR/RL$, and the summed density matrix by $LL/RR$.
Crucially, longitudinal polarization reweights the relative importance of the $s$- and $t$-channel amplitudes, and can selectively suppress the $t$-channel–dominated processes in $e^-_Le^+_L$ and $e^-_Re^+_R$ polarizations.
Since these components contribute predominantly separable spin states, their suppression reduces the incoherent separable background in the ensemble and enhances the observable entanglement.
This mechanism is explicitly manifest in the coefficient $C_{kk}$. The separable $t$-channel contribution contributes negatively to $C_{kk}$. Thus, based on the fractions of the helicity components shown in Fig.~\ref{fig:ee-FRrho}, one can understand the contrast in Fig.~\ref{fig:ee-cij&bi}: at the $Z$ pole, the $s$-channel dominates across the angular distribution, maintaining high correlations; at higher energies, the unpolarized state is overwhelmed by the $t$-channel contribution (yielding low or negative $C_{kk}$), while the polarized states $LR$ and $RL$ successfully suppress this classical noise, maintaining $C_{kk} \approx 1$ for all angles.

\begin{figure}[tb]
  \centering
  \includegraphics[width=0.32\linewidth]{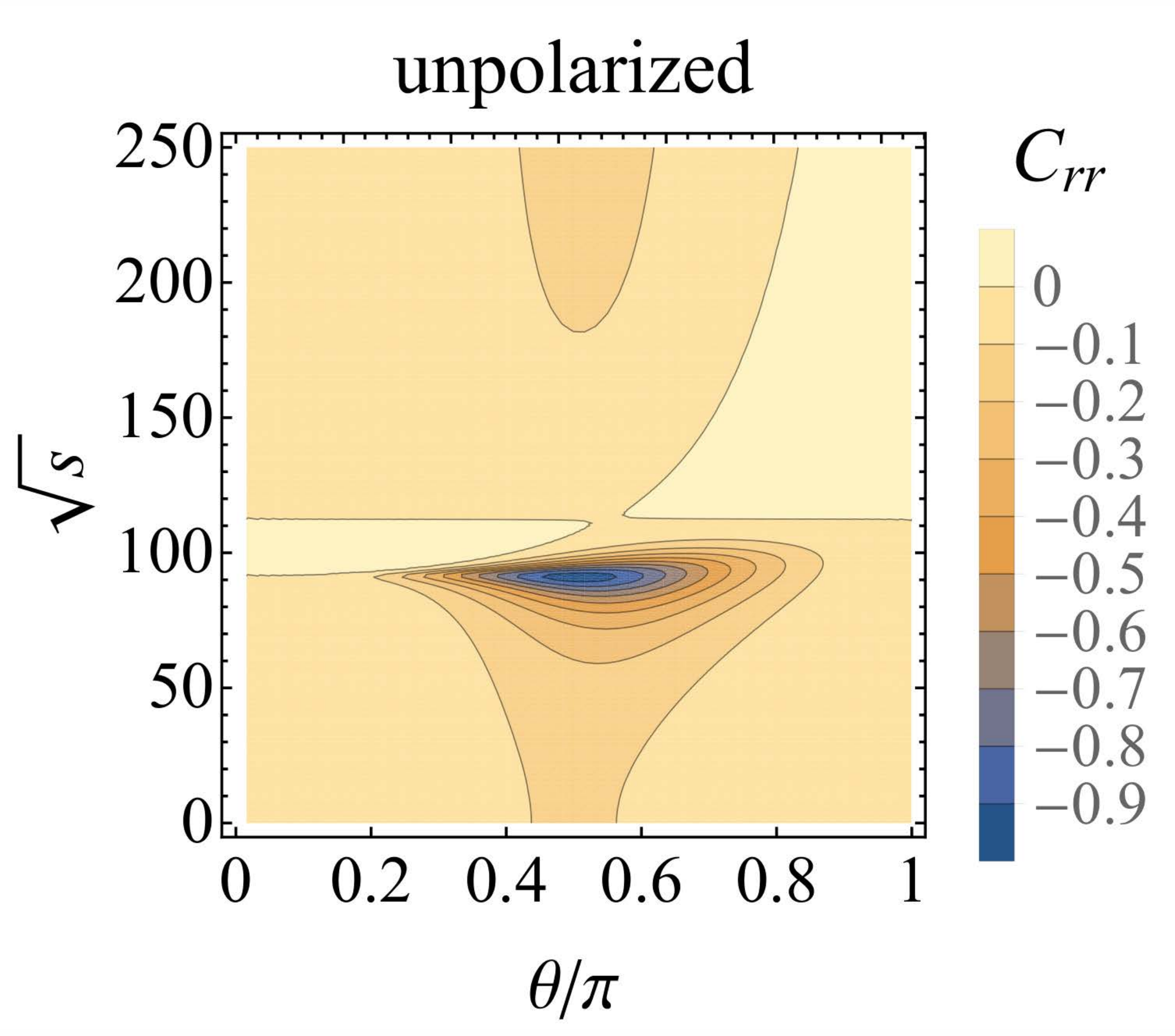}
  \includegraphics[width=0.32\linewidth]{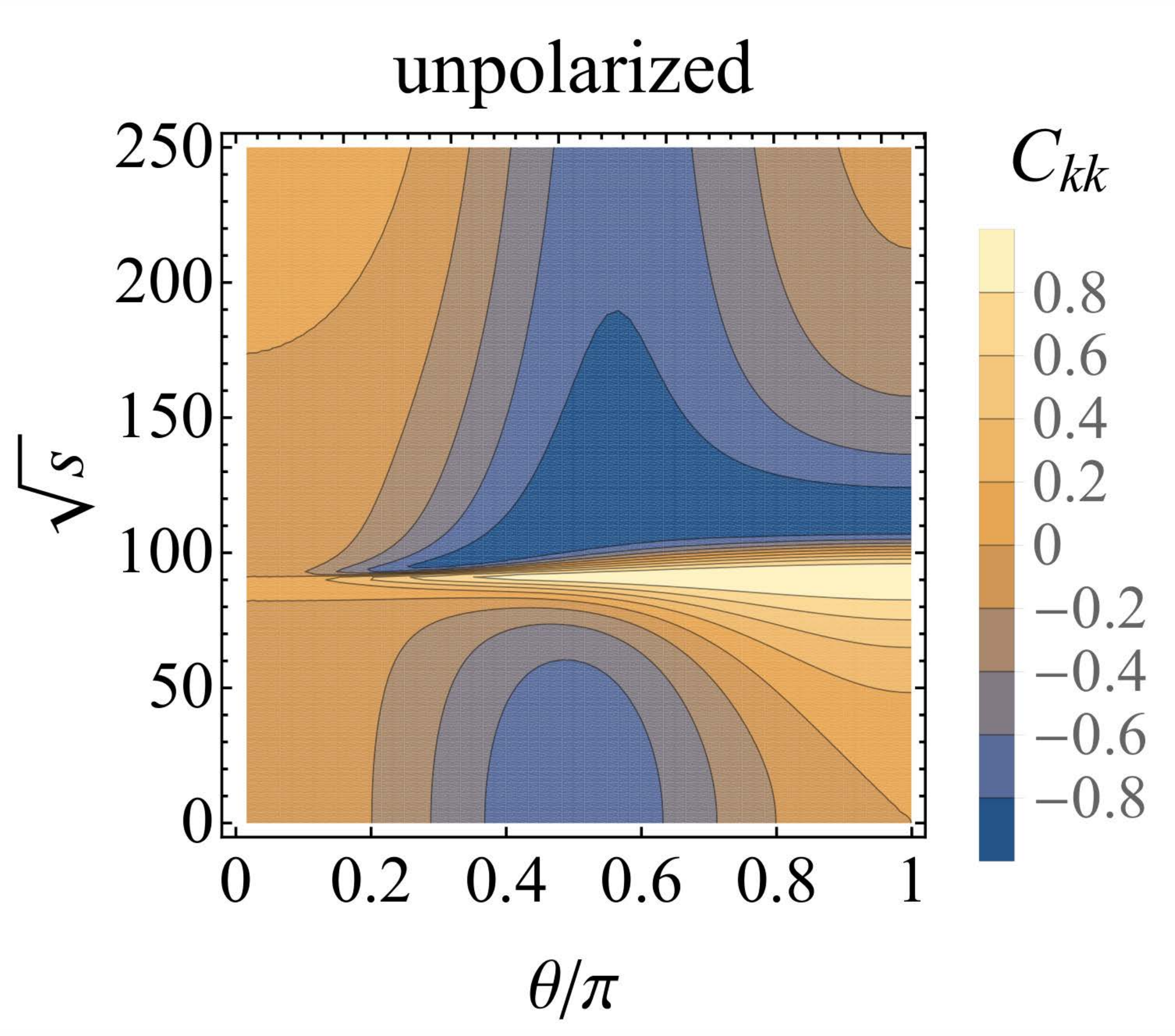}
  \includegraphics[width=0.33\linewidth]{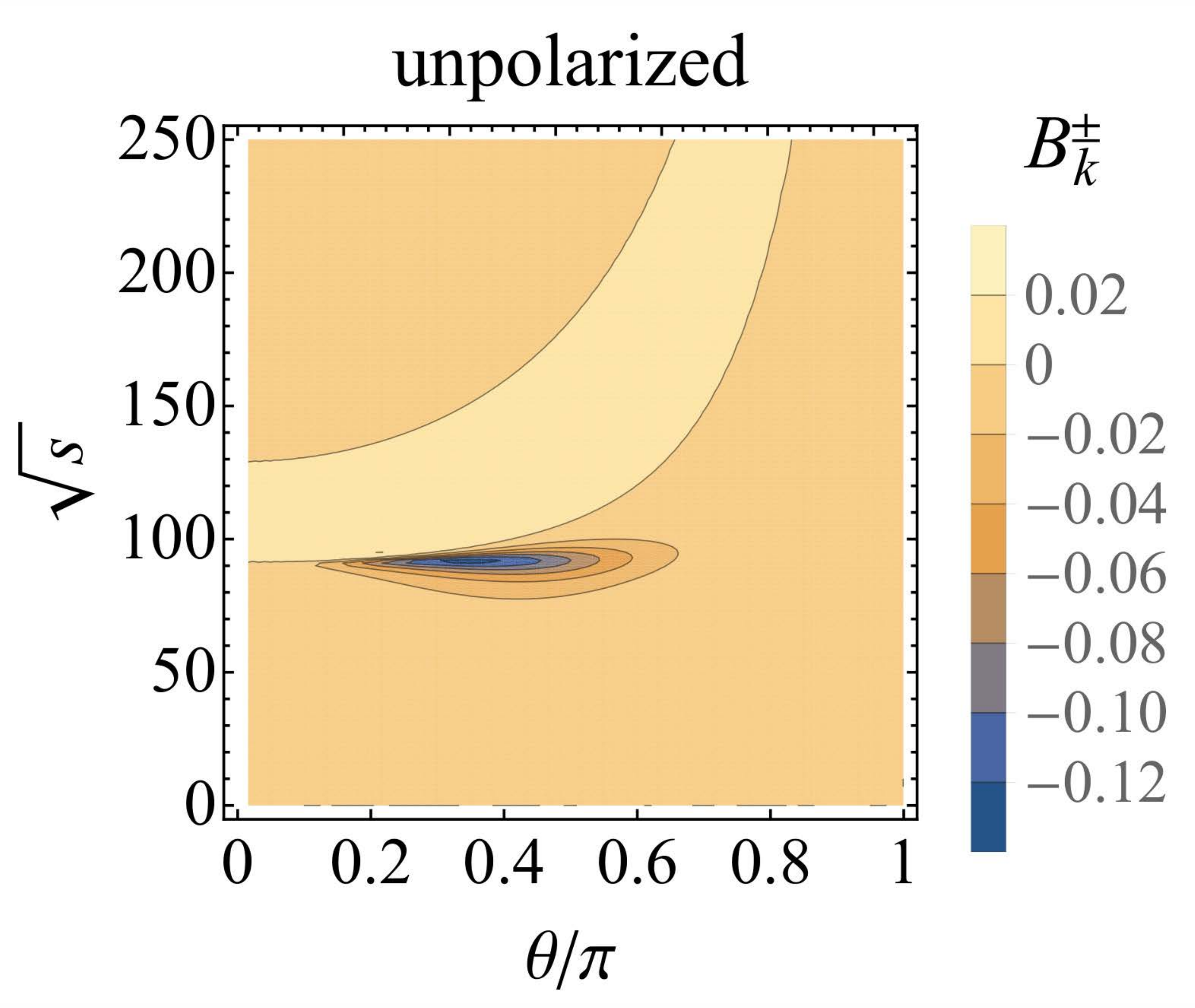}
  \centering
  \includegraphics[width=0.34\linewidth]{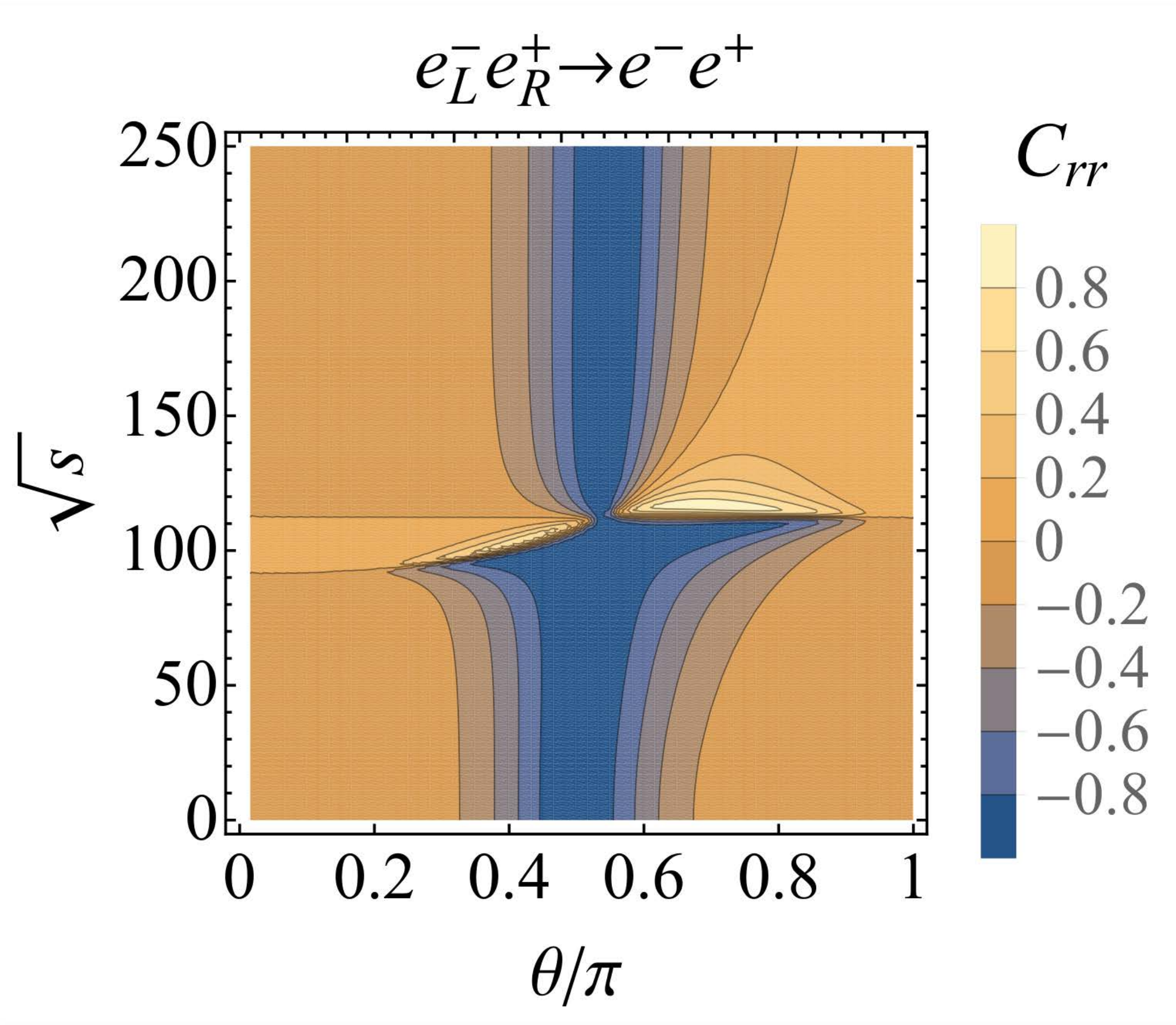} \qquad        \includegraphics[width=0.34\linewidth]{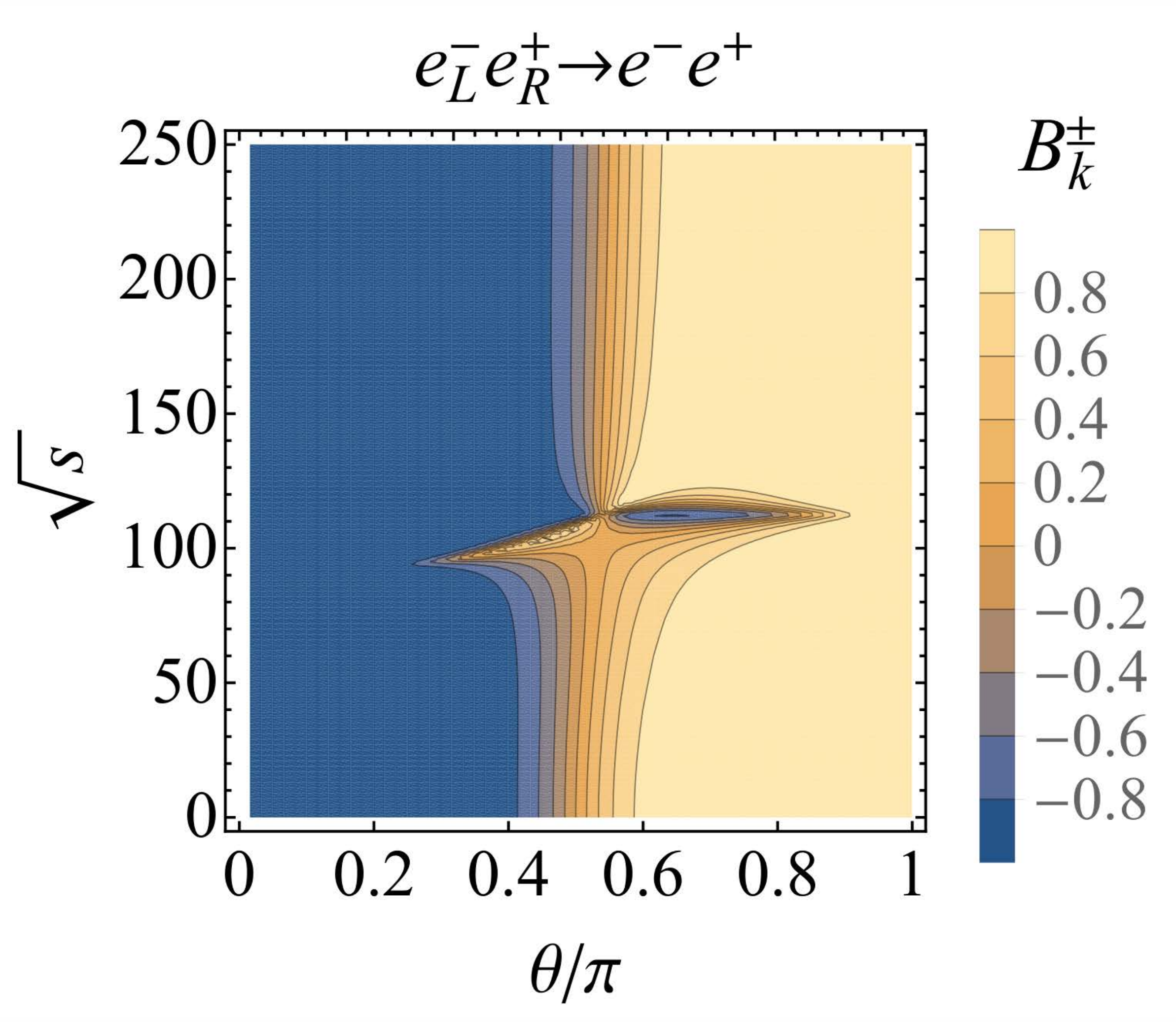}
  \caption{Contour plots for the Fano coefficients $C_{rr}$, $C_{kk}$ and $B_k^\pm$ in the plane of $\theta-\sqrt{s}$ (GeV) for the unpolarized process $e^-e^+ \to e^-e^+$ (upper panels), and $C_{rr}$ (left) and $B_k^\pm$ (right) for polarized configuration of $e^-_Le^+_R \to e^-e^+$ (lower panels).}
  \label{fig:ee-Fano}
\end{figure}

In the massless limit, kinematic constraints force the transverse polarization to vanish $B_{r,n}^\pm=0$. For fully polarized processes ($e^-_Re^+_L, \ e^-_Le^+_R$), $C_{kk}=1$. The non-trivial dynamics are thus captured by $C_{rr}$ (with $C_{nn}=-C_{rr}$) and $B^\pm_k$, along with $C_{kk}$ in the unpolarized case.
Figure~\ref{fig:ee-Fano} shows the contours of $C_{rr}$, $C_{kk}$ and $B_k^\pm$ for unpolarized Bhabha scattering (upper panels). For polarized scattering, the $C_{rr}$ distributions for the $LR$ and $RL$ configurations are nearly identical, whereas the longitudinal polarization vector $B_k^\pm$ changes sign between them.
We therefore present $C_{rr}$ and $B_k^\pm$ for the representative configuration $LR$ in Fig.~\ref{fig:ee-Fano} (lower panels).

The stark contrast between the entanglement properties at the $Z$ pole and at higher energies is governed by the relative weights of the helicity channels.
Near the $Z$ resonance, the entanglement-generating channels ($e^-_L e^+_R \to e^-e^+$ and $e^-_R e^+_L \to e^-e^+$) are significantly enhanced, dominating the cross-section, which leads to maximal entanglement.
In contrast, in the QED-dominated regions (low energy or high energy away from the peak), the separable channels ($e^-_L e^+_L$ and $e^-_R e^+_R$), originating purely from the contribution of the $t$-channel diagram, often dominate the total rate. Although entanglement can still be achieved in $LR/RL$ at high energies, the overwhelming dominance of the separable $LL/RR$ channels results in an overall state that is highly mixed and thus becomes separable for unpolarized beams.
\begin{figure}[tb]
  \centering
  \includegraphics[width=0.45\linewidth]{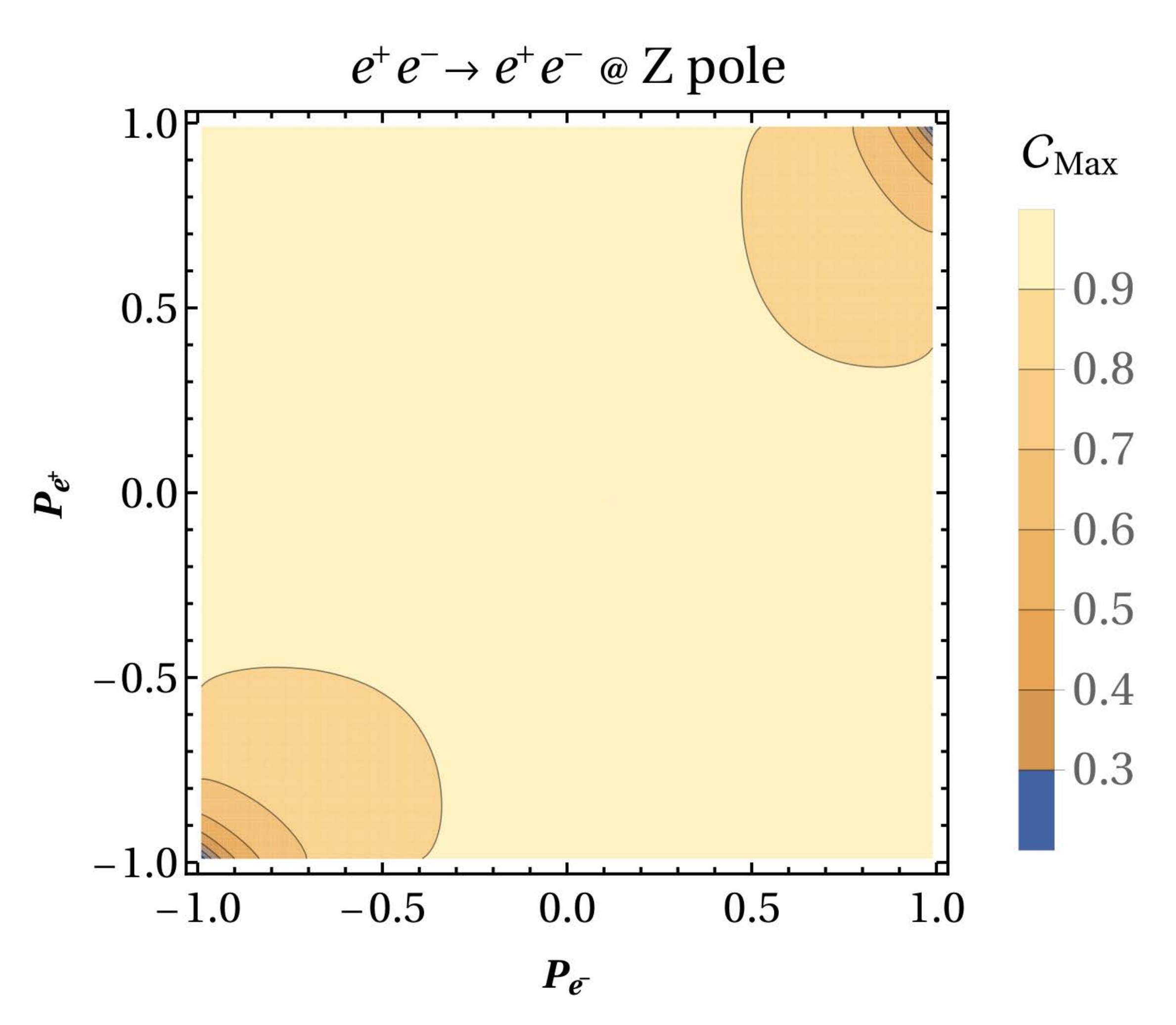} \qquad
  \includegraphics[width=0.45\linewidth]{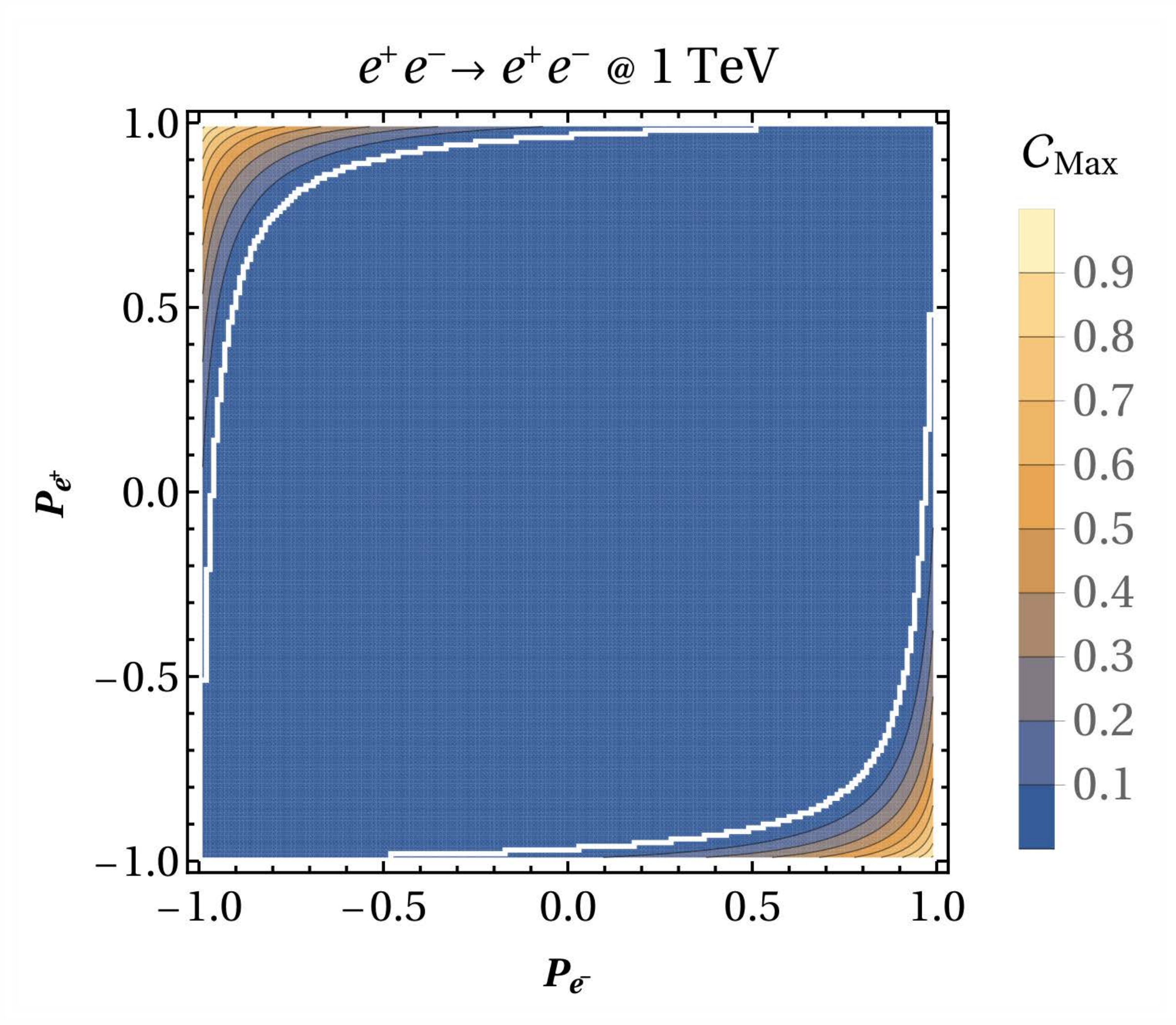}\\
  \includegraphics[width=0.45\linewidth]{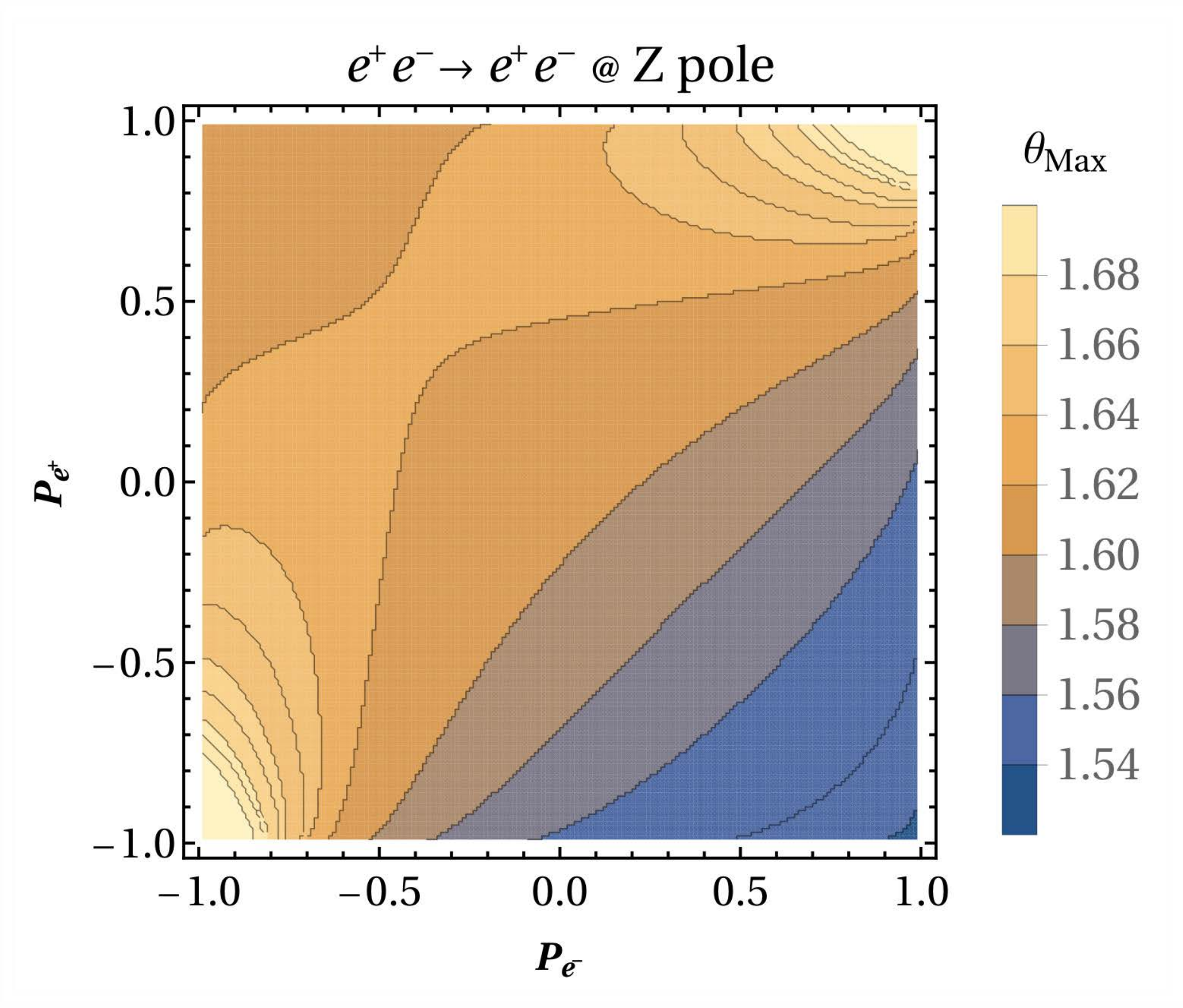} \qquad
  \includegraphics[width=0.45\linewidth]{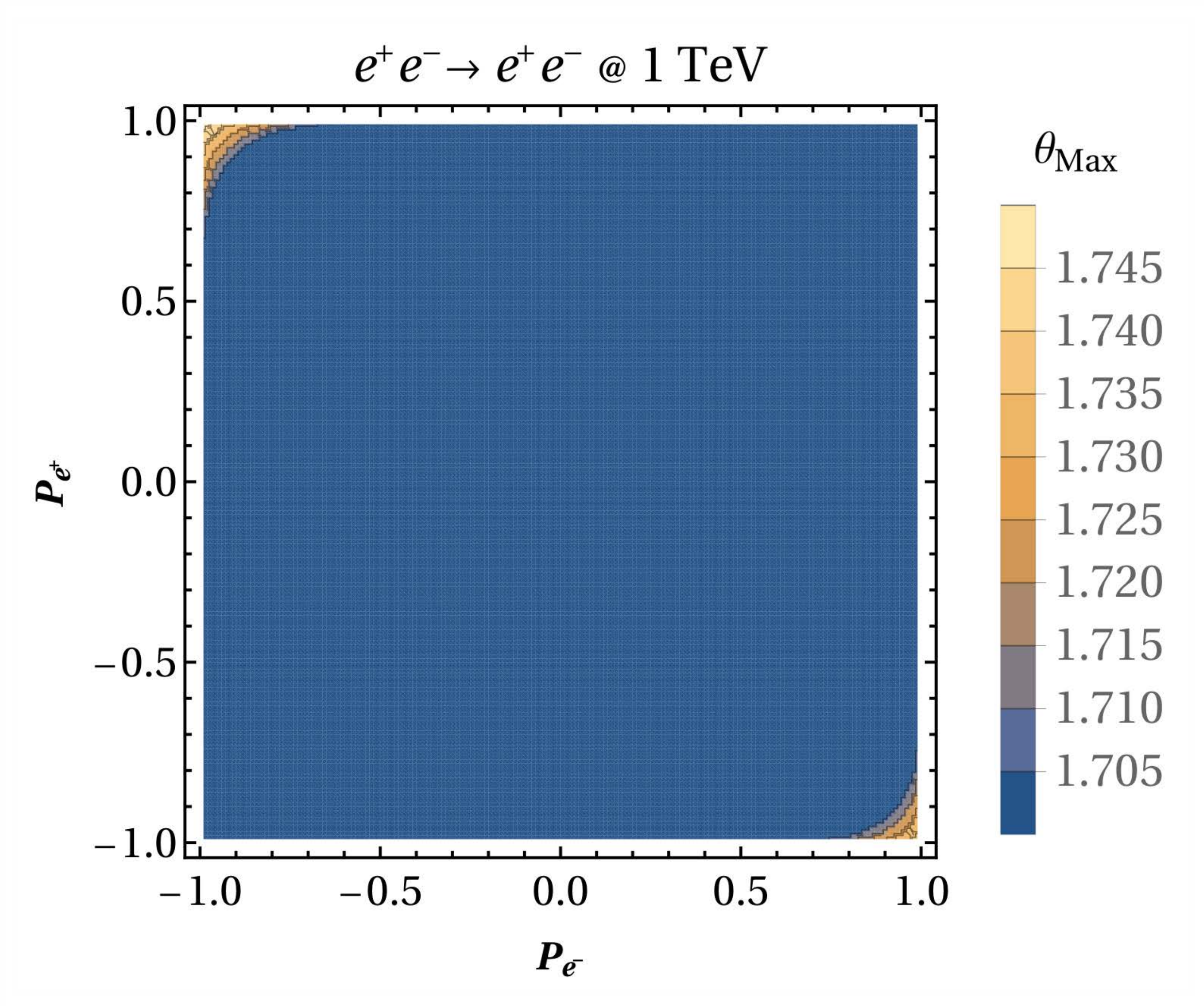}
  \caption{Contour plots for the maximum concurrence $\mathcal{C}_{\rm Max}$ (top panels) and the corresponding optimal scattering angle $\theta_{\rm Max}$ (bottom panels) for the $e^+e^-$ system, evaluated over the initial beam polarization parameters $(P_{e^-}, P_{e^+})$. The left (right) panels show the results at the $Z$ pole ($\sqrt{s}=1~\text{TeV}$), and the white contour line explicitly marks the $\mathcal{C}_{\rm Max}=0$ boundary. }
  \label{fig:eeCmax}
\end{figure}

Following the analysis of fully polarized and unpolarized Bhabha scattering, we now address the experimentally relevant question of achieving entanglement with realistic beam polarization and identifying the optimal polarization configurations. For this purpose, we examine the distributions of the maximum concurrence $\mathcal{C}_{\rm Max}$  (maximized over the scattering angle $\theta$), along with their corresponding optimal angles $\theta_{\rm Max}$, across the beam polarization plane $(P_{e^-}, P_{e^+})$ in Fig.~\ref{fig:eeCmax}. Similar behavior is exhibited for the maximum Bell parameter $\mathcal{B}_{\rm Max}$.

At the $Z$ pole, we find that entanglement is easily achieved, covering virtually the entire polarization space, including the unpolarized point $(0, 0)$, with most configurations yielding $\mathcal{C}_{\rm Max} > 0$. By contrast, at high energies ({\it e.g.}, $1~\rm{TeV}$),  while the fully polarized configurations ($LR$ or $RL$) can still reach maximal entanglement, the vast majority of the $(P_{e^-}, P_{e^+})$ plane fails to produce entanglement, with both $\mathcal{C}_{\rm Max}$ and $\mathcal{B}_{\rm Max}$ falling below their respective thresholds ($\mathcal{C}=0$). The $\mathcal{C}=0$ boundary, indicated by a white contour in Fig.~\ref{fig:eeCmax}, clearly delineates the narrow region where the entanglement survives.

The entanglement in Bhabha scattering at high energies can only be produced within an extremely narrow range of beam polarizations. This necessity prompts a critical consideration of whether the achievable polarization configurations in future $e^+e^-$ colliders are sufficient for measuring entanglement in high energy Bhabha scattering. According to the Technical Design Reports of the proposed facilities, the baseline design of the ILC provides the longitudinal electron/positron polarization of $P_{e^-}/P_{e^+} = 80\%/30\%$, with a possible eventual upgrade to $80\%/60\%$ \cite{Behnke:2013xla}. Similarly, the CLIC baseline accelerator is designed to deliver $80\%$ longitudinal polarization for the electron beam, without positron polarization~\cite{Roloff:2018dqu}. Accordingly, we adopt the configuration $(P_{e^-}, P_{e^+}) = \mathbf{(\pm0.8, \mp0.6)}$ as a representative example of a highly optimized polarization scheme achievable at future $e^+e^-$ colliders. As indicated in Fig.~\ref{fig:eeCmax}, this representative point falls into the $\mathcal{C}_{\rm Max}=0$ area at $1~\rm{TeV}$.
This explicitly demonstrates the need for high beam polarization in order to achieve the entanglement.

%%%%%%%%%%%%%%%%%%%%%%%%%%%%%%%%%%%%%%%%%%%%%%%%%%%%%%%%%%%
\subsubsection{Second Stabilizer R\'{e}nyi Entropy for \texorpdfstring{$e^+e^-$}{ee}} \label{sec:magic-ee}
%%%%%%%%%%%%%%%%%%%%%%%%%%%%%%%%%%%%%%%%%%%%%%%%%%%%%%%%%%%
We complete the analysis of quantum resources in the lepton sector by investigating the SSRE $\mathcal{M}_2$ in Bhabha scattering. This process introduces a unique feature that is absent in $t\bar{t}$ and $\mu^+\mu^-$ production: the significant interference between the $s$- and $t$-channels. This interference plays a crucial role in shaping the distribution of magic, particularly in regimes where entanglement is suppressed.

\begin{figure}[htbp]
  \centering
  \includegraphics[width=0.45\linewidth]{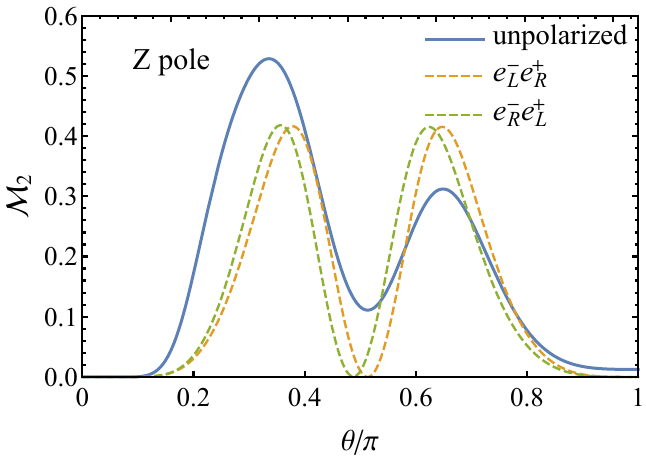}\qquad
  \includegraphics[width=0.43\linewidth]{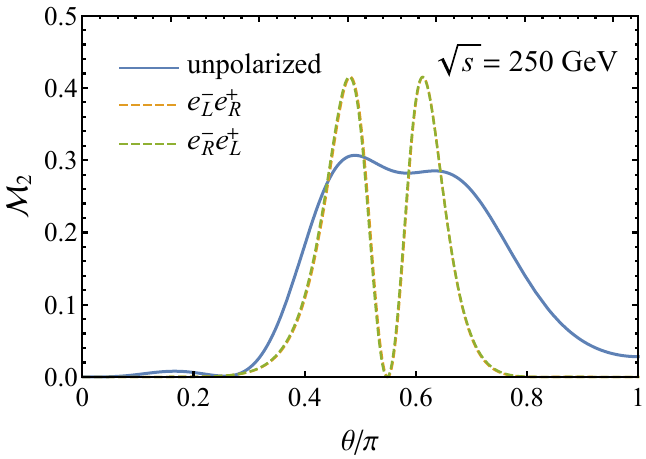}
  \includegraphics[width=0.43\linewidth]{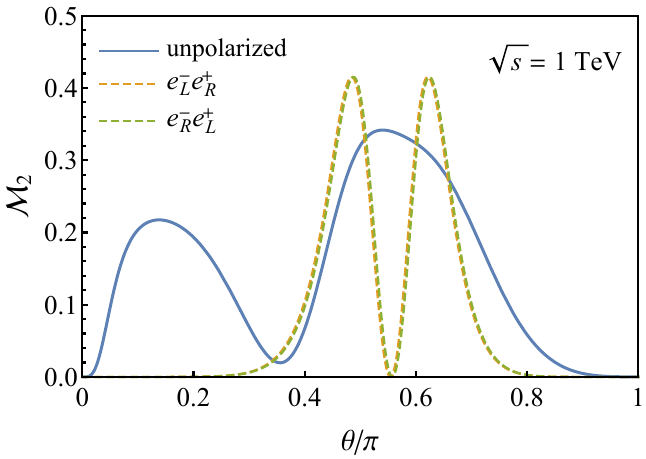}
  \caption{The SSRE $\mathcal{M}_2$ as a function of scattering angle $\theta$ for Bhabha Scattering at the $Z$ pole (left), 250 GeV (right) and 1 TeV (bottom) for unpolarized (solid), $LR$ (dashed), and $RL$ (dotted) beam polarization configurations.}
  \label{fig:magic-ee}
\end{figure}
\begin{figure}[htbp]
  \centering
  \includegraphics[width=0.45\linewidth]{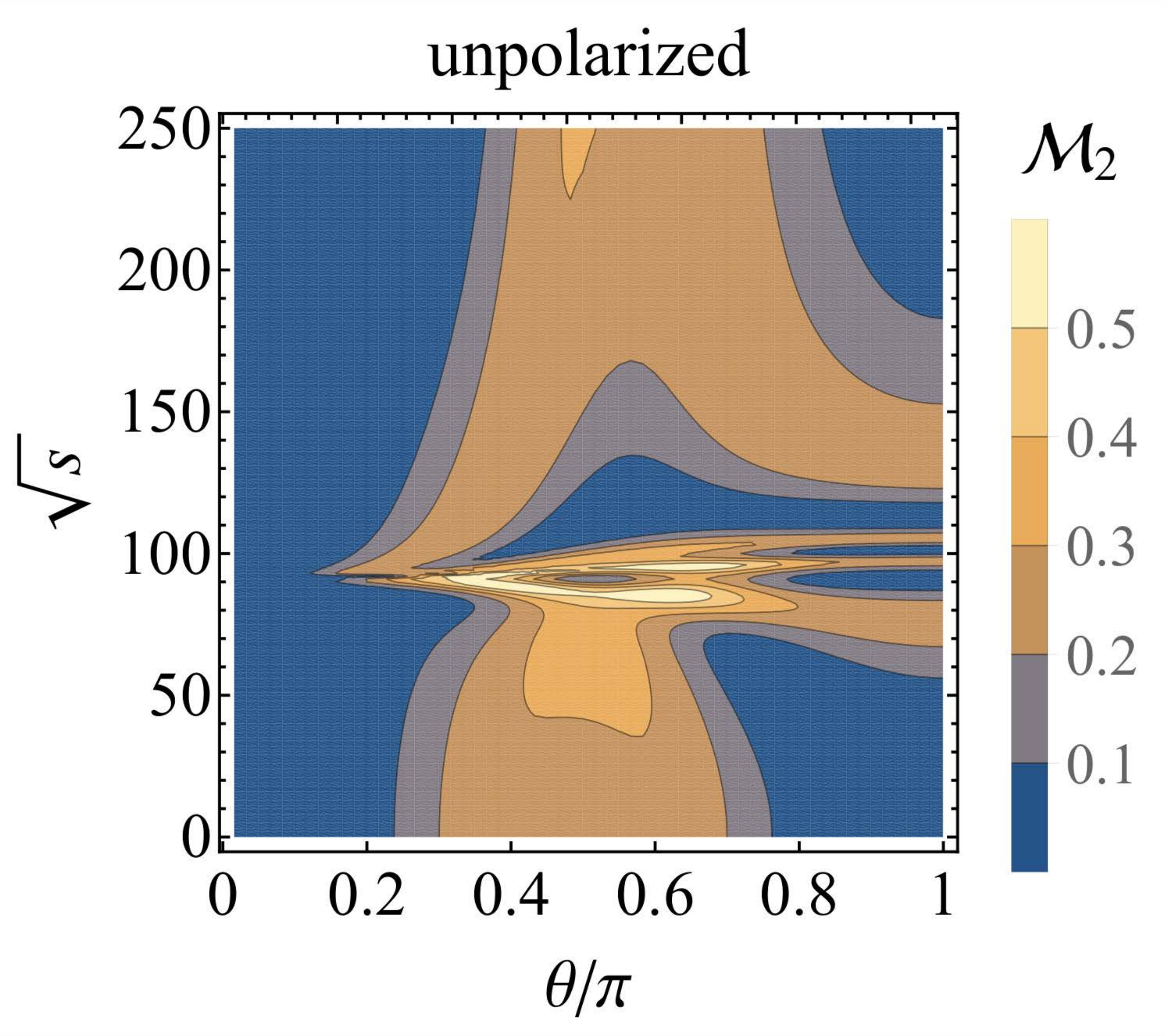}
  \includegraphics[width=0.45\linewidth]{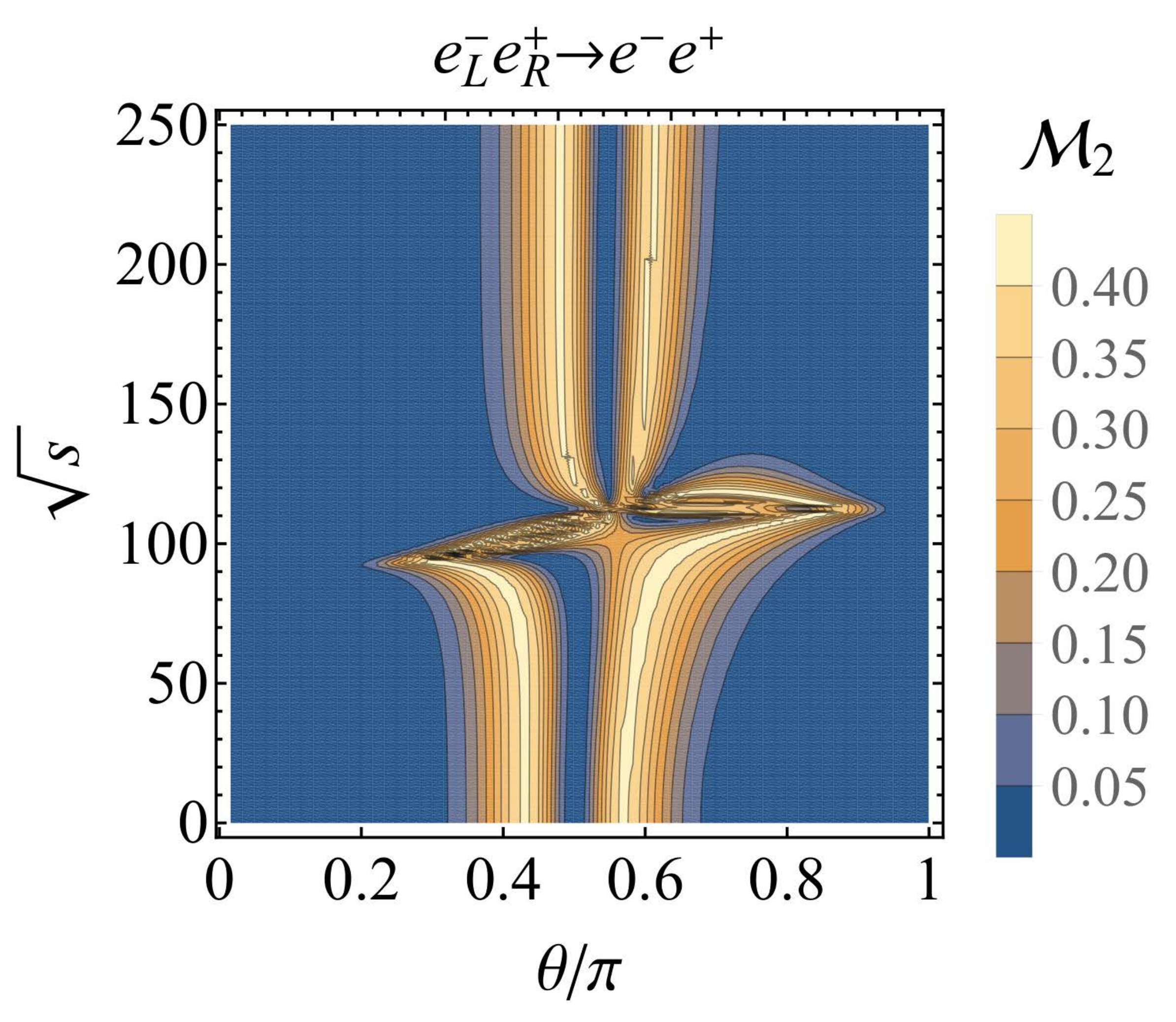}
  \caption{Contour plots for the SSRE $\mathcal{M}_2$ in the plane of $\theta-\sqrt{s}$ (GeV) for unpolarized (left), and $LR$ (right) polarized Bhabha Scattering. }
  \label{fig:ee-magic-contour}
\end{figure}

Figure~\ref{fig:magic-ee} illustrates the SSRE $\mathcal{M}_2$ of Bhabha scattering at $\sqrt{s}= m_Z$, 250 GeV and 1 TeV, comparing the effects of different beam polarizations, while Fig.~\ref{fig:ee-magic-contour} maps the full angular and energy dependence.
Similar to the $\mu^+\mu^-$ and $t\bar{t}$ systems, the polarized $e^+e^-$ system exhibits a twin-peak structure for $\mathcal{M}_2$, which contrasts with the single-peak behavior of $\mathcal{C}$ and $\mathcal{B}$. Notably, $\mathcal{M}_2$ can approach to zero in regions of both zero entanglement and maximal entanglement. This occurs because the Fano parameters in these corresponding regions approach the stabilizer limits of either $0$ or $\pm 1$.

However, Bhabha scattering distinguishes itself through its response to polarization. In $t\bar{t}$ and $\mu^+\mu^-$ production, the spin correlation matrix $C_{ij}$ is robust against polarization. In contrast, Bhabha scattering exhibits a strong polarization dependence in $C_{ij}$ itself, since the beam configuration reweights the relative contributions of the entangled $s$-channel and the separable $t$-channel. Consequently, the variation of $\mathcal{M}_2$ observed in Fig.~\ref{fig:ee-magic-contour} is governed by the combined modification of both polarization vectors $B_k^\pm$ and the correlation matrix $C_{ij}$.

The unpolarized case in Bhabha scattering is unique. At the $Z$ pole, where the $s$-channel dominates, $\mathcal{M}_2$ achieves its maximum value and displays a valley between two peaks, resembling the $\mu^+\mu^-$ system. This minimum coincides with the region of maximum entanglement, reflecting the proximity to a stabilizer Bell state.
The behavior at high energy differs fundamentally from the $t\bar{t}$ and $\mu^+\mu^-$ cases. As established in Fig.~\ref{fig:ee_contour_C&B}, the unpolarized $e^+e^-$ system exhibits entanglement only at the $Z$ pole, with $\mathcal{C}$ vanishing throughout the angular range at high energies.  Despite the total absence of entanglement, Fig.~\ref{fig:magic-ee} (bottom panel) reveals a significant non-zero magic distribution for unpolarized case at 1 TeV, peaking around $\theta \approx 0.2\pi$ and $0.6\pi$. At this energy scale, the single-particle polarization vectors $B^\pm_k$ are negligible. Thus, this ``magic without entanglement'' is driven entirely by the spin correlations $C_{rr}$, $C_{nn}$, and $C_{kk}$. Unlike pure $s$-channel processes where $C_{ij}$ values are simple, the $s/t$-channel interference in Bhabha scattering induces non-trivial values for these correlations (distinct from the stabilizer limits of $0$ or $\pm 1$), generating quantum magic even in a separable state.

\begin{figure}[hbpt]
  \centering
  \includegraphics[width=0.46\linewidth]{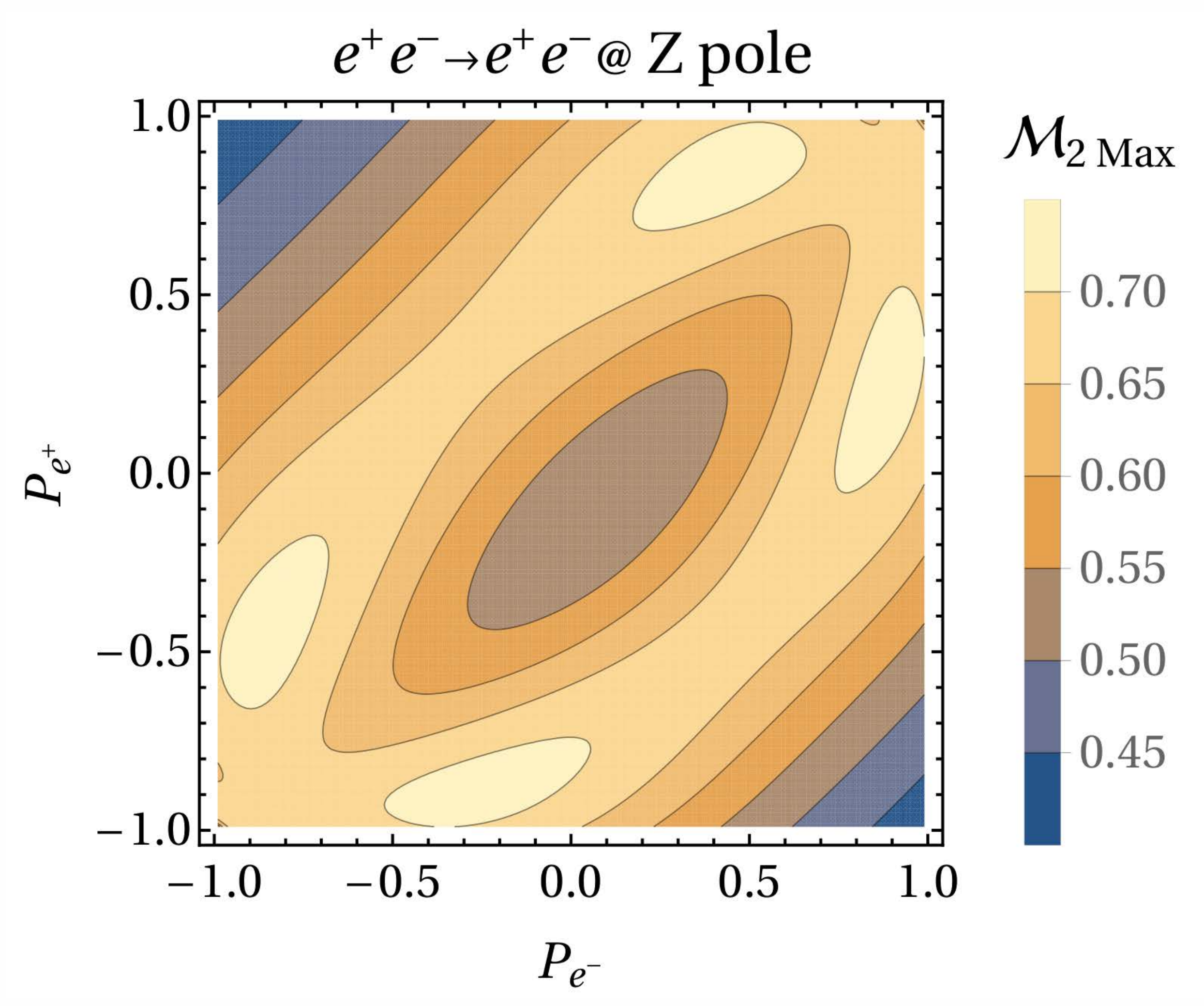} \qquad
  \includegraphics[width=0.44\linewidth]{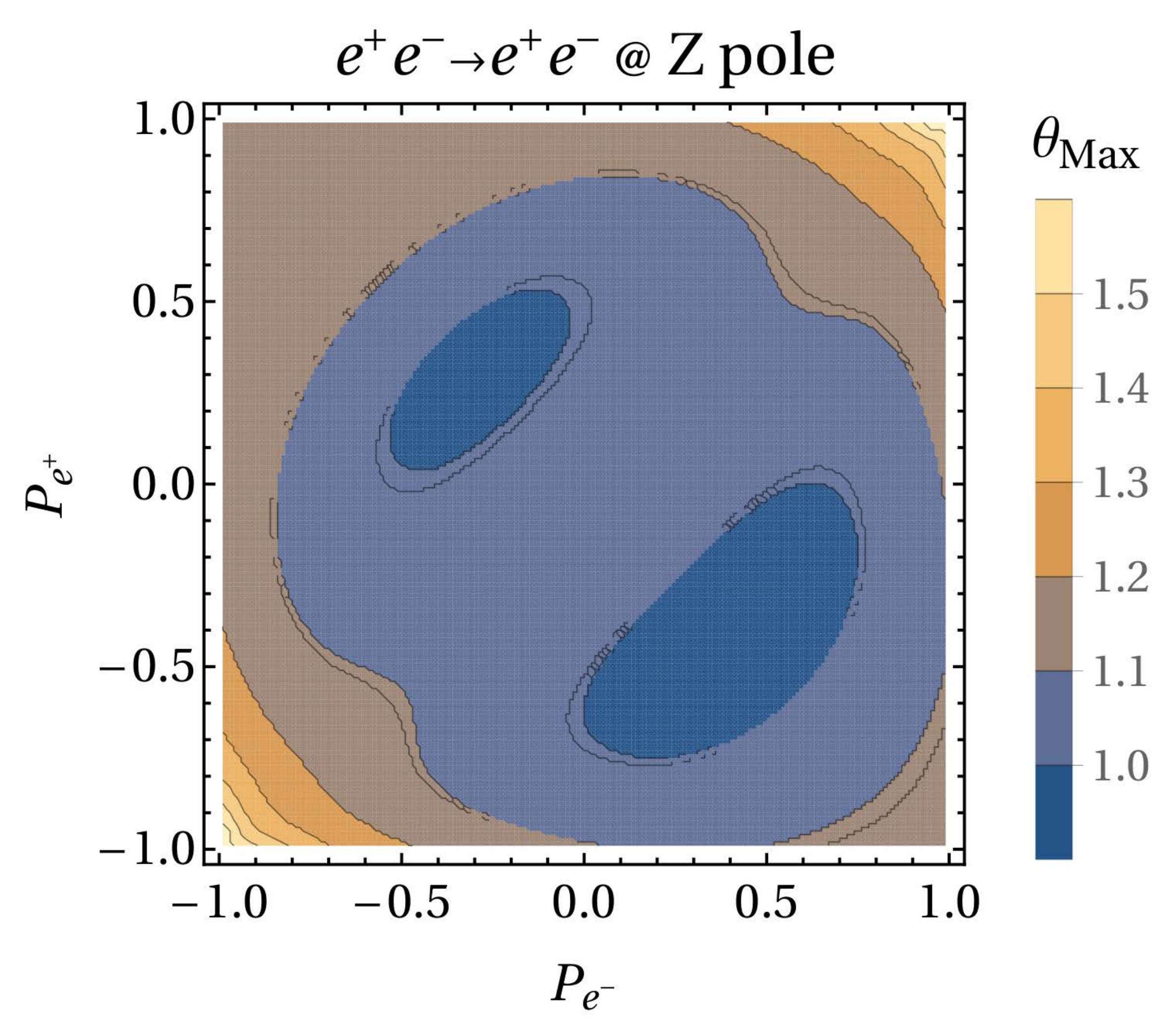}
  \caption{Contour plots for the maximum SSRE $\mathcal{M}_{2_{\rm Max}}$ (left) and the corresponding optimal scattering angle $\theta_{\rm Max}$ (right) for the $e^+e^-$ system at the $Z$ pole, evaluated over the polarization parameters $(P_{e^-}, P_{e^+})$.}
  \label{fig:eeM2max}
\end{figure}

Finally, we address the optimization of magic resources. Figure \ref{fig:eeM2max} presents the maximum value $\mathcal{M}_2$ that is achievable at the $Z$ pole across the beam polarization plane.
The maximal value of $\mathcal{M}_2$ is not found at fully polarized corners. Instead, the global maximum occurs in the partially polarized regions. This suggests that for quantum simulation tasks that rely on magic, the Bhabha scattering process offers a robust resource that does not strictly require the technically demanding fully polarized beams necessary for maximizing entanglement.

%%%%%%%%%%%%%%%%%%%%%%%%%%%%%%%%%%%%%%%%%%%%%%%%%%%%%%%%%%%
\subsection{Quantum Correlation and Magic in \texorpdfstring{$f\bar f$}{ff} Systems}
\label{subsec:SecIII-summary}
%%%%%%%%%%%%%%%%%%%%%%%%%%%%%%%%%%%%%%%%%%%%%%%%%%%%%%%%%%%
With the tomographic results for the $s$-channel massive state of $t\bar{t}$, $s$-channel massless state of $\mu^+\mu^-$, and $s/t$ channel massless production $e^+e^-$, that are governed by the SM electroweak gauge interactions, we now reiterate a few key aspects of the quantum resources of entanglement and magic and their dependence on beam polarization and kinematics.

\subsubsection{Behavior of Entanglement and Magic}
%%%%%%%%%%%%%%%%%%%%%%%%%%%%%%%%%%%%%%%%%%%%%%%%%%%%%%%%%%%%%%
In this work, we confirm the non-monotonic relationship between entanglement and magic.
As observed in the twin-peak structure of magic in Figs.~\ref{fig:Magic-tt} and \ref{fig:magic-contour-tt} for $t\bar t$, Figs.~\ref{fig:magic-mm} and \ref{fig:magic-contour-mm} for $\mu^+\mu^-$,
and Figs.~\ref{fig:magic-ee} and \ref{fig:ee-magic-contour} for $e^+e^-$, these two quantum resources are not simply correlated. This behavior is rooted in the stabilizer formalism. First, stabilizer states, which are efficiently simulatable classically and possess zero magic ($\mathcal{M}_2=0$), include two distinct classes relevant to collider physics:
\textit{Separable Quantum States} ({\it e.g.}, $\ket{\uparrow\uparrow}$ or $\ket{\downarrow\downarrow}$ with $\mathcal{C}=0$), and \textit{Maximally Entangled Bell States} ({\it e.g.}, $(\ket{\uparrow\uparrow} \pm \ket{\downarrow\downarrow})/\sqrt{2}$ with $\mathcal{C}=1$). Secondly, it is important to note the distinction between polarized and unpolarized ensembles.
While fully polarized beams lead to stabilizer product states in the forward/backward limits, unpolarized beams produce an incoherent mixture.
As seen in Fig.~\ref{fig:tt-cij&bi}, the averaged polarization parameter $B_k^\pm$ takes intermediate value ({\it e.g.}, $B_k^\pm \approx -0.3$). This classical mixing pulls the state away from the pure stabilizer sector.
Consequently, based on the interplay between entanglement and magic, the physical parameter space of collider processes can be classified into four distinct regimes:
\begin{itemize}
\item \textbf{The Stabilizer Limit of Separability ($\mathcal{C} \to 0, \mathcal{M}_2 \to 0$):} \\
This trivial regime corresponds to quantum states aligned with the stabilizer axes ({\it e.g.}, $\ket{\uparrow\uparrow}$), efficiently simulatable classically.
In the forward and backward scattering limits ($\theta \to 0, \pi$), angular momentum conservation enforces a strict selection of helicity states.
For fully polarized beams, the Fano coefficients approach discrete values ($B_k^\pm \to \pm 1$, $C_{kk}\to 1$, and $C_{nn,rr} \to 0$), driving the system into a stabilizer state.
Thus, both entanglement and magic vanish; see, {\it e.g.}, Figs.~\ref{fig:tt-C&B}(a) and \ref{fig:Magic-tt}(b).

\item \textbf{The Stabilizer Limit of maximum entanglement ($\mathcal{C} \to 1, \mathcal{M}_2 \to 0$):} \\
In the angular regions where the scattering amplitudes satisfy $|a| \approx |b|$,  typically near the maximum $\cos\theta \approx \pm f_A/f_V$, the system approaches a Bell state $(\ket{\uparrow\uparrow} \pm \ket{\downarrow\downarrow})/\sqrt{2}$, or as joint eigenstates of Pauli operators.
Tomographically, this corresponds to $C_{rr,kk} \to 1$, $C_{nn} \to -1$, and vanishing local polarization $B_k^\pm \to 0$.
Although the entanglement is maximal, the state returns to the stabilizer sector, resulting in the ``valleys'' of zero magic observed between the peaks; see, {\it e.g.}, Figs.~\ref{fig:mm-cij&bi} and \ref{fig:magic-mm}. Note that this limiting case could only be reached with massless fermions $\beta\to 1$.

\item \textbf{The Magic Transition Region ($0 < \mathcal{C} < 1, \mathcal{M}_2 > 0$):} \\
This is the general and common case for polarized scattering, where the system exhibits both quantum correlations and non-stabilizer characteristics.
Magic emerges essentially in the transition zones between the product-state limit and the Bell-state limit.
Here, the Fano coefficients assume fractional values ($0 < |C_{ij}|, |B_k| < 1$), implying coherent superpositions with nontrivial phase differences and weights that cannot be reduced to stabilizer structures. This explains the ``twin-peak'' pattern of magic flanking the entanglement maximum, as seen in Figs.~\ref{fig:Magic-tt},  \ref{fig:magic-mm}, and \ref{fig:magic-ee}.

\item \textbf{The Separable Magic region ($\mathcal{C} = 0, \mathcal{M}_2 > 0$):} \\
This behavior is observed, in particular, in the $t\bar t$ threshold region and in the unpolarized Bhabha scattering, where the reconstructed spin state is separable but non-stabilizer. This indicates that separability does not imply proximity to the stabilizer set.
For instance, there is no significant entanglement near the $e^-e^+ \to t\bar t$ threshold or in the forward/backward scatttering, yet the magic remains non-zero, as shown in Figs.~\ref{fig:Magic-tt} and \ref{fig:magic-contour-tt}.
This reflects the fact that, although the state is separable, it possesses non-stabilizer components arising from correlation $0<|C_{ij}|<1$ as well as the polarization $0<|B_i^{\pm}|<1$, as evidenced from Figs.~\ref{fig:tt-cij-contour} and \ref{fig:tt-b-contour}.
These non-stabilizer structures lead magic to emerge as predicted in Eq.~(\ref{eq:m2fano}), and as prominently seen in the two broad magic bands in the $\theta$--$\sqrt{s}$ plane of Fig.~\ref{fig:magic-contour-tt}.  Because  $B^\pm_i$ respond strongly to beam polarization while $C_{ij}$ change very little, polarization amplifies the magnitude of magic but does not substantially shift the locations of the stabilizer domains.
For the unpolarized Bhabha scattering, we have found that entanglement is only present near the $Z$ pole. Away from the pole, the reconstructed two-spin state becomes separable, and the concurrence vanishes. Nevertheless, Figs.~\ref{fig:magic-ee} and \ref{fig:ee-magic-contour} show that the same state retains nonzero $\mathcal{M}_2$ outside the $Z$-pole regime.  This highlights that the absence of entanglement constrains nonlocal quantum correlations, but it does not preclude non-stabilizer features in the spin subspace.
The origin of this behavior is from the interference between the amplitudes of the $s$- and $t$-channels.
Even when the overall state is separable, such interference changes the relative phases and weights of helicity components that enter the density-matrix reconstruction.
\end{itemize}

\subsubsection{Polarization Sensitivity: \texorpdfstring{$C_{ij}$}{Cij}, \texorpdfstring{$B_k$}{Bk}, and \texorpdfstring{$\mathcal{M}_2$}{M2}}
%%%%%%%%%%%%%%%%%%%%%%%%%%%%%%%%%%%%%%%%%%%%%%%%%%%%%%%%%%%%%%
The distinct response of quantum observables to beam polarization stems from their tomographic origins.
The concurrence $\mathcal{C}$ and Bell nonlocality $\mathcal{B}$ are primarily determined by the spin correlation matrix $C_{ij}$.
In $s$-channel dominated processes ($t\bar{t}$ and $\mu^+\mu^-$), the structure of $C_{ij}$ is governed by the ratio of axial-vector to vector couplings, $|f_A/f_V|$.
The approximate symmetry of this ratio for left- and right-handed fermions leads to the robustness of $C_{ij}$ against polarization changes.
Consequently, $\mathcal{C}$ and $\mathcal{B}$ are largely polarization-independent.

In contrast, magic $\mathcal{M}_2$ depends non-linearly on both $C_{ij}$ and the local polarization vectors $B_k^\pm$, as seen in Eq.~(\ref{eq:m2fano}).
While $C_{ij}$ is robust, $B_k^\pm$ is highly sensitive to the initial beam configuration, typically flipping sign and changing magnitude between $LR$ and $RL$.
For unpolarized beams, $B_k^\pm$ is partially diluted by averaging over initial helicities, but does not generically vanish in the presence of parity-violating electroweak couplings.
Beam polarization thus acts as a ``control knob'' that turns on the $B_k$ terms, driving the state away from the unpolarized mixed configuration and significantly enhancing the magic.

\subsubsection{The \texorpdfstring{$s/t$}{s/t}-Channel Interference}
%%%%%%%%%%%%%%%%%%%%%%%%%%%%%%%%%%%%%%%%%%%%%%%%%%%%%%%%%%%%%%
A comparison between Bhabha scattering ($e^+e^-$) and other processes highlights the role of multi-channel interference. For $e^+e^-\to t\bar t$ and $e^+e^-\to \mu^+\mu^-$, the dynamics is largely $s$-channel dominated, leading to a universal angular dependence of the reconstructed spin correlations.
However, Bhabha scattering receives an additional $t$-channel exchange that is strongly enhanced in the forward region and interferes with the $s$-channel, as shown in Fig.~\ref{fig:ee-FRrho}.
This kinematic reweighting increases the contribution of helicity sectors that carry weaker spin correlations, thereby diluting entanglement in the unpolarized ensemble, as evidenced from Fig.~\ref{fig:ee-C&B} and \ref{fig:ee_contour_C&B}.

Beam polarization in Bhabha scattering therefore does more than inducing nonzero polarizations.
By reweighting the underlying helicity amplitudes, it changes the relative importance of the $s$-channel and $t$-channel contributions, and thus reshapes both the reconstructed density matrix and the spin correlation matrix $C_{ij}$.
As a consequence, entanglement becomes highly polarization dependent away from the $Z$ pole, especially at high energies. The same interference mechanism also explains the ``separable magic'' region with $(\mathcal{C}=0,\mathcal{M}_2>0)$ as discussed above. In the unpolarized $e^-e^+\to e^-e^+$, as evidenced from Figs.~\ref{fig:ee-cij&bi} and \ref{fig:ee-Fano}, the approximate parity symmetry implies $B_k^\pm\simeq 0$, while the $s/t$ interference can still generate non-trivial Fano coefficients $C_{ij}$ away from the $Z$ pole.
Such spin correlations generically place the state outside the stabilizer set, yielding finite magic even when entanglement vanishes, as seen in Figs.~\ref{fig:magic-ee} and \ref{fig:ee-magic-contour}.

In summary, while high-energy colliders naturally produce maximally entangled states (Bell states) via chirality conservation, these states are magic-deficient. Accessing the resource of quantum magic requires operating away from the maximal entanglement points or utilizing beam polarization to navigate the non-stabilizer subspace of the full density matrix.
The kinematic and polarization regions identified in this section will directly guide the sensitivity estimates and optimization studies presented in next Section.

%%%%%%%%%%%%%%%%%%%%%%%%%%%%%%%%%%%%%%%%%%%%%%%%%%%%%%%%%%%
\section{Observability of Quantum Observables at Future \texorpdfstring{$e^+e^-$}{ee} Colliders}
\label{sec:Significance}
%%%%%%%%%%%%%%%%%%%%%%%%%%%%%%%%%%%%%%%%%%%%%%%%%%%%%%%%%%%
Having investigated the theoretical quantum structure of the $e^+e^-\to t\bar{t}$, $e^+e^-\to \mu^+\mu^-$, and $e^+e^-\to e^+ e^-$ processes, now we turn to the analyses for their experimental observability.
A fundamental distinction must be made between the theoretical quantum state and the experimentally reconstructed state.
At colliders, each observed event corresponds to a single measurement of a quantum sub-state, $\rho(\Omega)$, defined at a multi-dimensional phase-space point. The total spin density matrix for a physical state is constructed by performing the sum over phase-space $\Pi$,
\begin{equation} \label{eq:physicalstate}
\rho^{\rm phy}_{\alpha,\beta}(\Pi)=\sum_{\Omega\in \Pi} \rho(\Omega)_{\alpha,\beta},
\end{equation}
where each matrix element must be evaluated in the same fixed reference  frame. However, for practical reasons in high-energy experiments, especially involving event-by-event rotations, spin measurements are often performed using an event-dependent basis, such as the helicity basis  defined with respect to a particle momentum. With $N$ measurements, averaging these sub-states in an event-dependent basis yields a distinct construct termed a ``fictitious state'' \cite{Afik:2022kwm}:
\begin{align}
\label{eq:fictitious}
\overline{\rho}^{\rm fic}_{\alpha,\beta}(\Pi) =\frac{1}{N}\sum_{\Omega\in \Pi} \rho(\Omega)_{\alpha'(\Omega),\beta'(\Omega)} =\frac{1}{N}\sum_{\Omega\in \Pi} U^\dagger_{\alpha\alpha'(\Omega)}\rho(\Omega)_{\alpha,\beta}U_{\beta\beta'(\Omega)},
\end{align}
where $\alpha'(\Omega)$ and $\beta'(\Omega)$ denote the event-dependent spin quantization axes, and $U(\Omega)$ is the corresponding unitary transformation matrix (e.g., from a fixed frame to the helicity basis). Unlike a genuine quantum state, the resulting $\overline{\rho}^{\mathrm{fic}}$ is basis-dependent. Crucially, it has been shown that non-zero entanglement or Bell nonlocality observed in $\overline{\rho}^{\mathrm{fic}}$ nonetheless serves as a reliable witness, rigorously proving the existence of entanglement or Bell nonlocality in a subset of the actual quantum sub-states $\rho(\Omega)$~\cite{Cheng:2023qmz, Han:2023fci, Han:2025ewp}.\footnote{We leave the discussion of the magic of fictitious states to future work.}

Building upon this fictitious state formalism, we proceed to estimate the statistical significance of these quantum signals.
Throughout this section, we present baseline projections intended to compare channels and
polarization strategies rather than to provide a full experimental simulation. Unless stated
otherwise, we fold detector effects, event selection efficiencies, and residual backgrounds into
a single effective systematic uncertainty $\Delta_{\rm sys}$ on the extracted Fano coefficients.
A realistic study would additionally include initial state radiation/beamstrahlung, detector acceptance, and channel-specific
backgrounds; these effects are expected to shift the optimal angular windows but not the main
qualitative conclusions on polarization dependence.

%%%%%%%%%%%%%%%%%%%%%%%%%%%%%%%%%%%%%%%%%%%%%%%%%%%%%%%%%%%%%%%%%%%%%%
\subsection{\texorpdfstring{$e^+e^-\to t\bar{t}$}{ee>tt}}
%%%%%%%%%%%%%%%%%%%%%%%%%%%%%%%%%%%%%%%%%%%%%%%%%%%%%%%%%%%%%%%%%%%%%%

At $e^+e^-$ colliders, $t\bar t$ signal events can be classified by the decay modes of the $W^+W^-$ pair into three categories: (i) fully leptonic ($\ell\ell$), (ii) semi-leptonic ($\ell q$), and
(iii) fully hadronic ($qq$).
Each channel has its own pros and cons from the observational point of view. While the fully leptonic channel is cleaner with two energetic charged leptons in the final state, it suffers from a rather low branching fraction and difficult event construction involving two missing neutrinos. The fully hadronic channel has the largest branching fraction with up to $6$ indentifiable jets, it may encounter QCD backgrounds and combinatorial uncertainty in reconstructing the $t\bar t$ kinematics. The semi-leptonic channel, on the other hand, takes advantage of a larger branching fraction and cleaner and simpler kinematics of one charged lepton and one neutrino in the final state.
We thus choose to consider the semi-leptonic mode for our numerical analyses.

The leptonic decay channels $W \to \ell \nu$ with $\ell=e,\mu$ have an inclusive branching fraction  $Br(W \to \ell \nu)=21.34\%$~\cite{ParticleDataGroup:2024cfk}, while the hadronic branching fraction is $Br(W \to q\bar q)=67.41\%$~\cite{ParticleDataGroup:2024cfk}.
The semi-leptonic top-pair branching fraction into the $\ell q$ final state is therefore
\begin{equation}
  Br(t\bar t \to \ell+\text{jets})
  \;=\; 2\,Br(W\to \ell\nu)\,Br(W\to q\bar q)
  \;=\; 28.77\%,
\end{equation}
where the final state $\ell q$ represents $\ell^- \nu\,\bar b\, q\bar q'\, b$ or $q\bar q'\,\bar b\, \ell^+ \bar\nu\, b$, and both charge-conjugate modes are included.

With the decay approach~\cite{Afik:2020onf}, the double differential cross section describing the angular distributions of decay products of $t$ and $\bar t$ is
\begin{equation} \label{eq:diff2_xsec}
\frac{1}{\sigma}\frac{d^2\sigma}{d\cos\theta_{A,i} d\cos\theta_{B,j}}
= \frac{1}{4}\left( 1 + \kappa_A B^-_i \cos\theta_{A,i}
+ \kappa_B B^+_j \cos\theta_{B,j}
+ \kappa_A \kappa_B C_{ij} \cos\theta_{A,i} \cos\theta_{B,j} \right).
\end{equation}
Here $\kappa_{A,B}$ denote the spin–analyzing powers of the decay products.
The angle $\theta_{A,i}$ is the angle of the $t$ decay product in the $t$ rest frame relative to the axis $i$ which is the spin quantization axis.  The angle $\theta_{B,j}$ is the angle of the $\bar t$ decay product in the $\bar t$ rest frame relative to the axis $j$ which is the spin quantization axis.  Each decay product has its associated spin analyzing power $\kappa_A$ or $\kappa_B$. Finally, the coefficients $B^+_i$, $B^-_j$, and $C_{ij}$ are the same Fano coefficients as in Eq.~\eqref{eq:fano}.

Integrating out the other angular variables in Eq.~\eqref{eq:diff2_xsec}, one obtains the $t$ polarization $B^-_i$ and $\bar t$ the polarization $B^+_j$ from those simple linear relations
\begin{equation}
\label{eq:xsec_b-}
\frac{1}{\sigma}\frac{d \sigma}{d\cos\theta_{A,i}}
 = \frac{1}{2} \left( 1 + \kappa_A B^-_i \cos\theta_{A,i} \right), \quad
%\label{eq:xsec_b+}
 \frac{1}{\sigma}\frac{d \sigma}{d\cos\theta_{B,j}}
 = \frac{1}{2} \left( 1 + \kappa_B B^+_j \cos\theta_{B,j} \right),
\end{equation}
and the spin correlation matrix $C_{ij}$ from
\begin{equation}
\label{eq:xsec_cij}
\frac{1}{\sigma}\frac{d \sigma}{d(\cos\theta_{A,i} \cos\theta_{B,j})}
 =- \frac{1}{2} \left( 1 + \kappa_A \kappa_B C_{ij} \cos\theta_{A,i} \cos\theta_{B,j} \right) \log | \cos\theta_{A,i} \cos\theta_{B,j} |.
\end{equation}
Fano coefficients then can be readily extracted from the measured distributions by fitting, taking the asymmetry, or taking the mean. For example, using the mean, the coefficients are obtained by
\begin{align}
B^-_i = \frac{3 \langle \cos\theta_{A,i} \rangle}{\kappa_A},
\qquad
B^+_j = \frac{3 \langle \cos\theta_{B,j} \rangle}{\kappa_B},
\qquad
C_{ij} = \frac{9 \langle \cos{\theta_{A,i}}\cos{\theta_{B,j}} \rangle}{\kappa_A \kappa_B}.
\end{align}
Alternatively, the coefficients can be evaluated by forming the asymmetry of the distributions in Eqs.~\eqref{eq:xsec_b-}$-$\eqref{eq:xsec_cij} defined with respect to a linear variable $x$
\begin{equation}
A(x)=\frac{N(x>0)-N(x<0)}{N(x>0)+N(x<0)},
\end{equation}
where $N(x>0)$ is the number of events with $x>0$ and $N(x<0)$ is that with $x<0$. With the asymmetries, the coefficients are determined by
\begin{align}
\label{eq:asymmetries}
B^-_i = \frac{2}{\kappa_A}A(\cos{\theta_{A,i}}),
\qquad
B^+_j = \frac{2}{\kappa_B}A(\cos{\theta_{B,j}}),
\qquad
C_{ij}=\frac{4}{\kappa_A\kappa_B} A(\cos{\theta_{A,i}}\cos{\theta_{B,j}}).
\end{align}
The spin analyzing power $\kappa$ ranges from 0 to $\pm1$. A value of $\kappa=0$ indicates that there is no correlation between the spin of top  (or anti-top) and the direction of the decay product chosen, while a value of $\kappa=\pm1$ indicates the maximum correlation or anti-correlation.
With the SM electroweak interactions, the optimal spin analyzing powers of the leptonic and hadronic decay of the top quark \cite{Han:2023fci, Tweedie:2014yda} are
\begin{equation}
    |\kappa_{\rm lep}|=1,\quad\quad\quad |\kappa_{\rm had}|=0.64.
\end{equation}

From the analysis in Sec.~\ref{subsec:tt}, we recall that the maximal entanglement of the $t\bar t$ system is reached at the scattering angle $\theta_c = \arccos(\beta f_A^L/f_V^L)$ and $\arccos(-\beta f_A^R/f_V^R)$ for the $e^-_Le^+_R$ and $e^-_Re^+_L$ polarizations, which leads to a quantum state closest to a Bell state.
We thus select the events reconstructed in an angular window of size $\Delta \theta$ around center angle $\theta_c$
\begin{equation}
\label{eq:theta_cut}
\theta_c - \frac{\Delta \theta}{2} < \theta < \theta_c + \frac{\Delta \theta}{2} ,
\end{equation}
where $\theta$ is defined in the helicity basis in the $t\bar t$ c.m.~frame.
The value of $\Delta \theta$ can be optimized to achieve the optimal sensitivity based on the luminosity of the dataset.  A lower value of $\Delta \theta$ selects a state closer to a Bell state, but with lower statistics.  We note that $\Delta \theta$ may take a different value, and thus result in a different number of reconstructed events $N$, for the decay and kinematic approaches to achieve their corresponding optimal sensitivity.

With the decay approach, we measure the distributions of Eqs.~\eqref{eq:xsec_b-}$-$\eqref{eq:xsec_cij} using the angular selection described in Eq.~\eqref{eq:theta_cut}.
The values of $B^+_i$, $B^-_j$, and $C_{ij}$ are extracted by the asymmetries in Eq.~\eqref{eq:asymmetries}.  The statistical uncertainty on Fano coefficients using the asymmetry is
\begin{equation}
\label{eq:stat_decay}
\Delta_{\rm stat}(B^+_{i}) = \frac{4}{\sqrt{N}},
\qquad\qquad
\Delta_{\rm stat}(B^-_{j}) = \frac{4}{\sqrt{N}},
\qquad\qquad
\Delta_{\rm stat}(C_{ij}) = \frac{4}{\sqrt{N}},
\end{equation}
where $N$ is the number of reconstructed events in the analysis and the numeric factor of 4 is from Eq.~\eqref{eq:asymmetries}.

For top quark measurements, the target systematic uncertainty is at the one to a few percent level. For the leptonic final states of $\mu^+\mu^-$ and $e^+e^-$, the systematic error may be better controlled \cite{ILC:2013jhg, CLICdp:2018cto, CEPCStudyGroup:2018ghi, FCC:2018vvp}.
We therefore choose to show our results for the following benchmark values of the systematic uncertainty
\begin{equation}
 \Delta_{\rm sys} = \{ 0.1\%,\ 0.5\%,\ 1\%,\ 2\%\} .
\end{equation}
For simplicity, we take $\Delta_{\mathrm{sys}}$ as an overall uncertainty on the determination of the final Fano coefficients.

The statistical and systematic uncertainties of concurrence $\Delta (\mathcal{C})$ and Bell nonlocality $\Delta (\mathcal{B})$ depend on the uncertainty of $C_{ij}$ and $B_i^\pm$. The error–propagation formulas for observable $\Omega$ ($\mathcal{C}$, $\mathcal{B}$ and $\mathcal{M}_2$) is as follows,
\begin{equation}
    \Delta (\Omega)=\sqrt{\sum_{ij}\left(\frac{\partial\Omega}{\partial C_{ij}}\right)^2 \Delta^2(C_{ij})+\sum_i \left(\frac{\partial\Omega}{\partial B^+_i}\right)^2 \Delta^2(B_i^+)+\sum_j \left(\frac{\partial\Omega}{\partial B^-_j}\right)^2 \Delta^2(B_j^-)},
    \label{eq:error_prop}
\end{equation}
These uncertainties are propagated to the uncertainties on $\mathcal{C}$ and $\mathcal{B}$.
The final significance $\mathcal{S}$ is calculated as
\begin{equation}
\label{eq:significance}
\mathcal{S}(\mathcal{C}) = \frac{\mathcal{C}}{\sqrt{\Delta_{\rm stat}^2(\mathcal{C})+\Delta_{\rm sys}^2(\mathcal{C})}},
\qquad\qquad
\mathcal{S}(\mathcal{B}) = \frac{\mathcal{B} - 2}{\sqrt{\Delta_{\rm stat}^2 (\mathcal{B})+\Delta_{\rm sys}^2 (\mathcal{B})}},
\end{equation}
When it is relevant, we use a shorthanded notation $\Delta_{\rm tot}=\sqrt{\Delta_{\rm stat}^2+\Delta_{\rm sys}^2}$.

\begin{table*}[tb]
    \caption{Projected significance ($\mathcal{S}$) and precision ($\mathcal{S}^{-1}$) for observing entanglement ($\mathcal{C}$) and Bell nonlocality ($\mathcal{B}$) in $e^+e^-\to t \bar{t}$ at $\sqrt{s}=1~\mathrm{TeV}$ with an integrated luminosity $5\ \mathrm{ab^{-1}}$. The results are presented for systematic uncertainty $\Delta_{\rm sys}=1\%$ and $2\%$ with three polarization configurations $(P_{e^-}, P_{e^+})$. }
    \label{table:C&B-tt}
    \centering
    \tabcolsep=0.18cm
    \renewcommand\arraystretch{1.2}
    \resizebox{\textwidth}{!}{
        \begin{tabular}{|c|c||c|c|c|c|c||c|c|c|c|c|}
            \hline
            \multicolumn{2}{|c||}{\textbf{$\sqrt{s}=1$ TeV}} & \multicolumn{5}{c||}{\textbf{Entanglement Measure ($\mathcal{C}$)}} & \multicolumn{5}{c|}{\textbf{Bell Nonlocality ($\mathcal{B}-2$)}} \\
            \hline
            $\boldsymbol{(P_{e^-}, P_{e^+})}$ & $\Delta_{\rm sys}$  & $\mathcal{C}$ & $\Delta_{\rm tot}( \mathcal{C})$ & $\mathcal{S},\ \mathcal{S}^{-1}$ & $\theta_c$ & $\Delta\theta$ & $\mathcal{B}-2$ & $\Delta_{\rm tot}( \mathcal{B})$ & $\mathcal{S},\ \mathcal{S}^{-1}$ & $\theta_c$ & $\Delta\theta$\\
            \hline
            \multirow{2}[2]{*}{$\boldsymbol{(0, 0)}$} & 1$\%$ & $0.49$ & 0.022 & $>5\sigma$, $4.5\%$ & $119^\circ$ & $92^\circ$ & 0.37 & 0.090 & $4.2 \sigma$ & $119^\circ$ &$37^\circ$ \\
            \cline{2-12}
            & 2$\%$ & $0.55$ & 0.035 & $>5\sigma$, $6.4\%$ & $119^\circ$ & $75^\circ$ & 0.41 & 0.133 & $3.1 \sigma$ & $119^\circ$ & $29^\circ$ \\
            \cline{1-12}
            \multirow{2}[2]{*}{$\boldsymbol{(+0.8, -0.6)}$} &  $1\%$ & $0.41$ & 0.025 & $>5\sigma$, $6.1\%$ & $124^\circ$ & $110^\circ$ & 0.34 & 0.10 & $3.3 \sigma$ & $124^\circ$ &$38^\circ$ \\
            \cline{2-12}
            & $1\%$ & $0.41$ & 0.032 & $>5\sigma$, $7.9\%$ & $124^\circ$ & $110^\circ$ & 0.37 & 0.14 & $2.6 \sigma$ & $124^\circ$ & $31^\circ$ \\
            \cline{1-12}
            \multirow{2}[2]{*}{$\boldsymbol{(-0.8, +0.6)}$}
                & 1$\%$ & $0.49$ & 0.020 & $>5\sigma$, $4.2\%$ & $116^\circ$ & $101^\circ$ & 0.41 & 0.074 & $>5\sigma, 18\%$ & $116^\circ$ &$34^\circ$ \\
            \cline{2-12}
            & 2$\%$ & $0.49$ & 0.032 & $>5\sigma$, $6.5\%$ & $116^\circ$ & $101^\circ$ & 0.45 & 0.12 & $3.7 \sigma$ & $116^\circ$ & $25^\circ$ \\
            \hline
            \end{tabular} }
\end{table*}

The results for the significances $\mathcal{S}(\mathcal{C})$, $\mathcal{S}(\mathcal{B})$, and $\mathcal{S}(\mathcal{M}_2)$ in $e^+e^-\to t \bar{t}$ with polarization of $\mathbf{(0.8,-0.6)}$, $\mathbf{(-0.8,0.6)}$, and the unpolarized result for comparison, are shown in Table~\ref{table:C&B-tt} and Table~\ref{table:m2-tt}.
In these tables, we show the center value of scattering angle $\theta_c$ and an angular window of size $\Delta \theta$ for events selection, that would optimize the significance of the associated quantity.
Furthermore, when the significance $\mathcal{S}$ is above $5\sigma$, we show the expected precision, corresponding to $\mathcal{S}^{-1}$, achievable in the measurement.

\begin{table*}[tb]
    \caption{Projected significance ($\mathcal{S}$) and precision ($\mathcal{S}^{-1}$) for the magic observable ($\mathcal{M}_2$) in $e^+e^-\to t \bar{t}$ at $\sqrt{s}=1 ~\mathrm{TeV}$ with an integrated luminosity $5\ \mathrm{ab^{-1}}$. The results are presented for systematic uncertainty $\Delta_{\rm sys}=1\%$ and $2\%$, and three polarization configurations $(P_{e^-}, P_{e^+})$. The left half of the table presents results is for the central region, and the right half is for the backward region. }
    \label{table:m2-tt}
    \centering
    \tabcolsep=0.10cm
    \renewcommand\arraystretch{1.2}
    \resizebox{\textwidth}{!}{
        \begin{tabular}{|c|c||c|c|c|c|c||c|c|c|c|c|}
            \hline
            \multicolumn{2}{|c||}{\textbf{$\sqrt{s}=1$ TeV}} & \multicolumn{5}{c||}{\textbf{ Central Region}} & \multicolumn{5}{c|}{\textbf{ Backward Region}} \\
            \hline
            $\boldsymbol{(P_{e^-}, P_{e^+})}$ & $\Delta_{\rm sys}$  & $\mathcal{M}_2$ & $\Delta_{\rm tot}(\mathcal{M}_{2})$ & $\mathcal{S},\ \mathcal{S}^{-1}$ & $\theta_c$ & $\Delta\theta$ & $\mathcal{M}_2$ & $\Delta_{\rm tot}(\mathcal{M}_{2})$ & $\mathcal{S},\ \mathcal{S}^{-1}$ & $\theta_c$ & $\Delta\theta$\\
            \hline
            \multirow{2}[2]{*}{$\boldsymbol{(0, 0)}$}
                & 1$\%$ & 0.58 & 0.040 & $>5\sigma, 2.4\%$ & $91^\circ$ & $48^\circ$ & 0.66 & 0.058 & $>5\sigma, 3.0\%$ & $141^\circ$ & $48^\circ$\\
            \cline{2-12}
            & 2$\%$ & 0.60 & 0.061 & $>5\sigma, 3.5\%$ & $91^\circ$ & $35^\circ$ & 0.66 & 0.070 & $>5\sigma, 3.7\%$ & $141^\circ$ & $48^\circ$\\
            \cline{1-12}
            \multirow{2}[2]{*}{$\boldsymbol{(+0.8, -0.6)}$}
                & 1$\%$ & 0.81 & 0.035 & $>5\sigma, 1.5\%$ & $99^\circ$ & $48^\circ$ & 0.78 & 0.067 & $>5\sigma, 3.0\%$ & $145^\circ$ & $40^\circ$\\
            \cline{2-12}
            & 2$\%$ & 0.82 & 0.052 & $>5\sigma, 2.2\%$ & $99^\circ$ & $36^\circ$ & 0.78 & 0.077 & $>5\sigma, 3.4\%$ & $145^\circ$ & $40^\circ$\\
            \cline{1-12}
            \multirow{2}[2]{*}{$\boldsymbol{(-0.8, +0.6)}$}
                &1$\%$ & 0.79 & 0.029 & $>5\sigma, 1.3\%$ & $92^\circ$ & $45^\circ$ & 0.77 & 0.039 & $>5\sigma, 1.7\%$ & $137^\circ$ & $40^\circ$\\
            \cline{2-12}
            & 2$\%$& 0.79 & 0.049 & $>5\sigma, 2.1\%$ & $92^\circ$ & $34^\circ$ & 0.78 & 0.055 & $>5\sigma, 2.4\%$ & $137^\circ$ & $37^\circ$\\
            \cline{1-12}
            \hline
        \end{tabular}
    }
\end{table*}

As discussed in Sec.~\ref{subsec:tt}, entangled states cannot be observed near the $360$ GeV threshold.
This situation improves significantly as the collision energy increases and at $\sqrt{s}=1~\text{TeV}$ with an integrated luminosity $\mathcal{L}=5\ \mathrm{ab^{-1}}$, $5\sigma$ significance can easily be achieved.
Furthermore, applying an optimized angular window cut allows the measurement precision to reach below $5\%$.
Moreover, beam polarization does not enhance the maximal value of entanglement.
However, it can significantly modify the production cross section in $e^+e^- \to t\bar{t}$.
This change in the cross section leads to an improvement in the overall measurement precision.
A comparison of Tables~\ref{table:C&B-tt} shows that configuration $(P_{e^-},P_{e^+})=\mathbf{(-0.8,+0.6)}$ provides improved  significance for entanglement measurements.

The kinematic distribution of the quantum computational resource $\mathcal{M}_2$ requires dedicated measurements in different angular regions to maximize sensitivity. As shown in Fig.~\ref{fig:magic-mm}, $\mathcal{M}_2$ typically exhibits extremal behavior at central and backward scattering angles. Consequently, the results in Table~\ref{table:m2-tt} are presented by separately optimizing the cuts in the central and backward regions.
In $\sqrt{s}=1~\mathrm{TeV}$ with $\mathcal{L}=5\ \mathrm{ab^{-1}}$, the measured $\mathcal{M}_2$ universally achieves significance well exceeding $5\sigma$ in all polarization and angular configurations.
The introduction of longitudinal beam polarization significantly enhances the magnitude of the measured magic.
For the unpolarized case, the backward region yields a slightly higher magic value compared to the central region.
The sensitivity gain from beam polarization is even more pronounced here, increasing $\mathcal{M}_2$ from $0.58$ of the unpolarized process to $0.81$ of the polarized process in the central scattering region, which is a substantial enhancement of nearly $40\%$. The best precision ($\mathcal{S}^{-1} \approx 1.3\%$) is achieved in the central region with the $\mathbf{(-0.8, +0.6)}$ configuration.

%%%%%%%%%%%%%%%%%%%%%%%%%%%%%%%%%%%%%%%%%%%%%%%%%%%%%%%%%%%%%%%%%%%%%%
\subsection{\texorpdfstring{$e^+e^-\to\mu^+\mu^-$}{ee>mumu}}
%%%%%%%%%%%%%%%%%%%%%%%%%%%%%%%%%%%%%%%%%%%%%%%%%%%%%%%%%%%%%%%%%%%%%%
Since muons do not decay within the detector, we take advantage of it and use the kinematic approach~\cite{Cheng:2024rxi} to estimate statistical sensitivities.
In this approach, the event-by-event observables are functions of the measured scattering angle
$\theta$ and c.m. energy $\sqrt{s}$.
The Fano coefficients are then estimated by sample averages,
and the statistical uncertainty is determined by the distribution of the measured quantities and is derived from the variances of the observed distributions
\begin{equation}
    \Delta B_i^\pm=\frac{\sqrt{\operatorname{Var}\left(B_{i}^\pm\right)}}{\sqrt{N}}, \quad
    \Delta C_{i j}=\frac{\sqrt{\operatorname{Var}\left(C_{i j}\right)}}{\sqrt{N}}
\end{equation}

The the large projected luminosity is for electroweak precision measurements targeting accuracies of $\mathcal{O}(10^{-5})$ for key observables. Thus, we adopt more aggressive benchmark systematics $\Delta_{\rm sys}=0.1\%$ and $0.5\%$ motivated by the expected precision electroweak program in future $e^+e^-$ colliders.

\begin{table*}[tb]
    \caption{Projected significance ($\mathcal{S}$) and precision ($\mathcal{S}^{-1}$) for observing entanglement ($\mathcal{C}$) and Bell nonlocality ($\mathcal{B}$) in unpolarized and polarized $e^+e^-\to \mu^+\mu^-$. Results are shown for $\sqrt{s}=m_Z$ and 250 GeV, with an integrated luminosity $50\ \mathrm{ab^{-1}}$ and $5\ \mathrm{ab^{-1}}$,  respectively. Three polarization configurations $(P_{e^-}, P_{e^+})$ and corresponding benchmark systematic uncertainties ($\Delta_{\rm sys}$) are included. }
    \label{table:C&B-mm}
    \centering
    \tabcolsep=0.15cm
    \renewcommand\arraystretch{1.2}
    \resizebox{\textwidth}{!}{
        \begin{tabular}{|c|c|c||c|c|c||c|c|c|}
            \hline
            \multicolumn{3}{|c||}{$\boldsymbol{e^+e^-\to \mu^+\mu^-}$} & \multicolumn{3}{c||}{\textbf{Entanglement Measure ($\mathcal{C}$)}} & \multicolumn{3}{c|}{\textbf{Bell Nonlocality ($\mathcal{B}-2$)}} \\
            \hline
            $\sqrt{s}\ (\mathrm{GeV})$ & $\boldsymbol{(P_{e^-}, P_{e^+})}$ & $\Delta_{\rm sys}$  & $\mathcal{C}$ & $\Delta_{\rm tot}( \mathcal{C})$ & $\mathcal{S},\ \mathcal{S}^{-1}$  & $\mathcal{B}-2$ & $\Delta_{\rm tot}( \mathcal{B})$ & $\mathcal{S},\ \mathcal{S}^{-1}$ \\
            \hline
            \multirow{6}[6]{*}{$\boldsymbol{m_Z}$}
            & \multirow{2}[2]{*}{$\boldsymbol{(0, 0)}$}
                & $0.1 \%$ & $0.50 $ & $ 7\times10^{-7} $ & $>5\sigma,1.4\times10^{-6}$  & $0.23$ & $6\times10^{-7}$ & $>5\sigma, 2.6\times10^{-6}$  \\
            \cline{3-9}
            & & $0.5\%$ & $0.50$ & $1.8\times10^{-3}$ & $>5\sigma,0.35\%$ & $0.23$ & 0.009 & $>5\sigma,4\%$  \\
            \cline{2-9}
            & \multirow{2}[2]{*}{$\boldsymbol{(+0.8, -0.6)}$}
                & $0.1\%$ & $0.49$ & $ 7\times10^{-7} $ & $>5\sigma,1.4\times10^{-6}$  & $0.23$ & $6\times10^{-7}$ & $>5\sigma, 3\times10^{-6}$  \\
            \cline{3-9}
            & & $0.5\%$ & $0.49$ & $1.7\times10^{-3}$ & $>5\sigma,0.35\%$&  $0.23$ & 0.009 & $>5\sigma,4\%$  \\
            \cline{2-9}
            & \multirow{2}[2]{*}{$\boldsymbol{(-0.8, +0.6)}$}
                & $0.1\%$ & $0.51$ & $ 6\times10^{-7} $ & $>5\sigma,1.1\times10^{-6}$  & $0.25$ & $5\times10^{-7}$ & $>5\sigma, 2.1\times10^{-6}$  \\
            \cline{3-9}
            & & $0.5\%$ & $0.51$ & $1.8\times10^{-3}$ & $>5\sigma,0.35\%$& $0.25$ & 0.009 & $>5\sigma,4\%$ \\
            \hline\hline
            \multirow{6}[6]{*}{ \textbf{250}}
            & \multirow{2}[2]{*}{\textbf{(0, 0)}}
                & $1\%$ & $0.37$ & 0.0026 & $>5\sigma$, $0.7\%$& $0.13$ & 0.019 & $>5\sigma$, $14\%$ \\
            \cline{3-9}
            & & $2\%$ & $0.37$ & 0.0052 & $>5\sigma$, $1.4\%$& $0.13$ & 0.038 & $3.5 \sigma$ \\
            \cline{2-9}
            & \multirow{2}[2]{*}{$\boldsymbol{(+0.8, -0.6)}$}
                & $1\%$ & $0.38$ & 0.0027 & $>5\sigma$, $0.7\%$& $0.14$ & 0.019 & $>5\sigma$, $14\%$ \\
            \cline{3-9}
            & & $2\%$ & $0.38$ & 0.0053 & $>5\sigma$, $1.4\%$& $0.14$ & 0.038 & $3.7 \sigma$ \\
            \cline{2-9}
            & \multirow{2}[2]{*}{$\boldsymbol{(-0.8,\  +0.6)}$}
                & $1\%$ & $0.36$ & 0.0025 & $>5\sigma$, $0.7\%$&  $0.12$ & 0.019 & $>5\sigma$, $15\%$ \\
            \cline{3-9}
            & & $2\%$ & $0.36$ & 0.0051 & $>5\sigma$, $1.4\%$&  $0.12$ & 0.038 & $3.3 \sigma$ \\
            \hline
        \end{tabular}
    }
\end{table*}
Consequently, the measurement range of the scattering angle for the $\mu^+\mu^-$ system is evaluated based on the detector’s acceptance requirement,
\(10^\circ\le \theta \le 170^\circ\), and all tomographic observables are evaluated within this region. In this way, the reconstructed density matrix reflects the experimentally accessible ensemble rather than an idealized full–phase-space state.

We calculate the significance of observing the entanglement $\mathcal{S}(\mathcal{C})$ and Bell nonlocality $\mathcal{S}(\mathcal{B})$ in the unpolarized process $e^+e^- \to \mu^+\mu^-$ and two polarization schemes, $(P_{e^-},P_{e^+})=\mathbf{(0.8,-0.6)}$ and $\mathbf{(-0.8,0.6)}$, at the $Z$ pole with an integrated luminosity $\mathcal{L}=50\ \mathrm{ab^{-1}}$ and 250 GeV with $\mathcal{L}=5\ \mathrm{ab^{-1}}$, as summarized in Table~\ref{table:C&B-mm}.

The expected significance for observing entanglement in the process $e^+e^- \to \mu^+\mu^-$ reaches the $5\sigma$ level at operating energies of future $e^+e^-$ colliders.
The maximum values of both the concurrence and the Bell nonlocality are obtained at the $Z$ pole.
The analysis confirms that the underlying quantum structure of $e^+e^-\to\mu^+\mu^-$ remains robust, similar to the findings for $e^+e^- \to t\bar{t}$. The observed values for $\mathcal{C}$ and $\mathcal{B}$ show that beam polarization does not enhance the inherent quantum structure of the $\mu^+\mu^-$ system itself.
Table~\ref{table:C&B-mm} shows that all polarization choices achieve comparable precision in measuring maximal entanglement. Although the configuration $(P_{e^-},P_{e^+})=\mathbf{(-0.8,+0.6)}$ produces a larger production cross section and hence a smaller statistical uncertainty, the overall sensitivity in our study is dominated by the assumed systematic uncertainty. Nevertheless, given the polarization dependence of the cross section, the choice of $(P_{e^-},P_{e^+})=\mathbf{(-0.8,+0.6)}$ may produce improved entanglement measurements in this channel, once realistic configuration-dependent systematics are taken into account.

\begin{table}[tb]
    \caption{Projected significance ($\mathcal{S}$) and precision ($\mathcal{S}^{-1}$) for $\mathcal{M}_2$ in  unpolarized and polarized $e^+e^-\to \mu^+\mu^-$ at $\sqrt{s}=m_Z$ and 250 GeV, with an integrated luminosity $50\ \mathrm{ab^{-1}}$ and $5\ \mathrm{ab^{-1}}$, respectively. Results are presented across forward and backward scattering regions under varying systematic uncertainties ($\Delta_{\rm sys}$). }
    \label{table:m2-mm}
    \centering
    \tabcolsep=0.15cm
    \renewcommand\arraystretch{1.2}
    \resizebox{\textwidth}{!}{
    \begin{tabular}{|c|c|c||c|c|c||c|c|c|}
        \hline
        \multicolumn{3}{|c||}{$\boldsymbol{e^+e^-\to \mu^+\mu^-}$} & \multicolumn{3}{c||}{\textbf{Forward or Central Region }} & \multicolumn{3}{c|}{\textbf{Backward Region }} \\
        \hline
        $\sqrt{s}\ (\mathrm{GeV})$ & $\boldsymbol{(P_{e^-}, P_{e^+})}$ & $\Delta_{\rm sys}$ & $\mathcal{M}_2$ & $\Delta_{\rm tot}(\mathcal{M}_{2})$ & $\mathcal{S},\ \mathcal{S}^{-1}$ & $\mathcal{M}_2$ & $\Delta_{\rm tot}(\mathcal{M}_{2})$ & $\mathcal{S},\ \mathcal{S}^{-1}$ \\
        \hline
        \multirow{6}[6]{*}{{$\boldsymbol{m_Z}$}}
        & \multirow{2}[2]{*}{$\boldsymbol{(0, 0)}$}
            & $0.1 \%$ & $0.32$ & $ 0.0016 $ & $>5\sigma,0.5\%$& 0.21 & 0.002 &$>5\sigma,0.7\%$\\
        \cline{3-9}
        & & $0.5 \%$ & $0.32$ & $0.008$ & $>5\sigma,2.4\%$&0.21 &0.008 &$>5\sigma,3.7\%$\\
        \cline{2-9}
        & \multirow{2}[2]{*}{$\boldsymbol{(+0.8, -0.6)}$}
            & $0.1 \%$ & $0.44$ & $ 0.0015 $ & $>5\sigma,0.3\%$&0.32 &0.001 &$>5\sigma,0.4\%$\\
        \cline{3-9}
        & & $0.5 \%$ & $0.44$ & $0.007$ & $>5\sigma,1.7\%$&0.32 &0.007 &$>5\sigma,2.1\%$\\
        \cline{2-9}
        & \multirow{2}[2]{*}{$\boldsymbol{(-0.8, +0.6)}$}
            & $0.1 \%$& $0.41$ & $ 0.0014 $ & $>5\sigma,0.3\%$&0.42 &0.001 &$>5\sigma,0.3\%$\\
        \cline{3-9}
        & & $0.5 \%$& $0.41$ & $0.007$ & $>5\sigma,1.7\%$&0.42 &0.007 &$>5\sigma,1.7\%$\\
        \hline\hline
        \multirow{6}[6]{*}{\textbf{{250}}}
        & \multirow{2}[2]{*}{$\boldsymbol{(0, 0)}$}
            & $1 \%$& $0.13$ & 0.016 & $>5\sigma, 12.1\%$& 0.27 & 0.014 &$>5\sigma,5.1\%$\\
        \cline{3-9}
        & & $2 \%$& $0.13$ & 0.031 & $4.1\sigma$ & 0.27 & 0.027 &$>5\sigma,10.3\%$\\
        \cline{2-9}
        & \multirow{2}[2]{*}{$\boldsymbol{(+0.8, -0.6)}$}
            & $1 \%$& $0.27$ & 0.014 & $>5\sigma, 5.0\%$&0.42 &0.014 &$>5\sigma,3.3\%$\\
        \cline{3-9}
        & & $2 \%$& $0.27$ & 0.027 & $>5\sigma, 10.1\%$&0.42 &0.028 &$>5\sigma,6.7\%$\\
        \cline{2-9}
        & \multirow{2}[2]{*}{$\boldsymbol{(-0.8, +0.6)}$}
            & $1 \%$& $0.23$ & 0.014 & $>5\sigma, 6.0\%$&0.40 &0.014 &$>5\sigma,3.5\%$\\
        \cline{3-9}
        & & $2 \%$& $0.23$ & 0.028 & $>5\sigma, 12.0\%$&0.40 &0.028 &$>5\sigma,7.1\%$\\
        \hline
    \end{tabular}
    }
\end{table}

In contrast to $\mathcal{C}$ and $\mathcal{B}$, the quantum computational resource $\mathcal{M}_2$ is very sensitive to the initial beam polarization, confirming its utility as a unique and tunable quantum observable.
To optimize the measurement of $\mathcal{M}_2$ for the $e^+e^- \to \mu^+\mu^-$ process within the detector-accessible scattering range ($10^\circ \le \theta \le 170^\circ$), we divide the phase space into forward (or central) and backward regions, depending on the collider energies. We choose the forward/central region with the division at $\theta <0.47\pi$ for $Z$ pole, and shifting to $\theta <0.6\pi$ for $\sqrt{s}=250~\mathrm{GeV}$. Our results are shown in Table~\ref{table:m2-mm}.  The beam polarization can significantly modify the magnitude of $\mathcal{M}_2$. For example, at the $Z$ pole the configuration $\mathbf{(+0.8,-0.6)}$ increases $\mathcal{M}_2$ by about $34\%$ compared with the unpolarized case; at $\sqrt{s}=250~\mathrm{GeV}$, the enhancement is even more pronounced, with the polarized result being approximately a factor of two larger than the unpolarized one. Because of the larger scattering cross section, this configuration also provides a noticeable improvement in the achievable measurement sensitivity, yielding the most favorable expected precision in our benchmark scenario. On the other hand, similar to the results in Table~\ref{table:C&B-mm}, the expected precision in $\mathcal{M}_2$ shows only a weak dependence on the polarization configuration, largely due to the dominance of systematic uncertainty.

%%%%%%%%%%%%%%%%%%%%%%%%%%%%%%%%%%%%%%%%%%%%%%%%%%%%%%%%%%%%%%%%%%%%%%
\subsection{\texorpdfstring{$e^+e^-\to e^+e^-$}{ee>ee}}

Bhabha scattering $e^+e^-\to e^+e^-$ provides a unique platform for quantum information studies due to the inherent complexity introduced by the simultaneous presence of $s/t$-channel and $\gamma/Z$ exchanges. Analyzing this process allows us to probe the intricate interplay between the entanglement-generating $s$-channel amplitude and the entanglement-suppressing $t$-channel amplitude.
Based on the results in Sec.~\ref{subsec:ee}, we focus on measuring the QI observables at the $Z$ pole. With the kinematic approach, we compare the projected significance ($\mathcal{S}$) in an unpolarized process and two highly polarized configurations that can be achieved at future colliders, $\mathbf{(+0.8, -0.6)}$ and $\mathbf{(-0.8, +0.6)}$.
\begin{table*}[tb]
    \caption{Projected significance ($\mathcal{S}$) for observing entanglement ($\mathcal{C}$) and Bell nonlocality ($\mathcal{B}$) in Bhabha scattering at $Z$ pole with an integrated luminosity $50\ \mathrm{ab^{-1}}$. The results with systematic uncertainty $\Delta_{\rm sys}=0.1\%$ and $0.5\%$ are presented for three different initial beam polarization configurations $(P_{e^-}, P_{e^+})$. }
    \label{table:C&B-ee}
    \centering
    \tabcolsep=0.18cm
    \renewcommand\arraystretch{1.2}
    \resizebox{\textwidth}{!}{
        \begin{tabular}{|c|c||c|c|c|c||c|c|c|c|}
            \hline
            \multicolumn{2}{|c||}{$\boldsymbol{\sqrt{s}= m_Z}$} & \multicolumn{4}{c||}{\textbf{Entanglement Measure ($\mathcal{C}$)}} & \multicolumn{4}{c|}{\textbf{Bell Nonlocality ($\mathcal{B}-2$)}} \\
            \hline
            $\boldsymbol{(P_{e^-}, P_{e^+})}$ & $\Delta_{\rm sys}$  & $\mathcal{C}$ & $\Delta_{\rm tot}( \mathcal{C})$ & $\mathcal{S},\ \mathcal{S}^{-1}$ & $\theta$  & $\mathcal{B}-2$ & $\Delta_{\rm tot}( \mathcal{B})$ & $\mathcal{S},\ \mathcal{S}^{-1}$ & $\theta$ \\
            \hline
            \multirow{2}[2]{*}{$\boldsymbol{(0, 0)}$}
            & $0.1 \%$ & $0.56$ & 0.0006 & $>5\sigma,0.1\%$ & \multirow{6}[6]{*}{$(0.3\pi,0.9\pi)$}  & 0.35 & 0.0016 & $>5\sigma,0.5\%$ & \multirow{6}[6]{*}{$(0.4\pi,0.8\pi)$}  \\
            \cline{2-5}\cline{7-9}
            & $0.5 \%$ & $0.56$ & 0.0031 & $>5\sigma,0.6\%$ &  & 0.35 & 0.0082 & $>5\sigma,2.3\%$ &  \\
            \cline{1-5}\cline{7-9}
            \multirow{2}[2]{*}{$\boldsymbol{(+0.8, -0.6)}$}
            & $0.1 \%$ & $0.59$ & 0.0007 & $>5\sigma,0.1\%$ &  &  0.39 & 0.0017 & $>5\sigma,0.4\%$ &    \\
            \cline{2-5}\cline{7-9}
            & $0.5 \%$ & $0.59$ & 0.0034 & $>5\sigma,0.6\%$ & &  0.39 & 0.0085 & $>5\sigma,2.2\%$ &  \\
            \cline{1-5}\cline{7-9}
            \multirow{2}[2]{*}{$\boldsymbol{(-0.8, +0.6)}$}
                & $0.1 \%$ & $0.61$ & 0.0007 & $>5\sigma,0.1\%$ &   &  0.45 & 0.0017 & $>5\sigma,0.4\%$ &   \\
            \cline{2-5}\cline{7-9}
            & $0.5 \%$ & $0.61$ & 0.0035 & $>5\sigma,0.6\%$ & &  0.45 & 0.0085 & $>5\sigma,1.9\%$ &   \\
            \hline
        \end{tabular} }
\end{table*}

The results summarized in Table~\ref{table:C&B-ee} indicate that observing both entanglement and Bell nonlocality is highly feasible at the $Z$ pole with $\mathcal{L}=50~\mathrm{ab^{-1}}$. The projected significance exceeds $5\sigma$ for all polarization configurations and for all considered systematic-uncertainty benchmarks. To optimize the measured $\mathcal{C}$ and $\mathcal{B}$, the analysis is selected for the angular windows, $\theta \in (0.3\pi, 0.9\pi)$ for $\mathcal{C}$ and $\theta \in (0.4\pi, 0.8\pi)$ for $\mathcal{B}$. Within these selections, the reconstructed spin density matrix of the event ensemble is entangled and Bell nonlocal.

The Bhabha process exhibits a strong dependence on longitudinal beam polarization. The unpolarized result serves as a baseline, yielding $\mathcal{C}=0.56$ and a Bell violation margin ($\mathcal{B}-2$) of $0.35$. The application of polarization significantly enhances both the magnitude of the observables and the corresponding measurement precision.
The optimal $\mathbf{(-0.8, +0.6)}$ configuration maximizes the entanglement, $\mathcal{C}=0.61$ and $\mathcal{B}=2.45$. This represents an improvement of approximately $10\%$ for  $\mathcal{C}$ and a substantial increase of $28\%$ in $\mathcal{B}-2$, relative to the unpolarized baseline.
These results confirm that, unlike simple $s$-channel processes, Bhabha scattering is highly sensitive to initial state preparation, validating the role of polarization as a crucial mechanism for isolating and enhancing the desired quantum correlation signals.

\begin{table}[tb]
    \caption{Projected significance ($\mathcal{S}$) and precision ($\mathcal{S}^{-1}$) for the $\mathcal{M}_2$ in Bhabha scattering at $Z$ pole with an integrated luminosity $50\ \mathrm{ab^{-1}}$, measured for three different polarization configurations $(P_{e^-}, P_{e^+})$ with two benchmark systematic uncertainties $\Delta_{\rm sys}$. }
    \label{table:m2-ee}
    \centering
    \tabcolsep=0.15cm
    \renewcommand\arraystretch{1.2}
    \begin{tabular}{|c|c|c||c|c|c|c|}
        \hline
        \multicolumn{3}{|c||}{$\boldsymbol{e^+e^-\to e^+e^-}$} & \multicolumn{4}{c|}{\textbf{$\mathcal{M}_2$ Measurement}} \\
        \hline
        $\boldsymbol{\sqrt{s}}$ (GeV) & $\boldsymbol{(P_{e^-}, P_{e^+})}$ & $\Delta_{\rm sys}$ & $\mathcal{M}_2$ & $\Delta_{\rm tot}(\mathcal{M}_{2})$ & $\mathcal{S},\ \mathcal{S}^{-1}$ & $\theta$\\
        \hline
        \multirow{6}[6]{*}{$\boldsymbol{m_Z}$}
        & \multirow{2}[2]{*}{$\boldsymbol{(0, 0)}$}
            & $0.1 \%$& 0.53 & 0.0006 & $>5\sigma,0.1\%$&  \\
        \cline{3-6}
        & & $0.5 \%$& 0.53 & 0.0029 & $>5\sigma,0.5\%$& \multirow{2}[6]{*}{($\pi/4, \pi/2$)}\\
        \cline{2-6}
        & \multirow{2}[2]{*}{$\boldsymbol{(+0.8, -0.6)}$}
            & $0.1 \%$& 0.54 & 0.001 & $>5\sigma, 0.2\%$& \\
        \cline{3-6}
        & & $0.5 \%$& 0.54 & 0.005 & $>5\sigma, 1.0\%$& \\
        \cline{2-6}
        & \multirow{2}[2]{*}{$\boldsymbol{(-0.8, +0.6)}$}
            & $0.1 \%$& 0.54 & 0.001 & $>5\sigma, 0.2\%$&\\
        \cline{3-6}
        & & $0.5 \%$& 0.54 & 0.005 & $>5\sigma, 1.0\%$& \\
        \hline
    \end{tabular}
\end{table}
The magic observable $\mathcal{M}_2$ in Bhabha scattering is highly sensitive to longitudinal polarization. This sensitivity comes from the $s$-channel, which generates entanglement, and the $t$-channel, which suppresses it. Beam polarization acts as the primary experimental mechanism required to control the resulting quantum state.

To maximize the observable signal, the analysis is performed within the kinematically optimized angular window $\theta \in (\pi/4, \pi/2)$, where $\mathcal{M}_2$ reaches its maximum theoretical magnitude. The results in Table~\ref{table:m2-ee} confirm that the observable is present at the $Z$ pole with $\mathcal{L}=50~\mathrm{ab^{-1}}$ in all polarization scenarios, with a significance consistently exceeding $5\sigma$.

We find that using a partially polarized configuration, such as $\mathbf{(-0.8, +0.6)}$, maximizes the effective cross section for coherent production. This optimization results in a measurable increase in $\mathcal{M}_2$ by approximately $3\%$ compared to the unpolarized state.
More significantly, the presence of longitudinal polarization fundamentally enhances the non-stabilizer properties of the generated state. This improves the measurement precision and validates the role of polarization in the preparation of quantum computational resources.

%%%%%%%%%%%%%%%%%%%%%%%%%%%%%%%%%%%%%%%%%%%%%%%%%%%%%%%%%%%
\section{Summary and Conclusions}
\label{sec:summary}
%%%%%%%%%%%%%%%%%%%%%%%%%%%%%%%%%%%%%%%%%%%%%%%%%%%%%%%%%%%%%

The next generation of high-luminosity $e^+e^-$ colliders, including the ILC, FCC-ee, CEPC, and CLIC, will offer unprecedented opportunities to perform precision physics measurements and to measure quantum information observables at the energy frontier.
Longitudinal beam polarization provides a control parameter that can both improve statistical sensitivity and reshape the spin density matrix of the final state.

In this work, after a comprehensive introduction to the topics in Section \ref{sec:intro}, we first establish a unified tomographic framework for the production of fermion pairs $e^+e^-\to f\bar f$ as a two-qubit system in Section \ref{sec:observables}, focusing on three representative quantum  observables: entanglement $\mathcal{C}$, Bell nonlocality $\mathcal{B}$, and the second stabilizer R\'enyi entropy $\mathcal{M}_2$ as a proxy for magic in the spin subspace.
The formalism is applicable to other two-qubit states and easily generalizable to more complex systems and production mechanisms.

We next give a comprehensive presentation in Section \ref{sec:eecollider} to lay out the quantum behavior for the processes $e^+e^-\!\to t\bar{t}$, $e^+e^-\!\to \mu^+\mu^-$, and high-energy Bhabha scattering $e^+ e^- \to e^+ e^-$. These three processes represent distinct features: heavy mass threshold behavior in $t\bar t$ final state, $Z$ pole resonant production in the $\mu^+\mu^-$ and $e^+e^-$ processes, and the $s/t$ channel interplay in $e^+e^-$ Bhabha scattering. In particular, we show the significant effects of the polarizations of the initial beams on the QI structure. Our analysis reveals distinct mechanisms behind the polarization dependence of these observables, as summarized below.

\noindent
$\bullet$
For the $s$-channel processes $t\bar{t}$ and $\mu^+\mu^-$ studied in Section~\ref{subsec:tt} and \ref{subsec:mumu}, $\mathcal{C}$ and $\mathcal{B}$ are rather insensitive to longitudinal polarization, as seen in Figs.~\ref{fig:tt-C&B} and \ref{fig:mm-C&B}. This is because both are controlled mainly by the normalized spin–correlation matrix $C_{ij}$, whose structure is only weakly affected by the reweighting of the contributing helicity amplitudes by the polarizations
$(P_{e^-},P_{e^+})$.
In these channels, the beam polarization therefore acts primarily as a scaling factor for the total cross section, rather than as a handle that qualitatively reshapes the intrinsic spin correlations.

%\noindent
%$\bullet$
In contrast, the magic observable $\mathcal{M}_2$ is intrinsically sensitive to beam polarization, making it tunable by controlled polarization.
By construction, $\mathcal{M}_2$ depends nonlinearly on the full density matrix, including the single-particle polarization vectors $B_i^\pm$. Non-zero magic arises once the reconstructed state departs from the stabilizer-like structure; see results in Figs.~\ref{fig:Magic-tt}, \ref{fig:magic-contour-tt}, \ref{fig:magic-mm}, \ref{fig:magic-contour-mm}, \ref{fig:magic-ee}, \ref{fig:ee-magic-contour} and \ref{fig:eeM2max}.
Longitudinal polarization can significantly enhance $\mathcal{M}_2$ by driving the reduced spin state away from the stabilizer structure, leading to magic $\mathcal{M}_2 > 0$, as seen in Fig.~\ref{fig:Magic-tt}(b). In a different aspect, the kinematic regions with near-maximal entanglement approach stabilizer Bell states in this frame, characterized by vanishing single-particle polarizations $B_i^\pm$ and an approximately diagonal correlation matrix $C_{ij}$ with entries of magnitude close to one. In these regions, we find $\mathcal{M}_2\simeq 0$, as seen in Fig.~\ref{fig:magic-mm}.

\noindent
$\bullet$
The interplay of the $s$- and $t$-channel contributions in Bhabha scattering, $e^+e^-\!\to e^+e^-$, leads to quantum-information features that differ qualitatively from the $s$-channel processes. As discussed in Section~\ref{subsec:ee}, longitudinal polarization reweights the relative helicity amplitudes in the $s$- and $t$-channels, thereby reshaping the reconstructed spin density matrix and the full correlation tensor $C_{ij}$ ({\it e.g.} Figs.~\ref{fig:ee-cij&bi} and \ref{fig:ee-Fano}).
The $t$-channel contribution is strongly enhanced in the forward/backward region and dominates the cross section away from the $Z$ pole. As such, the unpolarized ensemble becomes decoherent due to the large contributions $LL/RR$ to a separable state.
However, for polarized configuration $RL/LR$ with $s$-channel and $s/t$-channel interference, the reconstructed two-spin state can still approach a maximally entangled Bell-like configuration in certain kinematic regions, as shown in Figs.~\ref{fig:ee-C&B} and \ref{fig:ee_contour_C&B}.

Regarding magic, Bhabha scattering shares the general pattern observed in $t\bar t$ and $\mu^+\mu^-$, namely, sizeable $\mathcal{M}_2$ is typically obtained away from Bell-like (stabilizer) points and can be efficiently accessed by tuning the beam polarization. In addition, the presence of $s/t$ interference makes a ``separable-magic'' domain with $\mathcal{C}=0$ but $\mathcal{M}_2>0$,
see Figs.~\ref{fig:magic-ee} and \ref{fig:ee-magic-contour}.

\noindent
$\bullet$
%(3).
In Sec.~\ref{subsec:SecIII-summary} we synthesize and interpret the key patterns identified in $t\bar t$, $\mu^+\mu^-$, and Bhabha scattering, providing a
unified picture of quantum resources across kinematics and polarization.
We demonstrate explicitly that entanglement and magic are generically non-monotonic. $\mathcal{M}_2$ exhibits a characteristic twin-peak pattern flanking the maximal entanglement region,  is typically obtained away from maximal entanglement points, and can be efficiently accessed by using beam polarization to shift the reconstructed density matrix into the non-stabilizer sector.

From a broader perspective, the observed polarization dependence is suggestive in the language of ``fermionic resource theories''  \cite{Sierant:2025fax,Leone:2024lfr,Chitambar:2018rnj,Emerson:2013zse}: polarization reweights the initial helicity composition and thus the relative importance of chiral electroweak couplings, leading to appreciable changes in the departure from classically simulable sets within the reduced spin sector.
In this sense, an $e^+e^-$ collision could be viewed as a controllable quantum channel acting on the spin density matrix, transforming a polarization-prepared two-qubit state into a final-state spin resource. We emphasize that connecting $\mathcal{M}_2$ to a fully fledged fermionic resource theory requires a dedicated study beyond this work.

Finally, as presented in Section \ref{sec:Significance}, we evaluate the observability of these quantum observables and find that future $e^+e^-$ colliders can provide a highly desirable environment for probing these QI observables under the design projections for luminosity and beam polarization. With the current designs of the collider center-of-mass energy and luminosity, and optimized scattering-angle windows, all three processes ($t\bar{t}$, $\mu^+\mu^-$, and $e^+e^-$) can achieve high expected statistical significance in the observability of entanglement, Bell nonlocality, and magic in the two-qubit state above $5\sigma$ significance and a few percent in the measurement accuracy as presented in Tables~\ref{table:C&B-tt}--\ref{table:m2-ee}.

This work lays the foundation for QI phenomenology in future lepton colliders and identifies
polarization and kinematic regions that optimize sensitivity to $\mathcal{B}, \mathcal{C}$, and $\mathcal{M}_2$. As demonstrated by a recent study to generate entangled states of a fermion pair in the electron-ion collider (EIC)  \cite{Cheng:2025zaw}, transverse beam polarizations become crucially important in $ep$ collisions. A natural extension of our current work would be to examine  the effects of transverse beam polarizations in $e^+e^-$ collisions, as well as the other systems.

\vskip 0.2cm

\noindent
{\bf Note added:}
When this manuscript was being finalized, another related paper appeared on arXiv \cite{Altakach:2026fpl}. While the general results are in agreement, these two papers have rather different emphases on physics topics and are complementary.

%%%%%%%%%%%%%%%%%%%%%%%%%%%%%%%%%%%%%%%%%%%%%%%%%%%%%%%%%%%

\begin{acknowledgments}
The authors thank Kun Cheng and Tong Arthur Wu for helpful discussions.
This work was supported in part by the U.S. Department of Energy under grant No. DE-SC0007914 and in  part by Pitt PACC. YCG was also supported by the China Scholarship Council.

\end{acknowledgments}

\newpage

\appendix
\section{Leading-Order Calculation for a two-qubit \texorpdfstring{$f\bar f$}{ff} system}
\label{sec:AppendixA}

Fermion pair production proceeds via the $s$-channel processes
\begin{equation}
 e^-e^+ \to \gamma^*, Z^* \to f_a \bar{f}_b\ \  (f\ne e),
\end{equation}
where $a,b$ are the spin indices for $f\bar{f}$. We define the production density matrix for each partonic process as
\begin{equation}
    R^{e^-e^+}_{ab,\bar{a}\bar{b}}(s,\theta) \propto \mathcal{M}(e^-e^+\to f_a \bar{f}_b) \mathcal{M}^*(e^-e^+\to f_{\bar{a}} \bar{f}_{\bar{b}}),
\end{equation}
where $s$ is the c.m.~energy squared and $\theta$ is scattering angle of $f$ with respect to the $e^-$ beam. The $4\times 4$ matrix $R$ has a decomposition
\begin{equation}
R(s,\theta) = \frac{1}{4} \bigg( \tilde{A}\, \mathbb{I}_2 \otimes \mathbb{I}_2
+ \sum_i \tilde{B}^{+}_i \sigma_i \otimes \mathbb{I}_2
+ \sum_j \tilde{B}^{-}_j \mathbb{I}_2 \otimes \sigma_j
+ \sum_{ij} \tilde{C}_{ij} \sigma_i \otimes \sigma_j \bigg),
\end{equation}
The spin density matrix $\rho$ in Eq.~(\ref{eq:fano}) is normalilzed by $R(s,\theta)/\tilde{A}$.

For $e^-e^+ \to f_a \bar{f}_b$ with an $s$-channel $\gamma$ and $Z$ exchange and averaging over initial $e^- e^+$ polarizations, the $\tilde{A}$, $\tilde{B}^\pm_i$, and $\tilde{C}_{ij}$ coefficients are given by
\begin{align}
\tilde{A} =& \,  \bigg\{ Q_e^2 Q_f^2 \big[ 2 - \beta^2 \sin^2\theta \big]
+ 2Q_e Q_f \operatorname{Re} \big[ \chi(s) \big]
\Big[ 2\beta g_A^e g_A^f \cos\theta + g_V^e g_V^f \big( 2 - \beta^2 \sin^2\theta \big) \Big] \notag \\
& + \big| \chi(s) \big|^2
\bigg[ \big( g_V^{e2} + g_A^{e2} \big) \Big( 2g_V^{f2} + 2\beta^2 g_A^{f2} - \beta^2
\big( g_V^{f2} + g_A^{f2}\big) \sin^2\theta \Big) + 8\beta g_V^e g_V^f g_A^e g_A^f \cos\theta \bigg]
\bigg\}, \\
\tilde{B}_k^{\pm}= & - 2 \Big\{Q_e Q_f \operatorname{Re} \big[ \chi(s) \big] \Big[\beta g_A^f g_V^e\big(1+\cos ^2 \theta\big)+2 g_A^e g_V^f \cos \theta\Big]\notag \\
& + \big| \chi(s) \big|^2 \Big[2 g_A^e g_V^e\big(\beta^2 g_A^{f 2}+g_V^{f 2}\big) \cos \theta+\beta g_A^f g_V^f\big(g_V^{e 2}+g_A^{e 2}\big)\big(1+\cos ^2 \theta\big)\Big]\Big\}, \\
\tilde{B}_r^{\pm}= & - 2 \sin \theta \sqrt{1-\beta^2}\Big\{Q_e Q_f \operatorname{Re} \big[ \chi(s) \big] \big[\beta g_A^f g_V^e \cos \theta+2 g_A^e g_V^f\big] \notag \\
& \hspace{7.5em} + \big| \chi(s) \big|^2 g_V^f\Big[\beta g_A^f\big(g_V^{e 2}+g_A^{e 2}\big) \cos \theta+2 g_A^e g_V^e g_V^f\Big] \Big\}, \\
\tilde{B}_n^{\pm}= & \,0,
\end{align}
\begin{align}
\tilde{C}_{n n}=&- \beta^2 \sin ^2 \theta\Big\{Q_e^2 Q_f^2+2 Q_e Q_f \operatorname{Re} \big[ \chi(s) \big] g_V^e g_V^f-\big| \chi(s) \big|^2 \big(g_V^{e 2}+g_A^{e 2}\big)\big(g_A^{f 2}-g_V^{f 2}\big)\Big\}, \\
\tilde{C}_{r r} =&  - \sin ^2 \theta \Big\{\big(\beta^2-2\big) Q_e^2 Q_f^2+2 Q_e Q_f \operatorname{Re} \big[ \chi(s) \big] g_V^e g_V^f\big(\beta^2-2\big) \notag \\
& \hspace{4em} +\big| \chi(s) \big|^2 \Big[\beta^2\big(g_A^{f 2}+g_V^{f 2}\big)-2 g_V^{f 2}\Big]\big(g_V^{e 2}+g_A^{e 2}\big)\Big\}, \\
\tilde{C}_{k k} = & \, \bigg\{Q_e^2 Q_f^2\Big[\big(\beta^2-2\big) \sin ^2 \theta+2\Big] \notag\\
& +2 Q_e Q_f \operatorname{Re} \big[ \chi(s) \big] \Big[2 \beta g_A^e g_A^f \cos \theta+g_V^e g_V^f\big((\beta^2-2) \sin ^2 \theta+2\big)\Big] \notag\\
& +\big| \chi(s) \big|^2 \bigg[8 \beta g_A^e g_A^f g_V^e g_V^f \cos \theta
+ \big(g_V^{e 2}+g_A^{e 2}\big)\Big(2 g_V^{f 2} \cos ^2 \theta-\beta^2 \big(g_A^{f 2}-g_V^{f 2}\big) \sin ^2 \theta+2 \beta^2 g_A^{f 2}\Big)\!\bigg]\! \bigg\}, \\
\tilde{C}_{k r} = &\, \tilde{C}_{r k}= -2 \sin \theta \sqrt{1-\beta^2} \Big\{Q_e^2 Q_f^2 \cos \theta+Q_e Q_f \operatorname{Re} \big[ \chi(s) \big] \left[\beta g_A^e g_A^f+2 g_V^e g_V^f \cos \theta\right] \notag \\
& \hspace{3em} +\big| \chi(s) \big|^2 \Big[2 \beta g_A^e g_A^f g_V^e g_V^f+g_V^{f 2}\big(g_V^{e 2}+g_A^{e 2}\big) \cos \theta\Big]\Big\}, \\
\tilde{C}_{nr} = &\, \tilde{C}_{rn} = \tilde{C}_{nk} =\tilde{C}_{kn}=0,
\label{eq:cij}
\end{align}
where $\beta = \sqrt{1-4m_f^2/s}$, and $Q_{\ell}=-1$, $Q_{t}=2/3$ are the electric charges, while $g_V^i$ and $g_A^i$ are the vector and axial-vector couplings given by
\begin{equation}
\label{gagv}
    g_V^i=\frac{1}{2}(g_L^i + g_R^i) = \frac{I_{i}^3}{2}- Q_i \sin ^2 \theta_W, \quad g_A^i= \frac{1}{2}(g_L^i - g_R^i) =\frac{I_{i}^3}{2},
\end{equation}
where $I_i^3$ is the third component of weak isospin: $-1/2$ for charged leptons and down-type quarks,
and $+1/2$ for neutrinos and up-type quarks.
The propagator factor $\chi(s)$ appearing in the interference between photon and $Z$ exchange is given by
\begin{align}
    \mathrm{Re}\!\left[\chi(s)\right]
    &= \frac{s\,(s - m_Z^2)}
       {\sin^2\theta_W \cos^2\theta_W
       \left[(s - m_Z^2)^2 + s^2 \Gamma_Z^2/m_Z^2\right]}, \\
    |\chi(s)|^2
    &= \frac{s^2}
       {\sin^4\theta_W \cos^4\theta_W
       \left[(s - m_Z^2)^2 + s^2 \Gamma_Z^2/m_Z^2\right]},
\end{align}
where $\theta_W$ is the weak mixing angle and $m_Z$ and $\Gamma_Z$ denote the mass and width of the $Z$ boson, respectively.

\newpage

\bibliographystyle{apsrev4-1}
\bibliography{refs}
\end{document}